\documentclass[journal]{IEEEtran}

\usepackage{amsmath}
\usepackage{amsfonts}
\usepackage{epsfig}
\usepackage{bar}
\usepackage{eepic}
\usepackage{verbatim}
\usepackage{pmat}
\usepackage{subfigure}
\usepackage{color}

    \newcommand{\vect}[1]{{\lowercase{\boldsymbol{#1}}}}
    \newcommand{\mat}[1]{{\uppercase{\boldsymbol{#1}}}}

\newcommand{\nn}{\nonumber}
\newcommand{\limzero}[1]{\lim_{#1\rightarrow 0}}
\newcommand{\liminfty}[1]{\lim_{#1\rightarrow \infty}}
\newcommand{ \divergence }[2]{ \mathsf{D} \left( #1 \| #2 \right) }
\newcommand{ \cnddiv }[3]{ \mathsf{D} \left( #1 \| #2 | #3 \right) }

\newcommand{\expb}[1]{ \exp \left[ #1 \right] } 
\newcommand{\trace}[1]{\mathsf{tr}\left\{#1\right\}}
\newcommand{\Exp}{\mathsf E}
\newcommand{\expect}[1]{{\Exp}\left\{#1\right\}}
\newcommand{\expcnd}[2]{{\Exp}\left\{\left. #1 \,\right|\, #2\right\}}

\newcommand{\inv}[1]{#1^{\scriptscriptstyle -\!1}}

\newcommand{\tran}[1]{#1^{\!\top\!}}

\newcommand{\herm}[1]{#1^{\scriptscriptstyle \text{H}\!}}
\newcommand{\diag}{\text{diag}}

\newcommand{\eref}[1]{(\ref{#1})}
\newcommand{\half}{\frac{1}{2}} 
\newcommand{\oneon}[1]{\frac{1}{#1}}

\newcommand{\e}{\vect{e}}

\newcommand{\s}{\vect{s}}

\newcommand{\x}{\vect{x}}
\newcommand{\y}{\vect{y}}

\newcommand{\eT}{\tran{\vect{e}}}

\newcommand{\sT}{\tran{\vect{s}}}

\newcommand{\A}{\mat{A}}

\newcommand{\I}{\mat{I}}

\newcommand{\N}{\mat{N}}

\newcommand{\Q}{\mat{Q}}

\renewcommand{\S}{\mat{S}}

\newcommand{\V}{\mat{V}}

\newcommand{\X}{\mat{X}}
\newcommand{\Y}{\mat{Y}}

\newcommand{\XH}{\herm{\X}} 

\newcommand{\ST}{\tran{\mat{S}}}

\newcommand{\XT}{\tran{\mat{X}}}

\def\myfigwidth{3.5in}
\def\myhalfwidth{3.4in}

\newtheorem{theorem}{Theorem}
\newtheorem{claim}{Claim}
\newtheorem{corollary}{Corollary}
\newtheorem{proposition}{Proposition}
\newtheorem{lemma}{Lemma}

\bstctlcite{BSTcontrol} 

\newcommand{\nsp}[1]{\hspace{-#1ex}}
\newcommand{\nind}{\hspace{-6ex}}

\newcommand{\pd}[1]{ {\frac{ \intd }{ \intd {#1} }} }
\newcommand{\mai}{I}
\newcommand{\fe}{{\mathcal{F}}} 
\newcommand{\energy}{{\mathcal{E}}} 
\newcommand{\entropys}{{\mathcal{S}}} 
\newcommand{\intd}{{\,\normalfont{\text d}}} 
\newcommand{\eff}{{ \mathsf{\eta} }} 
\newcommand{\mse}{{ \mathcal{E} }} 
\newcommand{\mmse}{ {\mathsf{mmse}} } 
\newcommand{\vrc}{{ \mathcal{V} }} 
\newcommand{\capacity}{{ \mathsf{C} }} 
\newcommand{\Csep}{ \capacity_{\normalfont{\text{sep}}} }
\newcommand{\Cjnt}{ \capacity_{\normalfont{\text{joint}}} }
\newcommand{\snr}{{ \mathsf{snr} }} 
\newcommand{\supn}{{ \normalfont{\text{(lmmse)}} }} 
\newcommand{\supb}{{ \normalfont{\text{(b)}} }} 
\newcommand{\sr}{{ \text{(r)} }} 
\newcommand{\si}{{ \text{(i)} }} 
\newcommand{\mSigma}{ \mathbf{\Sigma} } 
\newcommand{\amp}[1]{ \sqrt{\snr_{#1}} \, } 
\newcommand{\Snr}{ \A }

\newcommand{\XX}{ \underline{\X} } 
\newcommand{\Xa}{ \underline{\X}_a } 
\newcommand{\Iu}{ I^{(u)} }
\newcommand{\Gu}{ G^{(u)} }
\newcommand{\GuK}{ \Gu_K }
\newcommand{\muu}{ \mu^{(u)} }
\newcommand{\muuK}{ \muu_K }
\newcommand{\wu}[1]{ M^{(u)}{( #1 )} }
\newcommand{\wut}[1]{ \wu{#1} }

\newcommand{\tQ}{ \tilde{\Q} }

\newcommand{\tps}{ \left(2\pi\sigma^2\right) }
\newcommand{\oostp}{ \oneon{\sqrt{2\pi} } }
\newcommand{\loge}{\log e}

\newcommand{\sumK}{\sum^K_{k=1}}
\newcommand{\prodK}{\prod^K_{k=1}}

\newcommand{\sumu}{\sum^u_{a=1}}

\newcommand{\produ}{ \prod^u_{a=1} }

\newcommand{\sumKo}{\sum^{K_1}_{k=1}}

\newcommand{\prodab}{ \prod^u_{0\leq a\leq b} }

\newcommand{\pXYS}{p_{\X|\Y,\S}}
\newcommand{\ySx}{ \|\y-\S\x\|^2 }
\newcommand{\ySX}{ \|\y-\S\X\|^2 }
\newcommand{\ySXz}{ \|\y-\S\X_0\|^2 }
\newcommand{\ySXa}{ \|\y-\S\X_a\|^2 }

\newcommand{\exph}[1]{ \expb{ -\half #1 } }
\newcommand{\expqcnd}[2]{{\Exp_q}\left\{\left. #1 \,\right|\, #2\right\}}

\newcommand{\sm}[1]{\left\langle #1 \right\rangle}

\newcommand{\expbr}[1]{ \exp \left[ {#1} \right] } 
\newcommand{\expbs}[1]{ \expbr{ -\oneon{2\sigma^2} #1 } }

\newcommand{\Dzx}{ \cnddiv{ p_{Z|X,\snr;\eff} }{ p_{Z|\snr;\eff} }{ p_X } }
\newcommand{\pzi}[1]{ p_{#1}(z,\snr;\eff) }
\newcommand{\qzi}[1]{ q_{#1}(z,\snr;\xi) }
\newcommand{\pz}[1]{ p_{Z|\snr;\eff}(#1|\snr;\eff) }
\newcommand{\qz}[1]{ q_{Z|\snr;\xi}(#1|\snr;\xi) }

\begin{document}

\title{Randomly Spread CDMA: Asymptotics via Statistical Physics}

\author{
\parbox{.45\linewidth}{
  \begin{center}
    Dongning Guo\\
    Dept. of Electrical \& Computer Engineering \\
    Northwestern University \\
    2145 Sheridan Rd., Evanston, IL 60201, USA \\
    Email: dGuo@Northwestern.edu
  \end{center}}
\parbox{.05\linewidth}{}
\parbox{.45\linewidth}{
  \begin{center}
  Sergio Verd\'{u}\\
    Dept. of Electrical Engineering \\
    Princeton University \\
    Princeton, NJ 08544, USA \\
    Email: Verdu@Princeton.edu
  \end{center}}
\def\thefootnote{}
\thanks{Research partially supported by the U.S.\ National Science
  Foundation under Grant NCR-0074277, and through collaborative
  participation in the Communications and Networks Consortium
  sponsored by the U.S.\ Army Research Laboratory under the
  Collaborative Technology Alliance Program, Cooperative Agreement
  DAAD19-01-2-0011.  The U.S.\ Government is authorized to reproduce
  and distribute reprints for Government purposes notwithstanding any
  copyright notation thereon.}
\vspace{-5ex}
}

\maketitle%




\begin{abstract}
  This paper studies randomly spread code-division multiple access
  (CDMA) and multiuser detection in the large-system limit using the
  replica method developed in statistical physics.  Arbitrary input
  distributions and flat fading are considered.  A generic multiuser
  detector in the form of the posterior mean estimator is applied
  before single-user decoding.  The generic detector can be
  particularized to the matched filter, decorrelator, linear MMSE
  detector, the jointly or the individually optimal detector, and
  others.  It is found that the detection output for each user,
  although in general asymptotically non-Gaussian conditioned on the
  transmitted symbol, converges as the number of users go to infinity
  to a deterministic function of a ``hidden'' Gaussian statistic
  independent of the interferers.  Thus the multiuser channel can be
  decoupled: Each user experiences an equivalent single-user Gaussian
  channel, whose signal-to-noise ratio suffers a degradation due to
  the multiple-access interference.  The uncoded error performance
  (e.g., symbol-error-rate) and the mutual information can then be
  fully characterized using the degradation factor, also known as the
  multiuser efficiency, which can be obtained by solving a pair of
  coupled fixed-point equations identified in this paper.  Based on a
  general linear vector channel model, the results are also applicable
  to MIMO channels such as in multiantenna systems.
\end{abstract}


\newcounter{mytempeqncnt}

\section{Introduction}
\label{s:tni}

Consider a multidimensional Euclidean space in which each user (or
data stream) randomly selects a ``signature vector'' and modulates its
own information-bearing symbols onto it for transmission.  The
received signal is a superposition of all users' signals corrupted by
Gaussian noise.  Such a multiuser scheme, best described by a vector
channel model, is very versatile and is widely used in applications
that include code-division multiple access (CDMA),\index{Code-division
  multiple access (CDMA)} as well as certain multi-input multi-output
(MIMO) systems.\index{Multi-input multi-output (MIMO)} With knowledge
of all signature vectors, the goal is to estimate the transmitted
symbols and eventually recover the information intended for all or a
subset of the users.

This paper focuses on a paradigm of multiuser channels, known as
randomly spread CDMA \cite{Verdu98}.  In such a CDMA system, a number
of users share a common media to communicate to a single receiver
simultaneously over the same bandwidth.  Each user employs a randomly
generated spreading sequence (signature waveform) with a large
time-bandwidth product.  This multiaccess method has many advantages
particularly in wireless communications: frequency diversity,
robustness to channel impairments, ease of resource allocation, etc.
The price to pay is multiple-access interference\index{Multiple-access
  interference} (MAI) due to non-orthogonal spreading sequences from
all users.  Numerous multiuser detection\index{Multiuser detection}
techniques have been proposed to mitigate the MAI to various degrees.
This work is concerned with the performance of such multiuser systems
in two aspects: 1) Uncoded symbol-error-rate (or equivalently,
multiuser efficiency) and 2) Spectral efficiency, namely the total
information rate achievable by coded transmission and normalized by
the dimension of the multiuser channel.

\subsection{Gaussian or Non-Gaussian?}

The most efficient use of a multiuser channel is through jointly
optimal decoding, which is prohibitively complex with a large
population of users.  Although suboptimal, the philosophy of
separating the tasks of untangling the mutually interference streams
and exploiting the redundancy in the coded streams has received much
attention.  A multiuser detection front end supplies individual (hard
or soft) decision statistics to independent single-user decoders.
With the exception of decorrelating receivers, the multiuser detector
outputs are still contaminated by multiaccess interference, and their
statistical characterization is of paramount interest.

In \cite{Verdu83Mil, Verdu84PhD, Verdu86IT}, Verd\'u first used the
concept of multiuser efficiency\index{Multiuser efficiency|textbf} to
refer to the degradation of the output signal-to-noise ratio (SNR)
relative to a single-user channel calibrated at the same
bit-error-rate (BER) in binary (antipodal) uncoded transmission.  The
multiuser efficiencies of the single-user matched
filter,\index{Multiuser detector!matched filter}
decorrelator,\index{Multiuser detector!decorrelator} and linear
minimum mean-square error (MMSE) detector\index{Multiuser
  detector!linear MMSE detector} were found as functions of the
correlation matrix of the spreading sequences.  Particular attention
has been given to the asymptotic multiuser efficiency in the more
tractable region of high SNR.  Expressions for the optimum (high-SNR)
asymptotic multiuser efficiency were found in \cite{Verdu86IT,
  Verdu86ITa}.

In the large-system limit,\index{Large system!large-system limit}
where the number of users and the spreading factor both tend to
infinity with a fixed ratio, the dependence of system performance on
the sequences vanishes, and random matrix\index{Random matrix} theory
proves to be a capable tool for analyzing linear detectors.  The
limiting multiuser efficiency of the matched filter is trivial
\cite{Verdu98}.  The large-system multiuser efficiency of the linear
MMSE detector is obtained explicitly in \cite{Verdu98} for the
equal-power case (perfect power control), and in \cite{TseHan99IT} for
the case with flat fading as the solution to the Tse-Hanly fixed-point
equation.\index{Tse-Hanly equation} The efficiency of the decorrelator
is also known \cite{Verdu98, EldCha03IT, GuoVer02Blake}.  The success
of the multiuser efficiency analysis of the wide class of linear
detectors hinges on the fact that 1) the detection output is a sum of
independent components: the desired signal, the MAI and Gaussian
background noise, e.g., the decision statistic for user $k$ is
\begin{equation} \label{e:zl}
        \sm{X_k} = X_k + \mai_k + N_k;
\end{equation}
and 2) the multiple-access interference ($\mai_k$) is asymptotically
Gaussian (e.g., \cite{GuoVer02IT}).  As far as linear multiuser
detectors are concerned, regardless of the input distribution, the
performance is fully characterized by the noise enhancement associated
with the MAI variance.  Indeed, by regarding the multiuser detector as
part of the channel, an individual user experiences asymptotically a
single-user Gaussian channel with an SNR degradation equal to the
multiuser efficiency.

\begin{figure}
  \begin{center}
    \includegraphics[width=\myfigwidth]{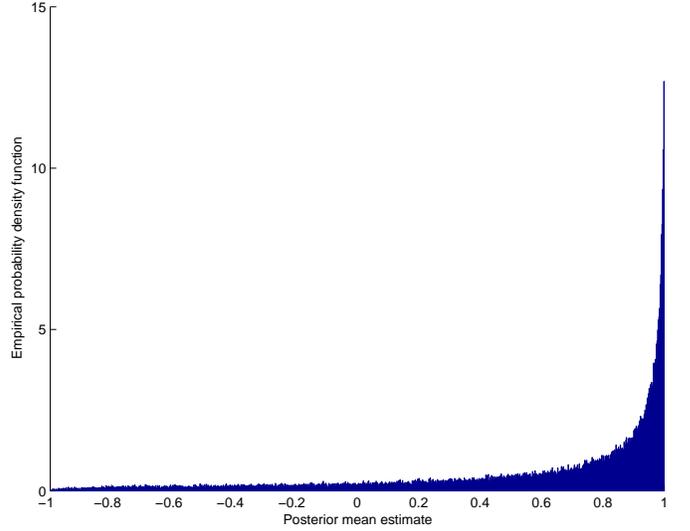}
    \caption{The empirical probability density function of the
      individually optimal soft detection output conditioned on $+1$
      being transmitted.  The system has 8 users, the spreading factor
      is 12, and SNR=2 dB.}
    \label{f:x}
  \end{center}
\end{figure}

The performance analysis of nonlinear detectors such as the optimal
ones is a hard problem.  The difficulty here is inherent to nonlinear
operations: The detection output cannot be decomposed as a sum of
independent components associated with the desired signal, the
interferences and the noise respectively.  Moreover, the detection
output is in general asymptotically non-Gaussian conditioned on the
input.  An extreme case is the maximum-likelihood multiuser detector
for binary transmission, the hard decision output of which takes only
two values.  The difficulty remains even if we consider soft detection
outputs.  Hence, unlike for a Gaussian output statistic, the
conditional variance of a general detection output does not lead to a
simple characterization of the multiuser efficiency or error
performance.  For illustration, Figure \ref{f:x} plots the approximate
probability density function obtained from the histogram of the soft
output statistic of the individually optimal detector conditioned on
$+1$ being transmitted.  The simulated system has 8 users with binary
inputs, a spreading factor of 12, and SNR=2 dB.  A total of 10,000
trials were recorded.  Note that negative decision values correspond
to decision error; hence the area under the curve on the negative half
plane gives the BER.  The distribution shown in Figure \ref{f:x} is
far from Gaussian.  Thus the usual notion of output SNR fails to
capture the essence of system performance.  In fact, much literature
is devoted to evaluating the error performance by Monte Carlo
simulation.

\begin{figure}
  \begin{center}
    \includegraphics[width=\myfigwidth]{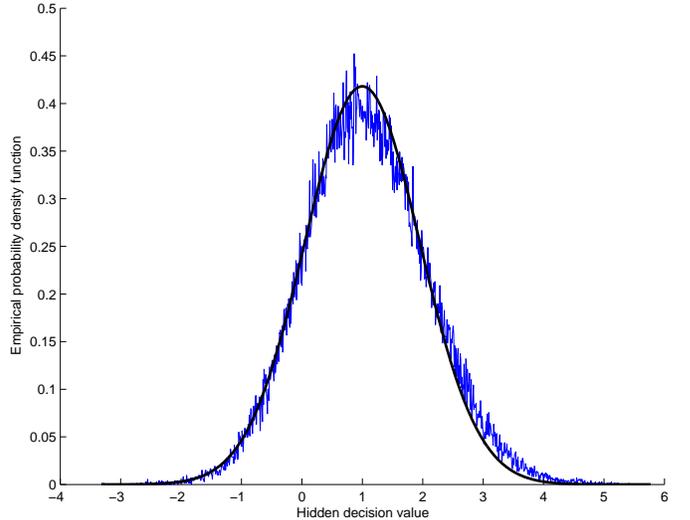}
    \caption{The empirical probability density function of the hidden
      equivalent Gaussian statistic conditioned on $+1$ being
      transmitted.  The system has 8 users, the spreading factor is
      12, and SNR=2 dB.  The asymptotic Gaussian distribution is also
      plotted for comparison.}
    \label{f:z}
  \end{center}
\end{figure}

This paper makes a contribution to the understanding of multiuser
detection in the large-system regime.  It is found under certain
assumptions that the output decision statistic of a nonlinear
detector, such as the one whose distribution is depicted by Figure
\ref{f:x}, converges in fact to a very simple monotone function of a
``hidden'' conditionally Gaussian random variable, i.e.,
\begin{equation} \label{e:zf}
        \sm{X_k} \rightarrow f( Z_k )
\end{equation}
where $Z_k =X_k + N'_k$ and $N'_k$ is Gaussian.  One may contend that
it is always possible to monotonically map a non-Gaussian random
variable to a Gaussian one.  What is surprisingly simple and useful
here is that 1) the mapping $f$ neither depends on the instantaneous
spreading sequences, nor on the transmitted symbols which we wish to
estimate in the first place; and 2) the statistic $Z_k$ is equal to
the desired signal plus an independent Gaussian noise.  Indeed, a few
parameters of the system determine the function $f$.  By applying an
inverse of this function to the detection output $\sm{X_k}$, an
equivalent conditionally Gaussian statistic $Z_k$ is recovered, so
that we are back to the familiar ground where the output SNR (defined
for the equivalent Gaussian statistic $Z_k$) completely characterizes
the system performance.  The multiuser efficiency is simply obtained
as the ratio of the output and input SNRs.  We will refer to this
result as the ``decoupling principle'' since asymptotically, after
applying $f^{-1}$, each user's data goes through an equivalent
single-user channel with an additive Gaussian noise which is
independent of the interferers' data.

Under certain assumptions, we show the decoupling principle to hold
for not only optimal detection, but also a broad family of generic
multiuser detectors, called the posterior mean estimators (PME), which
compute the mean value of the input conditioned on the observation
assuming a certain postulated posterior probability distribution.
Simply put, the generic detector is the optimal detector for a
postulated multiuser system that may be different from the actual one.
In case the postulated posterior is identical to the one induced by
the actual multiuser channel and input, the PME is a soft version of
the individually optimal detector.  The postulated posterior, however,
can also be chosen such that the resulting PME becomes one of many
other detectors, including but not limited to the matched filter,
decorrelator, linear MMSE detector, as well as the jointly optimal
detector.  Moreover, the decoupling principle holds for not only
binary inputs, but arbitrary input distributions with finite power.

For illustration of the new findings, Figure \ref{f:z} plots the
approximate probability density function obtained from the histogram
of the conditionally Gaussian statistic obtained by applying $f^{-1}$
to the non-Gaussian detection output in Figure~\ref{f:x}.  The
theoretically predicted Gaussian density function is also shown for
comparison.  The ``fit'' is good considering that a relatively small
system of 8 users with a processing gain of 12 is considered.  Note
that in case the multiuser detector is linear, the mapping $f$ is also
linear, and~\eref{e:zf} reduces to \eref{e:zl}.

By virtue of the decoupling principle, the mutual information between
the input and the output of the generic detector for each user
converges to the input-output mutual information of the equivalent
single-user Gaussian channel under the same input, which admits a
simple analytical expression.  Hence the large-system spectral
efficiency of several well-known linear detectors, first found in
\cite{VerSha99IT} and \cite{ShaVer01IT} with and without fading
respectively, can be recovered straightforwardly using the decoupling
principle.  New results on the spectral efficiency of nonlinear
detection and arbitrary inputs under both joint and separate decoding
are also obtained.  Furthermore, the additive decomposition of optimal
spectral efficiency as a sum of single-user efficiencies and a joint
decoding gain \cite{ShaVer01IT} applies under more general conditions
than originally thought.

As in random matrix spectrum analysis, our large-system results are
representative of the behavior of systems of moderate size.  As shown
in Figures \ref{f:x} and \ref{f:z}, a randomly spread system with as
few as 8 users can often be well approximated by the large-system
limiting results.

\subsection{Random Matrix vs.\ Spin Glass}
\index{Random matrix}

Much of the early success in the large-system analysis of linear
detectors relies on the fact that the multiuser efficiency of a
finite-size system can be written as an explicit function of the
eigenvalues of the correlation matrix of the random signature
waveforms, the empirical distributions of which converge to a known
function in the large-system limit \cite{TulVer04FT,
  Bai99SS}.\index{Large system!large-system limit} As a result, the
large-system multiuser efficiency can be obtained as an integral with
respect to the limiting eigenvalue distribution.  Indeed, this random
matrix technique is applicable to any performance measure that can be
expressed as a function of the eigenvalues.  Based on an explicit
expression for CDMA channel capacity in \cite{Verdu86Allerton},
Verd\'u and Shamai quantified the optimal spectral efficiency in the
large-system limit \cite{VerSha99IT, ShaVer01IT} (see also
\cite{GraAle98IT, RapPop00TC}).  The expression found in
\cite{VerSha99IT} also solved the capacity of single-user narrowband
multiantenna channels as the number of antennas grows---a problem that
was open since the pioneering work of Foschini \cite{Foschi96BLTJ} and
Telatar \cite{Telata99ETT}.  Unfortunately, few explicit expressions
of the efficiencies in terms of eigenvalues are available beyond the
above cases.  Much less success has been reported in the application
of random matrix theory when either the detector is nonlinear or the
inputs are non-Gaussian constellations.

A major consequence of random matrix theory is that the dependence of
performance measures on the spreading sequences vanishes as the system
size increases without bound.  In other words, the performance
measures are ``self-averaging.''\index{Self-averaging} In the context
of physical science, this property is nothing but a manifestation of a
fundamental law that the fluctuation of macroscopic properties of
certain many-body systems vanishes in the thermodynamic limit, i.e.,
when the number of interacting bodies becomes large.  This falls under
the general scope of statistical mechanics (aka statistical physics),
whose principal goal is to study the macroscopic properties of
physical systems from the principle of microscopic interactions.
Indeed, the asymptotic eigenvalue distribution of certain correlation
matrices can be derived via statistical physics (e.g.,
\cite{Rodger99SS}).  Tanaka pioneered the user of statistical physics
concepts and methodologies in multiuser detection and obtained the
large-system uncoded minimum BER (hence the optimal multiuser
efficiency) and spectral efficiency with equal-power binary inputs
\cite{Tanaka01NIPS, Tanaka01ISIT, Tanaka01EPL, Tanaka02IT}.  In
\cite{GuoVer02Blake} we further elucidated the relationship between
CDMA and statistical physics and generalized to the case of unequal
powers.  Inspired by \cite{Tanaka02IT}, M\"uller and Gerstacker
\cite{MulGer04IT} studied the channel capacity under separate decoding
and noticed that the additive decomposition of the optimum spectral
efficiency in \cite{ShaVer01IT} holds also for binary inputs.
M\"uller thus further conjectured the same formula to be valid
regardless of the input distribution \cite{Muller02WCIT}.

In this paper, we build upon Tanaka's ground-breaking contribution
\cite{Tanaka02IT} and present a unified treatment of Gaussian CDMA
channels and multiuser detection assuming an arbitrary input
distribution and flat fading characteristic.  A wide class of
multiuser detectors, optimal as well as suboptimal, are studied under
the same umbrella of posterior mean estimation.  The central results
are the decoupling principle for generic multiuser detection, the
characterization of multiuser efficiency via a pair of nonlinear
equations, as well as the spectral efficiencies of separate and joint
decoding.

The key tool in this paper, the replica method,\index{Replica method}
has its origin in spin glass theory\index{Spin glass!spin glass
  theory} \cite{EdwAnd75JPF}.  Analogies between statistical physics
and neural networks, coding, image processing, and communications have
long been noted (e.g., \cite{Nishim01, Sourla89Nature}).  There have
been many recent activities applying statistical physics wisdom to
sparse-graph error-correcting codes (e.g., \cite{KabSaa99EPL,
  Montan00EPJB, MurKab00PRE, MonSou00EPJB, KabSaa04JPA}).  Similar
techniques have also been used to study capacity of MIMO channels
\cite{MouSim03IT}.  Among others, mean field theory is used to derive
iterative detection algorithms \cite{Kabash03JPA, FabWinITsub}.  The
first application of the replica method to multiuser detection was
made in \cite{Tanaka02IT}.  In this paper, we draw a parallel between
the general statistical inference\index{Statistical inference} problem
in multiuser communications and the problem of determining the
configuration of random spins subject to quenched
randomness.\index{Quenched randomness} For the purpose of analytical
tractability, we will invoke common assumptions in the statistical
physics literature: 1) the self-averaging
property\index{Self-averaging!self-averaging principle} applies, 2)
the ``replica trick'' is valid, and 3) replica symmetry holds.  These
assumptions have been used successfully in many problems in
statistical physics as well as in neural networks and coding theory,
to name a few, while a complete justification of the replica method is
a notoriously difficult challenge in mathematical physics, which has
seen some important progress recently \cite{Talagr01TCS, Talagr03}.
The results in this paper are based on the aforementioned assumptions
and therefore the mathematical rigor is pending on breakthroughs in
those problems.  A set of easy-to-check sufficient conditions under
which the replica method is justified is yet to be found.  In
statistical physics it has been found that results obtained using the
replica method may still capture many of the qualitative features of
the system performance even when the key assumptions fail
\cite{MezPar87, Dotsen94}.  Furthermore, the decoupling principle
carries great practicality and finds convenient uses in finite-size
systems where the analytical asymptotic results are a good
approximation.

The remainder of this paper is organized as follows.  Section
\ref{s:mod} gives the model and summarizes the main results.  Relevant
statistical physics concepts and methodologies are introduced in
Section \ref{s:stat}.  Calculations based on a real-valued channel are
presented in Section \ref{s:prf}.  Complex-valued channels are
discussed in Section \ref{s:cc}, followed by some numerical examples
in Section \ref{s:nr}.  Some conclusions are drawn in Section
\ref{s:con}.

\section{Model and Summary of Results}
\label{s:mod}

\subsection{System Model}

\begin{figure}
  \begin{center}
    \begin{picture}(250,100)(0,0)
\thinlines
\put(0,10){\vector(1,0){15}}
\put(0,60){\vector(1,0){15}}
\put(0,90){\vector(1,0){15}}
\put(60,10){\vector(1,0){25}}
\put(60,60){\vector(1,0){25}}
\put(60,90){\vector(1,0){25}}
\put(65,0){$X_K$}
\put(65,50){$X_2$}
\put(65,80){$X_1$}
\put(35,30){$\vdots$}
\Thicklines
\put(15,0){\framebox(45,20){Encoder}}
\put(15,50){\framebox(45,20){Encoder}}
\put(15,80){\framebox(45,20){Encoder}}
\Thicklines
\put(85,0){\framebox(70,100){\parbox{70pt}{
      \begin{center}Multiuser\\ channel\\ $\Y=\S\X+\N$\end{center}}}}
\put(155,50){\vector(1,0){30}}
\put(165,38){$\Y$}
\Thicklines
\put(185,0){\framebox(50,100){\parbox{50pt}{
      \begin{center}Joint decoding\end{center}}}}
\thinlines
\put(235,10){\vector(1,0){15}}
\put(235,60){\vector(1,0){15}}
\put(235,90){\vector(1,0){15}}
\end{picture}
    \caption{A multiuser system with joint decoding.}
    \label{f:jd}
  \end{center}
\end{figure}
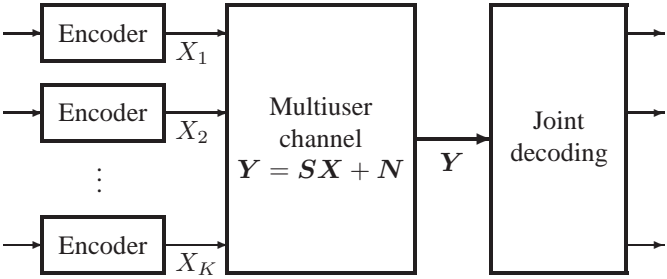

Consider the synchronous $K$-user CDMA system with spreading factor
$L$ as depicted in Figure \ref{f:jd}.  Each encoder maps its message
into a sequence of channel symbols.  All users employ the same type of
signaling so that at each interval the $K$ symbols are independent
identically distributed (i.i.d.) random variables with distribution
(probability measure) $P_X$, which has zero mean and unit variance.
Let $\X=\tran{ [X_1,\dots, X_K]}$ denote the vector of input symbols
from the $K$ users in one symbol interval.  For notational convenience
in the analysis, it is assumed that either a probability density
function or a probability mass function of $P_X$ exists, and is
denoted by $p_X$.\footnote{The results in this paper hold in full
  generality and do not depend on the existence of a probability
  density or mass function.}  Let also $p_\X(\x) =\prodK p_X(x_k)$
denote the joint (product) distribution.

Let the instantaneous SNR of user $k$ be denoted by $\snr_k$ and $\Snr
=\diag\{ \amp{1}, \dots ,\amp{K}\}$.  Denote the spreading sequence of
user $k$ by $\s_k=\oneon {\sqrt{L}} [S_{1k}, S_{2k} ,\dots, S_{Lk}
\tran{]}$, where $S_{nk}$ are i.i.d.\ random variables with zero mean
and finite moments.  Let the symbols and spreading sequences be
randomly chosen for each user and not dependent on the SNRs.  The
$L\times K$ channel ``state'' matrix is denoted by $\S=[\amp{1}
\s_1,\dots, \amp{K} \s_K]$.  The synchronous CDMA channel with flat
fading is described by:
\begin{eqnarray}
  \Y &=& \sum^K_{k=1} \sqrt{\snr_k}\, \s_k X_k + \N \\ &=& \S\X + \N
  \label{e:sch}
\end{eqnarray}
where $\N$ is a vector consisting of i.i.d.\ zero-mean Gaussian random
variables.  Depending on the domain that the inputs and spreading
chips take values, the input-output relationship \eref{e:sch}
describes either a real-valued or a complex-valued fading channel.

The linear system \eref{e:sch} is quite versatile.  In particular,
with $\snr_k =\snr$ for all $k$, it models the canonical
MIMO\index{Multi-input multi-output (MIMO)} channel in which all
propagation coefficients are i.i.d.  An example is single-user
communication with $K$ transmit antennas and $L$ receive antennas,
where the channel coefficients are not known to the transmitter.

\subsection{Posterior Mean Estimation}

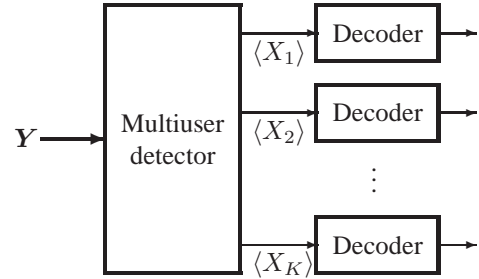
\begin{figure}
\begin{center}
\begin{picture}(180,100)(150,0)
\put(150,47){$\Y$}
\Thicklines
\put(160,50){\vector(1,0){25}}
\put(185,0){\framebox(50,100){\parbox{50pt}{
      \begin{center}Multiuser\\ detector\end{center}}}}
\put(265,0){\framebox(45,20){Decoder}}
\put(265,50){\framebox(45,20){Decoder}}
\put(265,80){\framebox(45,20){Decoder}}
\put(285,30){$\vdots$}
\thinlines
\put(235,10){\vector(1,0){30}}
\put(235,60){\vector(1,0){30}}
\put(235,90){\vector(1,0){30}}
\put(240,0){$\sm{X_K}$}
\put(240,50){$\sm{X_2}$}
\put(240,80){$\sm{X_1}$}
\put(310,10){\vector(1,0){15}}
\put(310,60){\vector(1,0){15}}
\put(310,90){\vector(1,0){15}}
\end{picture}
    \caption{Multiuser detection followed by independent single-user decoding.}
    \label{f:sd}
  \end{center}
\end{figure}

The information-bearing symbol (vector) $\X$ is drawn according to the
prior distribution $p_\X$.  The channel response to the input $\X$ is
an output $\Y$ generated according to a conditional probability
distribution $p_{\Y|\X,\S}$ where $\S$ is the channel state.  Upon
receiving $\Y$, the estimator would like to infer the transmitted
symbol $\X$ with knowledge of $\S$.


The most efficient use of the multiuser channel \eref{e:sch} is
achieved by optimal joint decoding as depicted in Figure \ref{f:jd}.
Due to the complexity of joint decoding, the processing is often
separated into multiuser detection\index{Multiuser detection} followed
by single-user error-control decoding as shown in Figure \ref{f:sd}.
A multiuser detector\index{Multiuser detector} front end estimates the
transmitted symbols given the received signal and the channel state,
without using any knowledge of the error-control codes employed by the
transmitters.  Conversely, each single-user decoder only observes the
sequence of decision statistics corresponding to one user, and does
not take into account the existence of any other users (in particular,
it does not use any knowledge of the spreading sequences).  By
adopting this separate decoding approach, the channel together with
the multiuser detector front end is viewed as a bank of coupled
single-user channels.  The detection output sequence for an individual
user is in general not a sufficient statistic for decoding this user's
own information.

To capture the intended suboptimal structure, one has to restrict the
capability of the multiuser detector; otherwise the detector could in
principle encode the channel state and the received signal $(\S,\Y)$
into a single real number as its output to each user, which is a
sufficient statistic for all users.  A plausible choice is the
(canonical) {\em posterior mean estimator},\index{Posterior mean
estimation (PME)!posterior mean estimator|textbf} which computes the
mean value of the posterior probability distribution $\pXYS$,
hereafter denoted by angle brackets $\sm{\cdot}$:
\begin{equation}  \label{e:smx}
  \sm{\X} = \expcnd{ \X }{ \Y, \S }.
\end{equation}
Also known as the conditional mean estimator, this estimator achieves
the minimum mean-square error for each user, and is therefore the
(nonlinear) MMSE detector.  We also regard it as a soft-output version
of the individually optimal multiuser detector (assuming uncoded
transmission).  The posterior probability distribution $\pXYS$ is
induced from the input distribution $p_\X$ and the conditional
Gaussian density function $p_{\Y|\X,\S}$ of the channel \eref{e:sch}
by the Bayes formula:
\begin{equation}
  \pXYS(\x|\y,\S) = \frac{ p_{\X}(\x) p_{\Y|\X,\S}(\y|\x,\S) }
  { \int p_{\X}(\x) p_{\Y|\X,\S}(\y|\x,\S) \,\intd \x }.
  \label{e:bf}
\end{equation}

The PME can be understood as an ``informed'' optimal estimator which
is supplied with the posterior distribution $\pXYS$ and then computes
its mean.  A generalization of the canonical PME is conceivable:
Instead of informing the estimator with the actual posterior $\pXYS$,
we can supply at will any other well-defined conditional distribution
$q_{\X|\Y,\S}$.  Given $(\Y,\S)$, the estimator can nonetheless
perform ``optimal'' estimation based on this postulated measure $q$.
We call this the {\em generalized posterior mean estimation}, which is
conveniently denoted as
\begin{equation}  \label{e:vxq}
  \sm{\X}_q = \expqcnd{ \X }{ \Y, \S }
\end{equation}
where $\Exp_q\{\cdot\}$ stands for the expectation with respect to the
postulated measure $q$.  For brevity, we will also refer to
\eref{e:vxq} by the name of the posterior mean estimator, or simply
the PME.  In view of \eref{e:smx}, the subscript in \eref{e:vxq} can
be dropped if the postulated measure $q$ coincides with the actual one
$p$.

In general, postulating $q\neq p$\ causes degradation in detection
performance.  Such a strategy may be either due to lack of knowledge
of the true statistics or a particular choice that corresponds to a
certain estimator of interest.  In principle, any deterministic
estimation can be regarded as a PME since we can always choose to put
a unit mass at the desired estimation output given $(\Y,\S)$.  We will
see in Section \ref{s:sd} that by postulating an appropriate measure
$q$, the PME can be particularized to many important multiuser
detectors.  As will also be shown in this paper, the generic
representation \eref{e:vxq} allows a uniform treatment of a large
family of multiuser detectors which results in a simple performance
characterization for all of them.

It is enlightening to introduce a new concept: the {\em retrochannel},
which is defined for a given channel and input as a companion channel
in the opposite direction characterized by a posterior distribution.
Given the multiuser channel $p_{\Y|\X,\S}$ with an input $p_\X$, we
have a (canonical) retrochannel defined by $p_{\X|\Y,\S}$ \eref{e:bf},
which, upon an input $(\Y,\S)$, generates a random output $\X$
according to $p_{\X|\Y,\S}$.  A retrochannel in the single-user
setting is similarly defined.  In general, any valid posterior
distribution $q_{\X|\Y,\S}$ can be regarded as a retrochannel.  Note
that the retrochannel samples from the Bayesian posterior distribution
(in general, the postulated one) in such a way that, conditioned on
the observation, the input to the channel and the output of the
retrochannel are independent.  It is clear that the PME output
$\sm{\X}_q$ is the expected value of the output of the retrochannel
$q_{\X|\Y,\S}$ given $(\Y,\S)$.

In this paper, the posterior $q_{\X|\Y,\S}$ supplied to the PME is
assumed to be the one that corresponds to a postulated CDMA system,
where the input distribution is an arbitrary $q_X$, and the
input-output relationship of the postulated channel differs from the
actual channel \eref{e:sch} by only the noise variance.  Precisely,
the postulated channel is characterized by
\begin{equation} \label{e:pch}
  \Y = \S\X' + \sigma\N'
\end{equation}
where the channel state matrix $\S$ is identical to that of the actual
channel \eref{e:sch}, and $\N'$ is statistically the same as the
Gaussian noise $\N$ in \eref{e:sch}.  The postulated input
distribution $q_X$ is assumed to have zero-mean and finite moments,
and $q_{\X|\Y,\S}$ is determined by $q_X$ and $q_{\Y|\X,\S}$ according
to the Bayes formula.  Here, $\sigma$ serves as a control parameter.
Indeed, the PME so defined is the optimal detector for a postulated
multiuser system with its input distribution and noise level different
from the actual ones.  In general, the assumed information about the
channel state $\S$ could also be different from the actual instances,
but this is out of the scope of this work, as we limit ourselves to
study the (rich) family of multiuser detectors that can be represented
as the PME parameterized by the postulated input and noise level
$(q_X, \sigma)$.

We note that PME under postulated posterior is known in the Bayes
statistics literature.  This technique was introduced to multiuser
detection by Tanaka in the special case of equal-power users with
binary or Gaussian inputs under the name of marginal-posterior-mode
detectors \cite{Tanaka01NIPS, Tanaka02IT}.  In this paper we pursue
further that direction to treat arbitrary input, arbitrary power
distribution, and generic multiuser detection.

\subsection{Specific Detectors}
\label{s:sd}

The rest of this section assumes the system model \eref{e:sch} to be
real-valued.  The inputs $X_k$, the spreading chips $S_{nk}$, and all
entries of $\N$ take real values and have unit variance.  The
characteristic of the actual channel is
\begin{equation}  \label{e:py}
  p_{\Y|\X,\S}(\y|\x,\S) = (2\pi)^{-\frac{L}{2}} \expb{ -\frac{\ySx}{2} },
\end{equation}
and that of the postulated channel is
\begin{equation}  \label{e:qy}
  q_{\Y|\X,\S}(\y|\x,\S) = \tps^{-\frac{L}{2}} 
  \expb{ -\frac{ \ySx }{ 2\sigma^2 } }.  
\end{equation}

We identify specific choices of the postulated input distribution
$q_X$ and noise level $\sigma$ under which the PME is particularized
to well-known multiuser detectors.

\subsubsection{Linear Detectors}
\label{s:ld}

Let the postulated input be standard Gaussian, $q_X\sim \mathcal{N}
(0,1)$.  The optimal detector (PME) for the postulated model
\eref{e:pch} with standard Gaussian inputs is a linear filtering of
the received signal $\Y$:
\begin{equation}
  \sm{\X}_q = \inv{ \left[\ST\S + \sigma^2\I\right] } \,\ST\,\Y.
  \label{e:xqn}
\end{equation}
The control parameter $\sigma$ can be tuned to choose from the
single-user matched filter, decorrelator, MMSE detector, etc.  If
$\sigma \rightarrow \infty$, the PME estimate \eref{e:xqn} is
consistent with the single-user matched filter output:\index{Multiuser
  detector!matched filter}
\begin{equation} \label{e:mfx}
  \sigma^2 \, \sm{X_k}_q \longrightarrow \sT_k\,\Y,
  \quad \text{in } L^2 \text{ as } \sigma\rightarrow \infty.
\end{equation}
If $\sigma=1$, \eref{e:xqn} is exactly the soft output of the linear
MMSE detector.\index{Multiuser detector!linear MMSE detector} If
$\sigma\rightarrow 0$, \eref{e:xqn} converges to the decorrelator
output.\index{Multiuser detector!decorrelator}

\subsubsection{Optimal Detectors}
\label{s:od}

Let the postulated $q_X$ be identical to the true one, $p_X$.  The
posterior is then
\begin{equation}
  q_{\X|\Y,\S}(\x|\y,\S) = \frac{p_\X(\x)}{Z(\y,\S)} \, \expbs{\ySx}
  \label{e:qxysp}
\end{equation}
where $Z(\y,\S)$ is a normalization factor.

Suppose that the postulated noise level $\sigma\rightarrow 0$, then
the probability mass of the distribution $q_{\X|\Y,\S}$ is
concentrated on a vector that minimizes $\|\y-\S\x\|$, which also
maximizes the likelihood function $p_{\Y|\X,\S}(\y|\x,\S)$.  The
 PME $\limzero {\sigma}\sm {\X}_q$ is thus equivalent to
that of jointly optimal (or maximum-likelihood) detection
\cite{Verdu98}.\index{Multiuser detector!optimal detectors!jointly
  optimal}\index{Multiuser detector!optimal
  detectors!maximum-likelihood}

Alternatively, if $\sigma=1$, then the postulated measure coincides
with the actual measure, i.e., $q=p$.  The PME output $\sm{\X}$ is the
mean of the marginal of the conditional posterior probability
distribution.  It is the nonlinear MMSE detector for the actual
system, and is seen as a soft version of the individually optimal
detector \cite{Verdu98}.\index{Multiuser detector!optimal
  detectors!individually optimal}

Also worth mentioning here is that, if $\sigma\rightarrow\infty$, the
 PME reduces to the single-user matched filter.  Indeed,
\eref{e:mfx} can be shown to hold by noticing from \eref{e:qxysp} that
\begin{equation}
  q_{\X|\Y,\S}(\x|\y,\S) = p_{\X}(\x) \left[ 1 -
  \frac{\ySx}{2\sigma^2} + \mathcal{O}\Big(\oneon{\sigma^4}\Big) \right].
\end{equation}

\subsection{Main Results}
\label{s:main}

This subsection gives the main results of this paper assuming the
real-valued system model.  The detailed replica analysis for obtaining
these results is relegated to Sections \ref{s:stat} and \ref{s:prf}.
Results for a complex-valued model are given in Section \ref{s:cc}.

Consider the multiuser channel $p_{\Y|\X,\S}$ given by \eref{e:py}
with input $\X\sim p_\X$, and the  posterior mean
estimator \eref{e:vxq} parameterized by $(q_X,\sigma)$.  Section
\ref{s:sd} illustrated the versatility of the PME encompassing many
well-known detectors.  The goal here is to quantify the optimal
spectral efficiency $\oneon{L} I(\X;\Y|\S)$, the quality of the
detection output $\sm{X_k}_q$ for each user $k$, as well as the
input-output mutual information $I(X_k;\sm{X_k}_q|\S)$.

Although these performance measures are all dependent on the
realization of the channel state, such dependence vanishes in the
large-system asymptote.  A {\em large system}\index{Large
  system|textbf} here refers to the limit that both the number of
users and the spreading factor tend to infinity but with their ratio,
known as the {\em system load}, converging to a positive number, i.e.,
$K/L\rightarrow\beta$, which may or may not be smaller than 1.  It is
also assumed that the SNRs of all users, $\{\snr_k\}^K_{k=1}$, are
i.i.d.\ with distribution $P_\snr$, hereafter referred to as the {\em
  SNR distribution}.\index{SNR distribution|textbf} All moments of the
SNR distribution are assumed to be finite.  Clearly, the empirical
distributions of the SNRs converge to the same distribution $P_\snr$
as $K\rightarrow \infty$.  Note that this SNR distribution captures
the (flat) fading characteristics of the channel.

Given $(\beta, P_\snr, p_X, q_X, \sigma)$, we express the large-system
limit of the multiuser efficiency and spectral efficiency under both
separate and joint decoding.

\subsubsection{The Decoupling Principle}

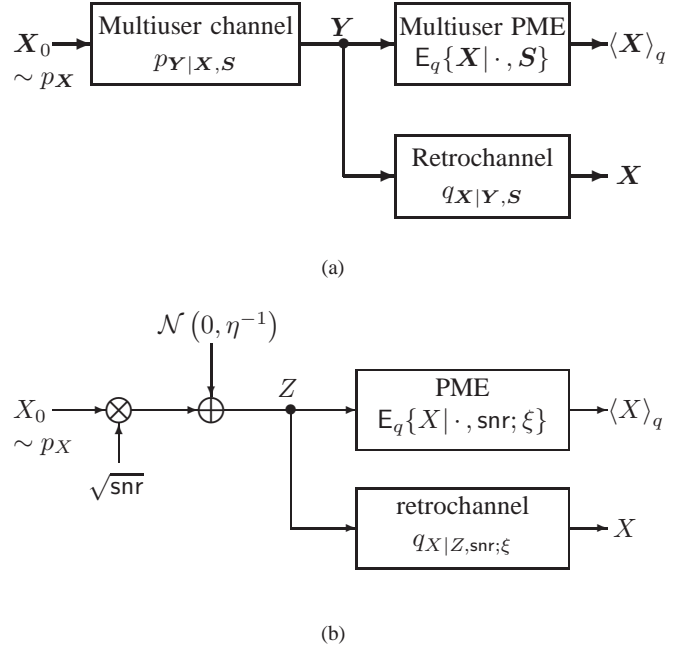
\begin{figure}
    \subfigure[]{
\begin{picture}(245,85)(0,0)
\thicklines
\put(15,65){\vector(1,0){15}}
\put(0,62){$\X_0$}
\put(0,50){$\sim p_\X$}
\put(30,50){\framebox(78,30){ \parbox{78pt}{ \vspace{-5pt}
      \center Multiuser channel \\ $p_{\Y|\X,\S} $ } } }
\put(108,65){\vector(1,0){37}}
\put(120,68){$\Y$}
\put(125,65){\makebox(0,0){$\bullet$}}
\put(125,15){\line(0,1){50}}
\put(125,15){\vector(1,0){20}}
\put(145,50){\framebox(65,30){ \parbox{65pt}{ \vspace{-5pt}
      \center Multiuser PME \\ $\Exp_q\{ \X | \,\cdot\, ,\S \}$ } } }
\put(210,65){\vector(1,0){15}}
\put(225,62){$\sm{\X}_q$}
\put(145,0){\framebox(65,30){ \parbox{65pt}{ \vspace{-5pt}
      \center Retrochannel \\ $q_{\X|\Y,\S}$ } } }
\put(210,15){\vector(1,0){15}}
\put(228,12){$\X$}
\end{picture}
\label{f:rch}
}
\subfigure[]{
    \begin{picture}(245,95)(0,0)
      \put(15,65){\vector(1,0){20}}
      \put(0,62){$X_0$}
      \put(0,50){$\sim p_X$}
      \put(40,65){\makebox(0,0){$\bigotimes$}}
      \put(25,33){ $\sqrt{\snr}$ }
      \put(40,45){\vector(0,1){15}}
      \put(45,65){\vector(1,0){25}}
      \put(75,65){\makebox(0,0){$\bigoplus$}}
      \put(75,90){\vector(0,-1){20}}
      \put(55,94){$\mathcal{N}\left(0,\eta^{-1}\right)$}
      \put(80,65){\vector(1,0){50}}
      \put(100,70){$Z$}
      \put(130,50){\framebox(80,30){\parbox{80pt}{ \vspace{-8pt}
            \center PME \\ $\Exp_q\{X|\,\cdot\,,\snr;\xi\}$}}}
      \put(225,62){$\sm{X}_q$}
      \put(227,17){$X$}
      \put(105,65){\makebox(0,0){$\bullet$}}
      \put(105,20){\line(0,1){45}}
      \put(105,20){\vector(1,0){25}}
      \put(210,65){\vector(1,0){15}}
      \put(210,20){\vector(1,0){15}}
      \put(130,5){\framebox(80,30){\parbox{80pt}{ \vspace{-8pt}
            \center retrochannel\\ $q_{X|Z,\snr;\xi}$ }} }
    \end{picture}
    \label{f:sur}
    }
    \caption{(a) The multiuser channel, the  (multiuser)
      PME, and the companion (multiuser) retrochannel.  (b) The
      equivalent single-user Gaussian channel, PME and retrochannel.}
    \label{f:ms}
\end{figure}

The multiuser channel $p_{\Y|\X,\S}$ and the multiuser posterior mean
estimator parameterized by $(q_X, \sigma)$ are depicted in Figure
\ref{f:rch}, together with the companion (multiuser) retrochannel
$q_{\X|\Y,\S}$.  Here the input to the multiuser channel is denoted by
$\X_0$ to distinguish from the output $\X$ of the retrochannel.  For
an arbitrary user $k$, the SNR is $\snr_k$, and $X_{0k}$, $X_k$ and
$\sm{X_k}_q$ denote the input symbol, the retrochannel output and the
PME output, all for user $k$.

In order to show the decoupling result, let us also consider the
composition of a Gaussian channel, a PME and a companion retrochannel
in the single-user setting as depicted in Figure \ref{f:sur}.  The
input and output are related by:
\begin{equation}
  Z = \sqrt{\snr} \, X_0 + \oneon{\sqrt{\eff}} \, N
  \label{e:zx0}
\end{equation}
where the input $X_0\sim p_X$, $\snr$ is the {\em input SNR}, $N\sim
\mathcal{N} (0,1)$ the noise independent of $X_0$, and $\eff>0$ the
{\em inverse noise variance}\index{Inverse noise variance}.  The
conditional distribution associated with the channel is
\begin{equation}
  p_{Z|X,\snr;\eff}(z|x,\snr;\eff) = \sqrt{\frac{\eff}{2\pi}}
      \expbr{ -\frac{\eff}{2} \left(z - \sqrt{\snr} \,x\right)^2 }.
      \label{e:pzx}
\end{equation}
Let $q_{Z|X,\snr;\xi}$ represent a Gaussian channel akin
to~\eref{e:zx0}, the only difference being that the inverse noise
variance is $\xi$ instead of $\eff$:
\begin{equation}
    q_{Z|X,\snr;\xi}(z|x,\snr;\xi) = \sqrt{\frac{\xi}{2\pi}}
      \expbr{ -\frac{\xi}{2} \left(z - \sqrt{\snr} \,x\right)^2 }.
      \label{e:qzx}
\end{equation}
Similar to that in the multiuser setting, by postulating the input
distribution to be $q_X$, a posterior probability distribution
$q_{X|Z,\snr;\xi}$ is induced by $q_X$ and $q_{Z|X,\snr;\xi}$ using
the Bayes rule.  Thus we have a single-user retrochannel defined by
$q_{X|Z,\snr;\xi}$, which outputs a random variable $X$ given the
channel output $Z$ (Figure \ref{f:sur}).  A (generalized) single-user
PME is defined naturally as (cf.~\eref{e:vxq}):
\begin{equation}
  \sm{X}_q = \expqcnd{ X }{ Z,\snr;\xi }.
  \label{e:xq}
\end{equation}
The probability law of the composite system depicted by Figure
\ref{f:sur} is determined by $\snr$ and two parameters $\eff$ and
$\xi$.  We define the mean-square error of the  PME as
\begin{equation}  \label{e:gmse}
  \mse(\snr;\eff,\xi) =
  \expcnd{ \left( X_0 - \sm{X}_q \right)^2 }{ \snr;\eff,\xi },
\end{equation}
and also define the variance of the retrochannel as
\begin{equation}  \label{e:vrc}
  \vrc(\snr;\eff,\xi) =
  \expcnd{ \left( X - \sm{X}_q \right)^2 }{ \snr;\eff,\xi }.
\end{equation}

The following is claimed.\footnote{Since as explained in Section
  \ref{s:tni}, rigorous justification for some of the key statistical
  physics tools (essentially the replica method) is still pending, the
  key results in this paper are referred to as claims.  Proofs are
  provided in Section \ref{s:prf} based on those statistical physics
  tools.}

\begin{claim}  \label{th:dp}
  Consider the multiuser channel \eref{e:sch} with input distribution
  $p_X$ and SNR distribution $P_\snr$.  Let its output be fed into the
  posterior mean estimator \eref{e:vxq} and a retrochannel
  $q_{\X|\Y,\S}$, both parameterized by the postulated input $q_X$ and
  noise level $\sigma$ (refer to Figure \ref{f:rch}).  Fix $(\beta,
  P_\snr$, $p_X, q_X, \sigma)$.  Let $X_{0k}$, $X_k$, and $\sm{X_k}_q$
  be the input, the retrochannel output and the  posterior
  mean estimate for user $k$ with input signal-to-noise ratio
  $\snr_k$.  Then,
  
  (a) The joint distribution of\, $( X_{0k}, X_k, \sm{X_k}_q )$
  conditioned on the channel state $\S$ converges in probability as
  $K\rightarrow \infty$ and $K/L\rightarrow \beta$ to the joint
  distribution of\, $(X_0, X, \sm{X}_q)$, where $X_0\sim p_X$ is the
  input to the single-user Gaussian channel \eref{e:pzx} with inverse
  noise variance $\eff$, $X$ is the output of the single-user
  retrochannel parameterized by $(q_X, \xi)$, and $\sm{X}_q$ is the
  corresponding  posterior mean estimate \eref{e:xq}, with
  $\snr=\snr_k$ (refer to Figure \ref{f:sur}).

  (b)  The parameter $\eff$, known as the multiuser efficiency, satisfies
  together with $\xi$ the coupled equations:
  \begin{subequations}    \label{e:ex}%
    \begin{eqnarray}
      \eff^{-1} &=& 1\; + \beta \, \expect{ \snr\cdot\mse(\snr;\eff,\xi) },
      \label{e:e} \\
      \xi^{-1} &=& \!\sigma^2 + \beta\,\expect{ \snr\cdot\vrc(\snr;\eff,\xi) },
      \label{e:x}
    \end{eqnarray}
  \end{subequations}
  where the expectations are taken over $P_\snr$.  In case of multiple
  solutions to \eref{e:ex}, $(\eff,\xi)$ is chosen to minimize the
  free energy\index{Free energy} expressed as\footnote{The base of
    logarithm is consistent with the unit of information measure in
    this paper unless stated otherwise.}\footnote{The integral with
    respect to $z$ is from $-\infty$ to $\infty$.  For notational
    simplicity we omit integral limits in this paper whenever they are
    clear from context.}
  \begin{equation}  \label{e:fel}
    \begin{split}
    \fe =& - \expect{ \int \pz{z} \, \log \qz{z} \intd\,z } \\
    & \quad  + \oneon{2\beta} [(\xi-1)\loge-\log\xi]
    -\half\log\frac{2\pi}{\xi} - \frac{\xi}{2\eff}\loge  \\
    & \quad + \frac{\sigma^2\xi(\eff-\xi)}{2\beta\eff} \loge
    + \oneon{2\beta}\log(2\pi) + \frac{\xi}{2\beta\eff} \loge.
  \end{split}
  \end{equation}
\end{claim}

Claim \ref{th:dp} reveals that, from an individual user's viewpoint,
the input-output relationship of the multiuser channel, PME and
companion retrochannel is increasingly similar to that under a simple
single-user setting as the system becomes large.  In other words,
given the three (scalar) input and output statistics, it is not
possible to distinguish whether the underlying system is in the
(large) multiuser or the single-user setting as depicted in Figures
\ref{f:rch} and \ref{f:sur} respectively.  It is also interesting to
note that the (asymptotically) equivalent single-user system takes an
analogous structure as the multiuser one.

Obtained using the replica method, the coupled equations \eref{e:ex}
may have multiple solutions.  This is known as phase
coexistence\index{Phase transition!phase coexistence} in statistical
physics.  Among those solutions, the thermodynamically dominant
solution is the one that gives the smallest value of the free energy
\eref{e:fel}.  This is the solution that carries relevant operational
meaning in the communication problem.  In general, as the system
parameters (such as the load) change, the dominant solution may switch
from one of the coexisting solutions to another.  This phenomenon is
known as {\em phase transition}\index{Phase transition|textbf} (refer
to Section \ref{s:nr} for numerical examples).

The single-user  PME \eref{e:xq} is merely a decision
function applied to the Gaussian channel output, which can be
expressed explicitly as
\begin{equation}
  \expqcnd{X}{Z,\snr;\xi} = \frac{ q_1(Z,\snr;\xi) }{ q_0(Z,\snr;\xi) }
  \label{e:qdf}
\end{equation}
where we define the following useful functions for all positive
integers $i=0,1,\dots$:
\begin{equation} \label{e:qzi}
  \qzi{i} = \expqcnd{ X^i \,
    q_{Z|X,\snr;\xi}(z|X,\snr;\xi) }{ \snr },
\end{equation}
where the expectation is taken over $q_X$.  Note that $\qzi{0}=
q_{Z|\snr;\xi} (z|\snr;\xi)$.  The decision function \eref{e:qdf} is
in general nonlinear.  Due to Claim \ref{th:dp}, although the
multiuser PME output $\sm{X_k}_q$ is in general non-Gaussian, it is in
fact asymptotically a function (the decision function \eref{e:qdf}) of
a conditional Gaussian random variable $Z$ centered at the actual
input $X_k$ scaled by $\sqrt{\snr_k}$ \, with a variance of
$\eff^{-1}$.

\begin{corollary} \label{cr:dp}
  In the large-system limit, the channel between the input $X_{0k}$ and
  the  multiuser posterior mean estimate $\sm{X_k}_q$ for
  user $k$ is equivalent to the Gaussian channel $p_{Z|X,\snr;\eff}$
  concatenated with the one-to-one decision function \eref{e:qdf} with
  $\snr=\snr_k$, where $\eff$ is the multiuser efficiency determined
  by Claim \ref{th:dp}.
\end{corollary}

As shown in Section \ref{s:jm}, for fixed $\snr$ and $\xi$, the
decision function \eref{e:qdf} is strictly monotone increasing in $Z$.
Therefore, in the large-system limit, given the detection output
$\sm{X_k}_q$, one can apply the inverse of the decision function to
recover an equivalent conditionally Gaussian statistic $Z$.  Note that
$\eff\in [0,1]$ from \eref{e:e}.  It is clear that, in the
large-system limit, the multiple-access interference is consolidated
into an enhancement of the thermal noise by $\eff^{-1}$, i.e., the
effective SNR is reduced by a factor of $\eff$, hence the term {\em
  multiuser efficiency}\index{Multiuser efficiency}.  Equal for all
users, the multiuser efficiency solves the coupled fixed-point
equations \eref{e:ex}.  Indeed, in the large-system limit, the
multiuser channel with the PME front end can be decoupled into a bank
of independent single-user Gaussian channels with the same degradation
in each user's SNR.  This is referred to as the {\em decoupling
  principle}.

Since the decision function is one-to-one, it is inconsequential from
both the detection and the information theoretic viewpoints.  Hence
the following result:

\begin{corollary}  \label{cr:i}
  In the large-system limit, the mutual information between input
  symbol and the output of the  multiuser posterior mean
  estimator for a particular user is equal to the input-output mutual
  information of the equivalent single-user Gaussian channel with the
  same input distribution and SNR, and an inverse noise variance
  $\eff$ equal to the multiuser efficiency given by Claim \ref{th:dp}.
\end{corollary}

According to Corollary \ref{cr:i}, the mutual information $I\left(
  X_k;\sm{X_k} | \S \right)$ for a user with signal-to-noise ratio
$\snr_k=\snr$ converges to a function of the effective SNR defined as
\begin{equation} \label{e:id}
  I(\eff\,\snr) = \Dzx,
\end{equation}
where $\cnddiv{\,\cdot\,}{\cdot}{\,\cdot\,}$ stands for conditional
(Kullback-Leibler) divergence, and $p_{Z|\snr;\eff}$ is the marginal
distribution of the output of the channel \eref{e:zx0}.  The overall
spectral efficiency under separate decoding is the sum of the
single-user mutual informations divided by the dimension of the
multiuser channel (spreading factor $L$), which is simply
\begin{equation}  \label{e:csep}
  \Csep(\beta) = \beta\, \expect{ I(\eff\,\snr) },
\end{equation}
where the expectation is over $P_\snr$.

In general, it is straightforward to determine the multiuser
efficiency $\eff$ (and the inverse noise variance $\xi$) by solving
the joint equations \eref{e:ex}.  Define the following functions akin
to~\eref{e:qzi}:
\begin{equation}  \label{e:pzi}
  \pzi{i} = \expcnd{ X^i \,
    p_{Z|X,\snr;\eff} (z \,|\, X,\snr;\eff ) }{\snr}.
\end{equation}
Some algebra leads to
\begin{equation}
  \begin{split}
  \mse(\snr;\eff,\xi) = 1 + \int &
  \pzi{0} \frac{ q_1^2(z,\snr;\xi) }{ q_0^2(z,\snr;\xi) } \\
  &\quad -2p_1(z,\snr;\eff) \frac{ q_1(z,\snr;\xi) }{ q_0(z,\snr;\xi) } \intd z
  \label{e:msez}
  \end{split}
\end{equation}
and
\begin{equation}
  \begin{split}
  \vrc(\snr;\eff,\xi) = \int &
  p_0(z,\snr;\eff) \frac{ q_2(z,\snr;\xi) }{ q_0(z,\snr;\xi) } \\
  &\quad -p_0(z,\snr;\eff) \frac{q_1^2(z,\snr;\xi)}{q_0^2(z,\snr;\xi)} \intd z.
  \label{e:vrcz}
  \end{split}
\end{equation}
Numerical integrations can be applied to evaluate \eref{e:msez} and
\eref{e:vrcz} in general.  It is then viable to find solutions to the
joint equations \eref{e:ex} numerically.  In case of multiple sets of
solutions, the ambiguity is resolved by choosing the one that
minimizes the free energy \eref{e:fel}.  Note that the mean-square
error and variance often admit simpler expressions than \eref{e:msez}
and \eref{e:vrcz} under certain practical inputs, which may ease the
computation significantly (see examples in Section \ref{s:kr}).

\subsubsection{Optimal Detection and Spectral Efficiency}
\label{s:js}
\index{Spectral efficiency}

Among all multiuser detection schemes, the individually optimal
detector has particular importance.  As we shall see, the optimal
spectral efficiency achievable by joint decoding is also tightly
related to the multiuser efficiency of optimal detection.

As shown in Section \ref{s:sd}, the soft individually optimal detector
can be regarded as a PME with a postulated measure that is exactly the
same as the actual measure, i.e., $q=p$.  Consider the channel, PME
and retrochannel in the multiuser setting as depicted in Figure
\ref{f:rch}.  It is clear that in case of optimal detection, the input
$\X_0$ to the multiuser channel and the retrochannel output $\X$ are
i.i.d.\ given $(\Y,\S)$.  The decoupling principle stated in Claim
\ref{th:dp} can be particularized in the case of $q=p$.  Easily, the
multiuser efficiency and the postulated inverse noise variance satisfy
joint equations:
\begin{subequations}
  \begin{eqnarray}
  \eff^{-1} &=& 1 + \beta\,\expect{ \snr\cdot\mse(\snr;\eff,\xi) }, \\
  \xi^{-1}  &=& 1 + \beta\,\expect{ \snr\cdot\vrc(\snr;\eff,\xi) }.
  \end{eqnarray}
  \label{e:exs}%
\end{subequations}
Due to the replica symmetry assumption, and noting that
$\mse(\snr;x,x) =\vrc(\snr;x,x)$ for all $x$, we take the solution
$\eff=\xi$.  It should be cautioned that \eref{e:exs} may have other
solutions with $\eff\neq\xi$ in the unlikely case that replica
symmetry does not hold for optimal detection.

In the equivalent single-user setting (Figure \ref{f:sur}), the above
arguments imply that the postulated channel is also identical to the
actual channel, and $X$ and $X_0$ are i.i.d.\ given $Z$.  The
posterior mean estimate of $X$ given the output $Z$ is
\begin{equation}  \label{e:xp}
  \sm{X} = \expect{ X | Z, \snr; \eff}.
\end{equation}
Clearly, $\sm{X}$ is also the (nonlinear) MMSE estimate, since it
achieves the minimum mean-square error:\index{Minimum mean-square
error (MMSE)}
\begin{equation}
  \mmse(\eff\,\snr) = \expcnd{ (X-\sm{X})^2 }{ \snr; \eff }.
  \label{e:mse}
\end{equation}
Indeed,
\begin{eqnarray}  \label{e:mv}
  \mse(\snr;x,x) = \vrc(\snr;x,x) = \mmse(x\,\snr), \quad\forall x.
\end{eqnarray}

The following is a special case of Corollary \ref{cr:dp} for the
individually optimal detector.
\begin{claim}  \label{th:c}
  In the large-system limit, the distribution of the output $\sm{X_k}$
  of the individually optimal detector for the multiuser channel
  \eref{e:sch} conditioned on $X_k=x$ being transmitted with
  signal-to-noise ratio $\snr_k$ is identical to the distribution of
  the posterior mean estimate $\sm{X}$ of the single-user Gaussian
  channel \eref{e:zx0} conditioned on $X_0=x$ being transmitted with
  $\snr =\snr_k$, where the optimal multiuser efficiency $\eff$
  satisfies a fixed-point equation:
  \begin{equation}
    \eff^{-1} = 1 + \beta\, \expect{ \snr\cdot\mmse(\eff\,\snr) }.
    \label{e:ecme}
  \end{equation}
\end{claim}

The single-user PME \eref{e:xp} is a (nonlinear) decision function
that admits an expression as \eref{e:qdf} with $q$ replaced by $p$.
The MMSE can be computed as
\begin{equation}
  \mmse(\eff\,\snr) = 1 - \int \frac{ p_1^2(z,\snr;\eff) }
      { p_0(z,\snr;\eff) } \intd z.
  \label{e:ma}
\end{equation}
Solutions to the fixed-point equation \eref{e:ecme} can in general be
found numerically.  There are cases in which \eref{e:ecme} has more
than one solution.  The ambiguity is resolved by taking the one that
minimizes the free energy \eref{e:fel} with $\xi=\eff$, or
equivalently, as we shall see next, the optimal spectral efficiency.

The single-user mutual information is given by \eref{e:id} due to
Corollary \ref{cr:i}, where the multiuser efficiency is now given by
Claim \ref{th:c}.  The optimal spectral efficiency under joint
decoding is greater than that under separate decoding \eref{e:csep},
where the increase is given by the following:

\begin{claim}  \label{th:x}%
  The spectral efficiency gain of optimal joint decoding over
  individually optimal detection followed by separate decoding of the
  multiuser channel \eref{e:sch} is determined, in the large-system
  limit, by the optimal multiuser efficiency as
  \begin{eqnarray}
    \Cjnt(\beta) - \Csep(\beta)
    &=& \half[ (\eff-1)\loge -\log\eff ] \\
    &=& \divergence{ \mathcal{N}(0,\eff) }{ \mathcal{N}(0,1) }.
    \label{e:gcc}%
  \end{eqnarray}%
  In other words, the spectral efficiency under joint decoding is
  \begin{equation}
    \Cjnt(\beta) = \beta\,\expect{ I(\eff\,\snr) }
    + \half[ (\eff-1)\loge -\log\eff ].
  \label{e:cj}
  \end{equation}
  In case of multiple solutions to \eref{e:ecme}, the optimal
  multiuser efficiency $\eff$ is the one that gives the smallest
  $\Cjnt$.
\end{claim}

Indeed, M\"uller's conjecture on the mutual information loss
\cite{Muller02WCIT} is true for arbitrary inputs and SNRs.
Incidentally, the loss is identified as a divergence between two
Gaussian distributions in \eref{e:gcc}.

Equal-power Gaussian input is the first known case that admits a
closed-form solution for the multiuser efficiency \cite[p.\ 
305]{Verdu98} and thus also the spectral efficiencies.  The spectral
efficiencies under joint and separate decoding were found for Gaussian
inputs with fading in \cite{ShaVer01IT}, and then found implicitly in
\cite{Tanaka02IT} and later explicitly in \cite{MulGer04IT} for
equal-power users with binary inputs.  Formula \eref{e:cj} is the
first general result for arbitrary input distributions and received
powers.

Interestingly, the spectral efficiencies under joint and separate
decoding are also related by an integral equation, given in
\cite[(160)]{ShaVer01IT} for the special case of Gaussian inputs.

\begin{theorem}  \label{th:js}
  Regardless of the input and power distributions,
  \begin{equation}
    \Cjnt(\beta) = \int_0^\beta \oneon{\beta'} \Csep(\beta') \intd \beta'.
    \label{e:js}
  \end{equation}
\end{theorem}

\begin{proof}
Since $\Cjnt(0)=0$ trivially, it suffices to show
\begin{equation}
  \beta \pd{\beta} \Cjnt(\beta) = \Csep(\beta).
  \label{e:db}
\end{equation}
By \eref{e:gcc} and \eref{e:cj}, it is enough to show
\begin{equation}
  \beta \pd{\beta} \expect{ I(\eff\,\snr) }
  + \half \pd{\beta} [ (\eff-1) \loge - \log\eff ] = 0.
  \label{e:ddz}
\end{equation}
Noticing that the multiuser efficiency $\eff$ is a function of the
system load $\beta$, \eref{e:ddz} is equivalent to
\begin{equation}
  \pd{\eff} \expect{ I(\eff\,\snr) }
  + \oneon{2\beta} \left( 1-\eff^{-1} \right) \loge = 0.
  \label{e:eie}
\end{equation}
By a recent formula that links the mutual information and MMSE in
Gaussian channels \cite{GuoSha05IT},\footnote{In fact, the proof of
  Theorem \ref{th:js} led us to the discovery of the general I-MMSE
  relationship in \cite{GuoSha05IT}.}
\begin{equation}
  \oneon{\loge} \pd{\eff} I(\eff\snr)
  = \frac{\snr}{2} \, \mmse(\eff\,\snr).
\end{equation}
Thus \eref{e:eie} holds as $\eff$ satisfies the fixed-point equation
\eref{e:ecme}.
\end{proof}

Theorem \ref{th:js} is an outcome of the chain rule of mutual
information, which holds for all inputs and arbitrary number of users:
\begin{equation}
  I(\X;\Y|\S) = \sum^K_{k=1} I(X_k;\Y|\S,X_{k+1},\dots,X_K).
  \label{e:isi}
\end{equation}
The left hand side of \eref{e:isi} is the total mutual information of
the multiuser channel.  Each mutual information in the right hand side
of \eref{e:isi} is a single-user mutual information over the multiuser
channel conditioned on the symbols of previously decoded users.  As
argued below, the limit of \eref{e:isi} as $K\rightarrow\infty$
becomes the integral equation \eref{e:js}.

Consider an interference canceler with PME front ends against yet
undecoded users that decodes the users successively in which reliably
decoded symbols are used to reconstruct the interference for
cancellation.  Since the error probability of intermediate decisions
vanishes with code block-length, the interference from decoded users
are asymptotically completely removed.  Assume without loss of
generality that the users are decoded in reverse order, then the PME
for user $k$ sees only $k-1$ interfering users.  Hence the performance
for user $k$ under such successive decoding is identical to that under
multiuser detection with separate decoding in a system with $k$
instead of $K$ users.  Nonetheless, the equivalent single-user channel
for each user is Gaussian by Corollary \ref{cr:dp}.  The multiuser
efficiency experienced by user $k$, $\eff(k/L)$, is a function of the
load $k/L$ seen by the PME for user $k$.  By Corollary \ref{cr:i}, the
single-user mutual information for user $k$ is therefore
\begin{equation}
  I\left(\eff(k/L) \,\snr_k \right).
\end{equation}
Since $\snr_k$ are i.i.d., the overall spectral efficiency under
successive decoding converges almost surely:
\begin{equation}
  \oneon{L} \sumK I\left( \eff (k/L) \, \snr_k \right)
  \rightarrow \expect{ \int_0^\beta I(\beta'\,\snr) \intd \beta' }.
  \label{e:ibc}
\end{equation}

Note that the above result on successive decoding is true for
arbitrary input distribution and arbitrary PME detectors.  In the
special case of individually optimal detection, for which the
postulated system is identical to the actual one, the right hand side
of \eref{e:ibc} is equal to $\Cjnt(\beta)$ by Theorem \ref{th:js}.  We
can summarize this principle as:
\begin{claim} \label{th:s}
  In the large-system limit, successive decoding with an individually
  optimal detection front end against yet undecoded users achieves the
  optimal CDMA channel capacity under arbitrary constraint on the
  input.
\end{claim}

Claim \ref{th:s} is a generalization of the result that a successive
canceler with a linear MMSE front end against undecoded users achieves
the capacity of the CDMA channel under \emph{Gaussian
  inputs}.\footnote{This principle, originally discovered by Varanasi
  and Guess \cite{VarGue97Asilomar}, has been shown with other proofs
  and in other settings \cite{VerSha99IT, RapHon98ISIT, Ariyav00TC,
    Muller01IT, GueVar04IT, Forney04Allerton}.}


\subsection{Recovering Known Results}
\label{s:kr}

As shown in \ref{s:sd}, several well-known multiuser detectors can be
regarded as appropriately parameterized PMEs.  Thus many previously
known results can be recovered as special case of the new findings in
Section \ref{s:main}.

\subsubsection{Linear Detectors}

Let the postulated prior $q_X$ be standard Gaussian so that the
 PME represents a linear multiuser detector.  Since the
input $Z$ and output $X$ of the retrochannel are jointly Gaussian
(refer to Figure \ref{f:sur}), the single-user PME is simply a linear
attenuator:
\begin{equation}
  \sm{X}_q = \frac{\xi\sqrt{\snr}}{1+\xi\snr} Z.
\end{equation}
From \eref{e:gmse}, the mean-square error is
\begin{eqnarray}
  \mse(\snr;\eff,\xi) \nsp{2}
  &=& \nsp{2} \expect{ \left[ X_0 - \frac{\xi\sqrt{\snr}}{1+\xi\snr} 
    \Big( \sqrt{\snr}X_0 + \frac{N}{\sqrt{\eff}} \Big) \right]^2 } \\
  \nsp{2} &=& \nsp{2} \frac{\eff+\xi^2\snr}{ \eff (1+\xi\snr)^2 }. 
\end{eqnarray}
Meanwhile, the variance of $X$ conditioned on $Z$ is independent of
$Z$.  Hence the variance \eref{e:vrc} of the retrochannel output is
independent of $\eff$:
\begin{equation}
  \vrc(\snr;\eff,\xi) = \oneon{1+\xi\snr}.
\end{equation}
From Claim \ref{th:dp}, one finds that $\xi$ is the solution to
\begin{equation}
  \xi^{-1} = \sigma^2 + \beta\, \expect{ \frac{\snr}{1+\xi\snr} },
  \label{e:xl}
\end{equation}
and the multiuser efficiency is determined as
\begin{equation}
  \eff = \xi + \xi\, (\sigma^2-1)\, \left[
    1+\beta\,\expect{\frac{\snr}{(1+\xi\snr)^2}} \right]^{-1}.
  \label{e:el}
\end{equation}
Clearly, the large-system multiuser efficiency of such a linear
detector is independent of the input distribution.

Suppose also that the postulated noise level $\sigma \rightarrow
\infty$.  The PME becomes the matched filter.  One finds $\xi\sigma^2
\rightarrow1$ by \eref{e:xl} and consequently, the multiuser
efficiency of the matched filter is \cite{Verdu98}
\begin{equation}
  \eff^{\text{(mf)}} = \oneon{1+\beta\,\expect{\snr}}.
\end{equation}

In case $\sigma=1$, one has the linear MMSE detector.  By \eref{e:el},
$\eff=\xi$ and by \eref{e:xl}, the multiuser efficiency $\eff^\supn$
satisfies
\begin{equation}
  \eff^{-1} = 1 + \beta\,\expect{\frac{\snr}{1+\eff\snr}},
  \label{e:th}
\end{equation}
which is the Tse-Hanly equation \cite{TseHan99IT,
  VerSha99IT}.\index{Tse-Hanly equation|textbf} The fixed-point
equation \eref{e:th} has a unique positive solution.

By letting $\sigma \rightarrow0$ one obtains the decorrelator.  If
$\beta<1$, then \eref{e:xl} gives $\xi\rightarrow\infty$ and
$\xi\sigma^2\rightarrow 1-\beta$, and the multiuser efficiency is
found as $\eff=1-\beta$ by \eref{e:el} regardless of the SNR
distribution (as shown in \cite{Verdu98}).  If $\beta>1$, and assuming
the generalized form of the decorrelator as the Moore-Penrose inverse
of the correlation matrix \cite{Verdu98}, then $\xi$ is the unique
solution to
\begin{equation}
  \xi^{-1} = \beta\, \expect{ \frac{\snr}{1+\xi\snr} }
\end{equation}
and the multiuser efficiency is found by \eref{e:el} with $\sigma=0$.
In the special case of identical SNRs, an explicit expression is found
\cite{EldCha03IT,GuoVer02Blake}
\begin{equation}
  \eff^{\text{(dec)}} = \frac{\beta-1}{\beta + \snr(\beta-1)^2},
  \quad\quad \beta>1.
\end{equation}

By Corollary \ref{cr:dp}, the mutual information with input
distribution $p_X$ for a user with $\snr$ under linear multiuser
detection is equal to the input-output mutual information of the
single-user Gaussian channel \eref{e:zx0} with the same input:
\begin{equation}
  I(X;\sm{X}_q|\snr) = I(\eff\,\snr), 
\end{equation}
where $\eff$ depends on which type of linear detector is in use.
Gaussian priors are known to achieve the capacity:
\begin{equation}
  \capacity(\snr) = \half \log(1+\eff\,\snr).
  \label{e:cns}
\end{equation}
By Corollary \ref{th:x}, the total spectral efficiency under Gaussian
inputs is expressed in terms of the linear MMSE multiuser efficiency:
\begin{equation}
  \begin{split}
  \Cjnt^\text{(Gaussian)}
  = & \frac{\beta}{2} \, \expect{\log\left(1+\eff^\supn\snr\right)} \\
  & \quad+ \half \left[ \left(\eff^\supn-1\right)\loge - \log\eff^\supn\right].
  \label{e:cn}
  \end{split}
\end{equation}
This is Shamai and Verd\'u's result for fading channels
\cite{ShaVer01IT}.

\subsubsection{Optimal Detectors}
\label{s:opt}

Using the actual input distribution $p_X$ as the postulated prior of
the  PME results in optimum multiuser detectors.  In case
of the jointly optimal detector, the postulated noise level
$\sigma=0$, and \eref{e:ex} becomes
\begin{subequations}
  \begin{eqnarray}
  \eff^{-1} &=& 1 + \beta\,\expect{ \snr\cdot\mse(\snr;\eff,\xi) }, \\
  \xi^{-1}  &=& \beta\,\expect{ \snr\cdot\vrc(\snr;\eff,\xi) },
  \end{eqnarray}
  \label{e:exz}%
\end{subequations}
where $\mse(\cdot)$ and $\vrc(\cdot)$ are given by \eref{e:msez} and
\eref{e:vrcz} respectively with $q_i(z,\snr;x) =p_i(z,\snr;x)$,
$\forall x$.  The parameters can then be solved numerically.

In case of the individually optimal detector, one sets $\sigma=1$ so
that $q=p$.  The optimal multiuser efficiency $\eff$ is the solution
to the fixed-point equation \eref{e:ecme} given in Claim \ref{th:c}.

It is of practical interest to find the spectral efficiency under the
constraint that the input symbols are antipodally modulated as in the
popular BPSK.  In this case, the probability mass function
$p_X(x)=1/2$, $x=\pm1$, maximizes the mutual information.  It can be
shown that
\begin{equation}
  \mmse(\gamma) = 1 - \int \frac{ e^{-\frac{z^2}{2}} }{\sqrt{2\pi}}
  \tanh \left( \gamma - z\sqrt{\gamma} \right) \intd z.
\end{equation}
By Claim \ref{th:c}, The multiuser efficiency, $\eff^\supb$, where the
superscript (b) stands for binary inputs, is a solution to the
fixed-point equation \cite{GuoVer02Blake}:
\begin{equation}
  \oneon{\eff} = 1 + \beta\,\expect{ \snr
    \bigg[ 1 - \int \frac{ e^{-\frac{z^2}{2}} }{\sqrt{2\pi}} \tanh
      \left( \eff\snr - z\sqrt{\eff \snr} \right) \intd z\bigg] }
  \label{e:meb}
\end{equation}
which is a generalization of an earlier result assuming equal-power
users due to Tanaka \cite{Tanaka02IT}.  The single-user channel
capacity for a user with signal-to-noise ratio $\snr$ is the same as
that obtained by M\"uller and Gerstacker \cite{MulGer04IT} and is
given by
\begin{equation}
  \begin{split}
  \capacity^\supb(\snr) = &
  - \int \frac{ e^{-\frac{z^2}{2}} }{\sqrt{2\pi}}
    \log\,\cosh \left( \eff^\supb \snr
    - z \sqrt{\eff^\supb \snr} \right) \intd z \\
  & \quad + \eff^\supb \, \snr \loge.
  \label{e:cbs}
  \end{split}
\end{equation}
The total spectral efficiency of the  CDMA channel subject to
binary inputs is thus
\begin{equation}
  \begin{split}
    \Cjnt^\supb = & \beta \, \expect{ -
    \int \frac{ e^{-\frac{z^2}{2}} }{\sqrt{2\pi}} \log\,\cosh
    \left( \eff^\supb \snr - z \sqrt{\eff^\supb \snr} \right) \intd z} \\
    & + \beta \, \eff^\supb \, \Exp{\snr} \loge
    + \half \left[ \left(\eff^\supb-1\right)\loge-\log\eff^\supb\right],
  \end{split}
\label{e:cb}
\end{equation}
which is also a generalization of Tanaka's implicit result
\cite{Tanaka02IT}.

\section{Communications and Statistical Physics}
\label{s:stat}

This section briefs the reader with concepts and methodologies that
will be needed to prove the results summarized in Section
\ref{s:main}.  Although one can work with the mathematical framework
only and avoid foreign concepts, we believe it is more enlightening to
draw an equivalence between multiuser communications and many-body
problems in statistical physics.  Such an analogy is seen in a
embryonic form in \cite{Tanaka02IT} and will be developed to a full
generality here.

\subsection{A Note on Statistical Physics}
\label{s:note}
\index{Statistical physics|textbf}

Consider the physics of a many-body system, the microscopic state of
which is described by the configuration of some $K$ variables as a
vector $\x$.  The state of the system evolves over time according to
some physical laws.  Let the energy associated with the state, called
the {\em Hamiltonian},\index{Hamiltonian|textbf} be denoted by the
function $H(\x)$.  Let $p(\x)$ denote the probability that the system
is found in configuration $\x$.  Then, at thermal equilibrium, the
{\em energy} of the system
\begin{equation}  \label{e:eph}
  \energy = \sum_{\x} p(\x) H(\x)
\end{equation}
is preserved, while the Second Law of Thermodynamics dictates that the
{\em entropy} (disorder) of the system
\begin{equation}  \label{e:etp}
  \entropys = -\sum_\x p(\x) \log p(\x)
\end{equation}
is maximized.  Although we are unable to follow the exact trajectory
of the configuration, e.g., we do not know the exact configuration
$\x$ at a given time, the probability distribution of the
configuration can be determined using the Lagrange multiplier method.
Indeed, using \eref{e:eph} and \eref{e:etp}, the equilibrium
probability distribution $p(\x)$ is found to be negative exponential
in the Hamiltonian, which is known as the {\em Boltzmann
  distribution}:\index{Boltzmann distribution|textbf}
\begin{equation}
  p(\x) = Z^{-1} \, \exp\left[ -\oneon{T} H(\x) \right]
  \label{e:bz}
\end{equation}
where
\begin{equation}
  Z = \sum_\x \exp\left[ -\oneon{T} H(\x) \right]
\label{e:zt}
\end{equation}
is the {\em partition function},\index{Partition function|textbf} and
the {\em temperature} $T\ge0$ is determined by the energy constraint
\eref{e:eph}.  The most probable configuration is the ground state
which has the minimum Hamiltonian.  Generally speaking, statistical
physics is a theory that studies macroscopic properties (e.g.,
pressure, magnetization) of such a system starting from the
Hamiltonian by taking the above probabilistic viewpoint.  One
particularly useful macroscopic quantity of the thermodynamic system
is the {\em free energy}:\index{Free energy|textbf}
\begin{equation}
  \fe = \energy - T\,\entropys.
\end{equation}
Using \eref{e:eph}--\eref{e:zt}, one finds that the free energy at
equilibrium can also be expressed as
\begin{equation}
  \fe = -T\,\log Z.
\end{equation}
Indeed, at thermal equilibrium, the temperature and energy of the
system remain constant, the entropy is the maximum possible, and the
free energy is at its minimum.  The free energy is often the starting
point for calculating macroscopic properties of a thermodynamic
system.

\subsection{Multiuser Communications and Spin Glasses}
\label{s:sg}

The communication problem faced by the detector is to infer
statistically the information-bearing symbols given the received
signal and knowledge about the channel state.  Naturally, the
posterior probability distribution plays a central role.  In the
multiple-access channel \eref{e:sch}, the channel state consists of
the spreading sequences and the SNRs, collectively represented by the
matrix $\S$.  The channel is described by the Gaussian density
$p_{\Y|\X,\S}$ given by \eref{e:py}.  By postulating an input $q_X$
and a channel \eref{e:qy} which differs from the actual one only in
the noise level, the postulated posterior distribution can be obtained
by using the Bayes formula (cf.\ \eref{e:bf}) as
\begin{equation}  \label{e:qxys}
    q_{\X|\Y,\S}(\x|\y,\S) = \frac{ \tps^{-\frac{L}{2}} q_\X(\x) }{
      q_{\Y|\S}(\y|\S) } \expb{ -\frac{ \ySx }{2\sigma^2} }
\end{equation}
where
\begin{equation}
  q_{\Y|\S}(\y|\S) =
  \tps^{-\frac{L}{2}} \expqcnd{ \expb{ -\frac{\ySX}{2\sigma^2} } }{ \S }
  \label{e:qys}
\end{equation}
and the expectation in \eref{e:qys} is taken conditioned on $\S$ over
$\X$ with distribution $q_\X$.

In order to take advantage of the statistical physics methodologies,
we create an artificial thermodynamic system, called spin glass, that
is equivalent to the communication problem.  In certain special cases,
this connection is found in \cite{Tanaka02IT}, while we now draw this
analogy in the general setting.  A {\em spin glass}\index{Spin
  glass|textbf} is a system consisting of many directional spins, in
which the interaction of the spins is determined by the so-called {\em
  quenched random variables}\index{Quenched randomness!quenched random
  variable} whose values are determined by the realization of the spin
glass.  An example is a system consisting molecules with magnetic
spins that evolve over time, while the positions of the molecules that
determine the amount of interactions are random (disordered) but
remain fixed for each concrete instance as in a piece of glass.  Let
the microscopic state of a spin glass be denoted by a $K$-dimensional
vector $\x$, and the quenched random variables by $(\y,\S)$.  The
system can be understood as $K$ random spins sitting in quenched
randomness $(\y,\S)$, and its statistical physics described as in
Section \ref{s:note} with a parameterized Hamiltonian $H_{\y,\S}(\x)$.

Indeed, suppose the temperature $T=1$ and that the Hamiltonian of a
piece of spin glass is defined as
\begin{equation}  \label{e:hrs}
  H_{\y,\S} (\x) = \frac{\ySx}{2\sigma^2} - \log q_\X(\x)
        + \frac{L}{2} \log\tps,
\end{equation}
then the configuration distribution of the spin glass at equilibrium
is given by \eref{e:qxys} and its corresponding partition function by
\eref{e:qys} (cf.\ \eref{e:bz} and \eref{e:zt}).  Precisely, the
probability that the transmitted symbol is $\X=\x$ under the
postulated model, given the observation $\Y$ and the channel state
$\S$, is equal to the probability that the spin glass is found at
configuration $\x$, given the quenched random variables $(\Y,\S)$.
Note that Gaussian distribution is a natural Boltzmann distribution
with squared Euclidean norm as the Hamiltonian.

The richness of the system is encoded in the quenched randomness
$(\Y,\S)$.  In the communication channel described by~\eref{e:sch},
$(\Y,\S)$ takes a specific distribution, i.e., it is a realization of
the received signal and channel state matrix according to the prior
and conditional distributions that underlie the ``original'' spins.
Indeed, the communication system depicted in Figure~\ref{f:rch} can be
also understood as a spin glass $\X$ subject to physical law $q$
sitting in the quenched randomness caused by another spin glass $\X_0$
subject to physical law $p$.  The channel corresponds to the random
mapping from a given spin glass configuration to an induced quenched
randomness.  Conversely, the retrochannel corresponds to the random
mechanism that maps some quenched randomness into an induced spin
glass configuration distribution.


The free energy of the thermodynamic (or communication) system
normalized by the number of users is ($T=1$)
\begin{equation}
  - \frac{T}{K} \log Z(\Y,\S) = -\oneon{K} \log q_{\Y|\S}(\Y|\S).
  \label{e:logp}
\end{equation}
Due to the self-averaging assumption, the randomness of \eref{e:logp}
vanishes as $K\rightarrow\infty$.  As a result, the free energy per
user converges in probability to its expected value over the
distribution of the quenched random variables $(\Y,\S)$ in the
large-system limit, which is denoted by $\fe$,
\begin{equation}
  \fe = - \liminfty{K} \expect{ \oneon{K} \log q_{\Y|\S}(\Y|\S) }.
  \label{e:fez}
\end{equation}
Hereafter, by the free energy we refer to the large-system limit
\eref{e:fez}, which will be calculated in Section \ref{s:prf}.

The reader should be cautioned that for disordered systems,
thermodynamic quantities may or may not be self-averaging
\cite{Comets98ib}.  The self-averaging property remains to be proved
or disproved in the CDMA context.  This is a challenging problem on
its own.  Buttressed by numerical examples and associated results
using random matrix theory, in this work the self-averaging property
is assumed to hold.

The self-averaging property resembles the asymptotic equipartition
property (AEP) in information theory \cite{CovTho91}.\index{Asymptotic
equipartition property (AEP)} An important consequence is that a
macroscopic quantity of a thermodynamic system, which is a function of
a large number of random variables, may become increasingly
predictable from merely a few parameters independent of the
realization of the random variables as the system size grows without
bound.  Indeed, such a macroscopic quantity converges in probability
to its ensemble average in the thermodynamic limit.

In the CDMA context, the self-averaging property leads to the strong
consequence that for almost all realizations of the received signal
and the spreading sequences, macroscopic quantities such as the BER,
the output SNR and the spectral efficiency, averaged over data,
converge to deterministic quantities in the large-system
limit.\index{Large system!large-system limit} Previous work (e.g.
\cite{TseHan99IT, GuoVer02IT, VerSha99IT}) has shown convergence of
performance measures for almost all spreading sequences.  The
self-averaging property results in convergence of certain empirical
performance measures, which holds for almost all realizations of the
data as well as noise.

\subsection{Spectral Efficiency and Detection Performance}

Consider the multiuser channel, the multiuser PME and the companion
retrochannel as depicted in Figure \ref{f:rch}.  Equipped with the
statistical physics concepts introduced in \ref{s:note} and
\ref{s:sg}, this subsection associates the spectral efficiency and
detection performance of such a system with more tangible quantities
for calculation.

\subsubsection{Spectral Efficiency and Free Energy}

For a fixed input distribution $p_X$, the total input-output mutual
information of the multiuser channel is
\begin{eqnarray}
  I(\X;\Y|\S)
  \nsp{2}&=&\nsp{2} \expcnd{ \log \frac{ p_{\Y|\X,\S}(\Y|\X,\S) }
    { p_{\Y|\S}(\Y|\S) } }{ \S } \label{e:mi} \\
  \nsp{2}&=&\nsp{2} \expcnd{ \log p_{\Y|\S}(\Y|\S) }{\S}
  - \frac{L}{2} \log (2\pi e).
  \label{e:mie}
\end{eqnarray}
where the simplification to \eref{e:mie} is because $p_{\Y|\X,\S}$
given by \eref{e:py} is an $L$-dimensional Gaussian density.
Calculating \eref{e:mie} is formidable for an arbitrary realization of
$\S$.  However, due to the self-averaging
property,\index{Self-averaging!self-averaging property} it suffices to
evaluate its expectation over the spreading sequences.  In view of
\eref{e:fez}, the large-system spectral efficiency is affine in the
free energy with a postulated measure $q$ identical to the actual
measure $p$:
\begin{eqnarray}
  \capacity
  \nsp{1}&=&\nsp{1} \oneon{L} I(\X;\Y|\S) \\
  \nsp{1}&=&\nsp{1} - \beta \,
  \expcnd{ \oneon{K} \log p_{\Y|\S}(\Y|\S) }{ \S } -\half\log(2\pi e) \\
  \label{e:cz}
  &\rightarrow& \beta \, \fe|_{q=p} - \half\log(2\pi e).
  \label{e:cf}
\end{eqnarray}

Relationship \eref{e:cf} is a full generalization of a previous
observation \cite[(82)]{Tanaka02IT} in some special cases.  In fact,
the analogy between free energy and information-theoretic quantities
has also been noticed in belief propagation \cite{YedFre01NIPS,
  PakAna02CISS}, coding \cite{MouSaa02PRE} and optimization problems
\cite{ChiBoy04IT}.

\subsubsection{Detection Performance and Moments}

In case of a multiuser detector front end, one is interested in the
quality of the detection output for each user, which is completely
described by the distribution of the detection output conditioned on
the input.  Let us focus on an arbitrary user $k$, and let $X_{0k}$,
$\sm{X_k}_q$ and $X_k$ be the input, the PME output, and the
retrochannel output, respectively (cf.\ Figure \ref{f:rch}).  Instead
of the conditional distribution $P_{\sm{X_k}_q|X_{0k}}$, we solve a
more ambitious problem: the joint distribution of $( X_{0k},
\sm{X_k}_q, X_k)$ conditioned on the channel state $\S$ in the
large-system limit.

Our approach is to calculate the joint moments
\begin{equation}
  \expcnd{ X_{0k}^i \, X_k^j\, \sm{X_k}_q^l}{\S},   \quad i,j,l=0,1,\dots
  \label{e:edij}
\end{equation}
By the self-averaging property,\index{Self-averaging!self-averaging
  property} each moment, as a function of the channel state $\S$,
converges to the same value for almost all realizations of $\S$.  Thus
it suffices to calculate
\begin{equation}
  \expect{ X_{0k}^i \, X_k^j \, \sm{X_k}_q^l }
  \label{e:mmt}
\end{equation}
as $K\rightarrow\infty$, which is viable by studying the free energy
associated with a modified version of the partition function
\eref{e:qys}.  More on this later.

The joint distribution becomes clear once all the moments \eref{e:mmt}
are determined, so does the relationship between the detection output
$\sm{X_k}_q$ and the input $X_{0k}$.  It turns out the large-system
joint distribution of $( X_{0k}, \sm{X_k}_q, X_k)$ is exactly the same
as that of the input, PME output and retrochannel output associated
with a single-user Gaussian channel with the same input distribution
but with a degradation in the SNR.  In other words, the subchannel seen by
an individual user is essentially equivalent to a single-user Gaussian
channel in the large-system limit.  The mutual information between the
input and the detection output for user $k$ is expressed as
\begin{equation}
  I( X_{0k};\sm{X_k}_q \,|\, \S ),
  \label{e:ik}
\end{equation}
which can be obtained once the input-output relationship is known.  It
will be shown that conditioning on the channel state $\S$ becomes
superfluous as $K\rightarrow \infty$.

We have distilled our problems under both joint and separate decoding
to finding some ensemble averages, namely, the free energy
\eref{e:fez} and the joint moments \eref{e:mmt}.  In order to
calculate these quantities, we resort to a powerful technique
developed in the theory of spin glass, the heart of which is sketched
in the following subsection.

\subsection{Replica Method}
\index{Replica method|textbf}

\begin{figure}
  \begin{center}
\begin{picture}(250,95)(0,0)

\thicklines

\put(0,25){\framebox(30,30){$p_\X$}}
\put(30,40){\vector(1,0){20}}
\put(33,45){$\X_0$}

\put(50,25){\framebox(40,30){$p_{\Y|\X,\S}$}}
\put(70,15){\vector(0,1){10}}
\put(67,5){$\S$}

\put(90,40){\vector(1,0){20}}
\put(95,43){$\Y$}

\put(110,20){\framebox(70,40){
    \parbox{70pt}{\center Retrochannel 1 $q_{\X|\Y,\S}$} }}
\put(120,70){\line(1,0){70}}
\put(120,60){\line(0,1){10}}
\put(190,30){\line(0,1){40}}
\put(180,30){\line(1,0){10}}
\put(140,90){\line(1,0){70}}
\put(140,80){\line(0,1){10}}
\put(210,55){\line(0,1){35}}

\put(145,75){Retrochannel $u$}

\put(152,0){$\S$}

\put(180,40){\vector(1,0){25}}
\put(190,50){\vector(1,0){25}}
\put(210,70){\vector(1,0){25}}
\put(205,36){$\X_1$}
\put(215,46){$\X_2$}
\put(235,66){$\X_u$}

\put(155,10){\vector(0,1){10}}
\put(152,0){$\S$}

\end{picture}
    \caption{The replicas of the retrochannel.}
    \label{f:vrchs}
  \end{center}
\end{figure}
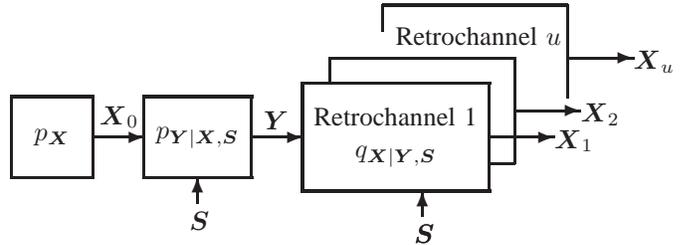

Direct calculation of the free energy in \eref{e:cf} is hard.  In
1975, S.~F.~Edwards and P.~W.~Anderson \cite{EdwAnd75JPF} invented the
replica method to study the free energy of magnetic and disordered
systems, which has since become a standard technique in statistical
physics \cite{MezPar87}.  The replica method was introduced to the
field of multiuser detection by Tanaka \cite{Tanaka02IT} to analyze
the optimal detectors under equal-power Gaussian or binary input (see
also \cite{Muller04ESPC}).  Concurrent to our work
\cite{GuoVer02Blake, GuoVer02ISIT, GuoVer03ITW, Guo04PhD}, the replica
method has also been used to analyze large dual antenna systems
\cite{Muller03TSP} and belief propagation decoding of CDMA
\cite{Caire02JWCC, Kabash03JPA, CaiMul04IT, TanOka05IT}.

Essentially, the replica method takes the following steps:
\begin{enumerate}
\item Reformulate \eref{e:fez} as
\begin{equation}
  \fe = - \liminfty{K} \oneon{K} \limzero{u} \frac{\partial}
      {\partial u} \log \expect{ Z^u(\Y,\S) }
  \label{e:fku}
\end{equation}
where $Z(\Y,\S)=q_{\Y|\S}(\Y|\S)$.  The equivalence of \eref{e:fez}
and \eref{e:fku} can be verified by noticing that for all $\Theta>0$,
\begin{equation}
  \limzero{u} \frac{\partial}{\partial u} \log \expect{ \Theta^u }
  = \limzero{u} \frac{ \expect{\Theta^u \log \Theta} }{ \expect{\Theta^u} }
  = \expect{ \log \Theta }.
\end{equation}

\item For an arbitrary positive integer $u$, calculate
\begin{equation}
  - \liminfty{K} \oneon{K} \log \expect{ Z^u(\Y,\S) }
  \label{e:logz}
\end{equation}
by introducing $u$ replicas of the system (hence the name ``replica''
method).

\item Assuming the resulting expression from Step 2 to be valid for
  all real-valued $u$ at the vicinity of $u=0$, take its derivative at
  $u=0$ to obtain the free energy \eref{e:fku}.  It is also assumed
  that the limits in \eref{e:fku} can be interchanged.
\end{enumerate}

Note that the validity of the replica method hinges on the two
assumptions made in Step 3.  We now elaborate on how to perform Step
2, i.e., how to calculate \eref{e:logz} for an integer $u$, henceforth
referred to as the {\em replica number}.  

For an arbitrary positive integer $u$, we introduce $u$ independent
replicas of the retrochannel (or the spin glass) with the same
received signal $\Y$ and channel state $\S$ as depicted in Figure
\ref{f:vrchs}.  The partition function\index{Partition function} of
the replicated system is
\begin{equation}
  Z^u(\y,\S) = \expqcnd{ \produ q_{\Y|\X,\S}(\y|\X_a,\S) }{ \S }
  \label{e:zu}
\end{equation}
where the expectation is taken over the replicas $\{X_{ak}|
a=1,\dots,u,\, k=1,\dots,K\}$.  Here, $X_{ak}$ are i.i.d.\ (with
distribution $q_X$) since $(\Y,\S)$ are given.  With the new
expression \eref{e:zu} using the replicas, we proceed as follows.
Since $q_{\Y|\X_a,\S}$ is a conditional Gaussian density, their
product in \eref{e:zu} is a scaled version of another Gaussian density
conditioned on $\S$ and all $\X_a$.  By taking the integral with
respect to $\y$ first and then averaging over the spreading sequences,
one finds that
\begin{equation}
  \oneon{K} \log \expect{ Z^u(\Y,\S) } = \oneon{K}
  \log \expect{ \expbr{ \frac{K}{\beta} G_K^{(u)}( \Snr, \XX ) } }
  \label{e:fex}
\end{equation}
where $G_K^{(u)}$ is some function of the SNRs and the transmitted
symbols and their replicas, collectively denoted by a $K\times (u+1)$
matrix $\XX=[\X_0,\dots,\X_u]$.

The replica method then exploits the symmetry in $\XX$ in order to
evaluate \eref{e:fex}.  Instead of calculating the expectation
\eref{e:fex} with respect to $\XX$ all at once, we do it by first
conditioning on the correlation matrix $\Q=(1/K) \tran{\XX}\Snr^2\XX$.
It turns out that conditioned on the replica correlation matrix $\Q$,
the expectation with respect to $\XX$ is equivalent to an integral
over a multivariate Gaussian distribution due to the central limit
theorem, which helps to reduce \eref{e:fex} to:
\begin{equation}
  \oneon{K} \log \int \expbr{ \frac{K}{\beta} G^{(u)}(\Q) }
  \muuK(\intd \Q) + \mathcal{O}\left( \oneon{K} \right)
  \label{e:feq}
\end{equation}
where $G^{(u)}$ is some function (independent of $K$) of the $(u+1)
\times (u+1)$ random correlation matrix $\Q$, and $\muuK$ is the
probability measure of $\Q$.

Since for each pair $(a,b)$, $Q_{ab}= \oneon{K} \sumK \snr_k X_{ak}
X_{bk}$ is a sum of independent random variables, the probability
measure $\muuK$ satisfies the large deviations property.  Indeed, by
Cram\'er's Theorem \cite{Ellis85}, there exists a rate function $\Iu$
such that the measure $\muuK$ satisfies
\begin{equation}
  -\liminfty{K} \oneon{K} \log \muuK(\mathcal{A})
  = \inf_{\Q\in\mathcal{A}} \Iu(\Q) \,\log e
\end{equation}
for all measurable sets $\mathcal{A}$ of $(u+1)\times (u+1)$ matrices.
The rate function $\Iu$ is obtained through the Legendre-Fenchel
transform of the cumulant generating function of $\muuK$.  A key
observation is that as $K\rightarrow \infty$, the mass of the integral
in \eref{e:feq} concentrates on a particular subshell of $\Q$.  Using
Varadhan's theorem \cite{Ellis85},\index{Varadhan's Theorem}
\eref{e:feq} is found to converge to
\begin{equation} \label{e:sq}
  \sup_{\Q} \left[\oneon{\beta} \Gu(\Q)-\Iu(\Q)\right] \,\loge.
\end{equation}

Seeking the extremum \eref{e:sq} over a $(u+1)^2$-dimensional space is
hard.  It turns out that in many problems the supremum in $\Q$
satisfies {\em replica symmetry}, namely, that the supremum in $\Q$ is
identical over all replicated dimensions.  Assuming replica symmetry
holds, the supremum is over merely a few order parameters, and the
free energy can be obtained analytically.  The validity of replica
symmetry can be checked by calculating the Hessian of $\big[\beta^{-1}
\Gu-\Iu\big]$ at the replica symmetric supremum \cite{Nishim01}.  If
the Hessian is positive definite, then the replica symmetric solution
is stable against replica symmetry breaking, and it is the unique
solution because of the convexity of the function $\big[\beta^{-1}
\Gu-\Iu\big]$.  Under equal-power binary input and individually
optimal detection, \cite{Tanaka02IT} showed that if the system
parameters satisfy certain condition, the replica-symmetric solution
is stable against replica symmetry breaking (see also
\cite{NisShe01AIP}).  In some other cases, replica symmetry can be
broken \cite{Kabash03JPA}.  Unfortunately, there is no known general
condition for replica symmetry to hold.  The replica-symmetric
solution, assumed for analytical tractability in this paper, is
consistent with numerical results in the experiments shown in Section
\ref{s:nr}.

At any rate, the supremum \eref{e:sq} can be obtained as a function of
the replica number $u$.  The final step is to continue the expression
to real-valued $u$ and take the derivative at $u=0$.  The free energy
\eref{e:fku} is thus found and the mutual information obtained by
\eref{e:cz}.

The replica method is also used to calculate the moments \eref{e:mmt}.
Clearly, $\X_0$---$(\Y,\S )$---$[\X_1,\dots,\X_u]$ is a Markov chain.
The moments \eref{e:mmt} are equivalent to some moments under the
replicated system:
\begin{equation}
  \liminfty{K} \expect{ X_{0k}^i \, X_{mk}^j \, \prod^l_{a=1} X_{ak} }
  \label{e:dpd}
\end{equation}
where we choose $m>l$, which can be readily evaluated by working with
a modified partition function akin to~\eref{e:zu}.

We remark that the essence of the replica method here is its
capability of converting a difficult expectation (e.g., of a
logarithm) with respect to a given large system to an expectation of a
simpler form with respect to the replicated system.  Quite different
from conventional techniques is the emphasis of large systems and
symmetry from the beginning, where the central limit theorem and large
deviations help to calculate the otherwise intractable quantities.
The fact that certain statistics converge to a Gaussian distribution
in the thermodynamic limit is central to the application of replica
theory and to practical algorithms based upon the fixed-disorder
equivalent of replica theory (i.e., the TAP approach \cite{Nishim01}).
Another technique that takes advantage of the asymptotic normality is
the so-called ``cavity method'' in \cite{MezPar87}.

Following the replica recipe outlined above, a more detailed analysis
of the real-valued channel is carried out in Section \ref{s:prf}.  The
complex-valued counterpart is discussed in Section \ref{s:cc}.  As
previously mentioned, while the replica trick and replica symmetry are
assumed to be valid as well as the self-averaging property, their
rigorous justification is still an open problem in mathematical
physics.


\section{Proofs Using the Replica Method}
\label{s:prf}
\index{Replica method}

This section proves Claims \ref{th:dp}--\ref{th:x} using the replica
method.  The free energy \eref{e:fez} is first obtained and then the
spectral efficiency under joint decoding is derived.  The joint
moments \eref{e:mmt} are then found and it is demonstrated that the
multiuser channel can be effectively decoupled.  For notational
convenience, natural logarithms are assumed throughout this section.

\subsection{Free Energy}
\label{s:fe}

We will find the free energy\index{Free energy} by \eref{e:fku} and
then the spectral efficiency\index{Spectral efficiency} follows
immediately from \eref{e:cf}.  From \eref{e:py}, \eref{e:qy} and
\eref{e:zu},
\begin{eqnarray}
  \nsp{2}&& \nind \expect{ Z^u(\Y,\S) } \nn \\
  \nsp{2}&=&\nsp{2} \expect{ \int p_{\Y|\X,\S}(\y|\X_0,\S)
    \produ q_{\Y|\X,\S}(\y|\X_a,\S) \intd \y } \label{e:ezu} \\
  \nsp{2}&=&\nsp{2} \Exp \left\{ \int (2\pi)^{-\frac{L}{2}}
    (2\pi\sigma^2)^{-\frac{uL}{2}} \, \exph{ \ySXz } \right. \nn \\
  \nsp{2}&& \quad\times \left. \produ \expbs{ \ySXa }
    \,\intd \y \right\}.
  \label{e:ezx}
\end{eqnarray}
where the expectations are taken over the channel state matrix $\S$,
the original symbol vector $\X_0$ (i.i.d.\ entries with distribution
$p_X$), and the replicated symbols $\X_a$, $a=1,\dots,u$ (i.i.d.\ 
entries with distribution $q_X$).  Note that $\S$, $\X_0$ and $\X_a$
are independent in \eref{e:ezx}.  Let $\XX= [\X_0,\dots,\X_u]$.  From
the fact that the $L$ dimensions of the CDMA channel are independent
and statistically identical, we write \eref{e:ezx} as
\begin{equation}
  \begin{split}
    & \nsp{3} \expect{ Z^u(\Y,\S) } \\
    =& \; \Exp \Bigg\{ \vphantom{\left[\expect{\produ}\right]^L}
    \left[  \left( 2\pi\sigma^2 \right)^{-\frac{u}{2}} \int
        \Exp\left\{ \expb{ -\frac{( y-\tilde{\S}\Snr\X_0 )^2}{2} }
        \right.\right. \\
        & \; \times \produ\left.\left.\left. \nsp{1}
          \expb{ -\frac{ ( y-\tilde{\S}\Snr\X_a )^2 }{2\sigma^2} }
          \right| \Snr, \XX \right\}
        \frac{\intd y}{\sqrt{2\pi}} \right]^L \Bigg\}
    \label{e:ezs}
  \end{split}
\end{equation}
where the inner expectation in \eref{e:ezs} is taken over $\tilde
{\S}= [S_1,\dots, S_K]$, a vector of i.i.d.\ random variables each
taking the same distribution as the random spreading chips $S_{nk}$.
Define the following variables:
\begin{equation}
  V_a = \oneon{\sqrt{K}} \sumK \amp{k} S_k X_{ak}, \quad a=0,1,\dots,u.
  \label{e:va}
\end{equation}
Clearly, \eref{e:ezs} can be rewritten as
\begin{equation}
  \expect{ Z^u(\Y,\S) } =
  \expect{ \expbr{ L \, \GuK\left( \Snr,\XX \right) } }
  \label{e:zud}
\end{equation}
where
\begin{equation}
  \begin{split}
    &\nsp{3}\GuK\left( \Snr,\XX \right) \\
    = & - \frac{u}{2} \log\tps
    + \log \int  \nsp{.5} \Exp \left\{
    \expb{ -\frac{\big( y-\sqrt{\beta}\,V_0 \big)^2}{2} } \right. \\
    & \quad \times \produ \left.\left.
    \expb{ -\frac{\big( y-\sqrt{\beta}\,V_a\big)^2}{2\sigma^2} } \right|
    \Snr, \XX \right\} \frac{\intd y}{\sqrt{2\pi}}.
  \end{split}
  \label{e:gud}
\end{equation}
Note that given $\Snr$ and $\XX$, each $V_a$ is a sum of $K$ weighted
i.i.d.\ random chips.  Due to a vector version of the central limit
theorem, $\V$ converges to a zero-mean Gaussian random vector as
$K\rightarrow \infty$.  For $a,b = 0,1,\dots,u$, define
\begin{equation}
  Q_{ab} = \expcnd{V_a V_b}{\Snr,\XX}
  = \oneon{K} \sumK \snr_k X_{ak} X_{bk},
  \label{e:qab}
\end{equation}
Although inexplicit in notation, $Q_{ab}$ is a function of $\{\snr_k,
X_{ak}, X_{bk}\}^K_{k=1}$.  The random vector $\V$ in \eref{e:gud} can
be replaced by a zero-mean Gaussian vector with covariance matrix
$\Q$.  The reader is referred to \cite[Appendix B]{Tanaka02IT} or
\cite{Guo04PhD} for a justification of the change through the
Edgeworth expansion.  As a result,
\begin{equation}
  \expbr{ \GuK(\Snr,\XX) } = \expbr{ G^{(u)}(\Q) + \mathcal{O}(K^{-1}) }
  \label{e:gd}
\end{equation}
where the integral of the Gaussian density in \eref{e:gud} can be
simplified to obtain (refer to \cite{Guo04PhD} for details)
\begin{equation}
  \begin{split}
  \Gu(\Q) = -\half & \log \det(\I+\mSigma\Q)
  - \half \log\left(1+\frac{u}{\sigma^2}\right) \\
  & - \frac{u}{2} \log\left( 2\pi\sigma^2 \right)
  \label{e:guq}
  \end{split}
\end{equation}
where $\mSigma$ is a $(u+1)\times (u+1)$ matrix:\footnote{The indexes of
  all $(u+1)\times(u+1)$ matrices in this paper start from 0.}
\begin{equation}
  \mSigma = \frac{\beta}{\sigma^2+u}
  \pmatset{5}{13pt}
  \pmatset{6}{5pt}
  \begin{pmat}[{|.}]
    u\; & \;-\eT \cr \-
    -\e\; & \;\left(1+\frac{u}{\sigma^2}\right)\I-\oneon{\sigma^2}\e\eT \cr
  \end{pmat}
  \label{e:sab}
\end{equation}
where $\e$ is a $u\times 1$ column vector whose entries are all 1.
It is clear that $\mSigma$ is invariant if two nonzero indexes are
interchanged, i.e., $\mSigma$ is symmetric in the replicas.

By \eref{e:zud} and \eref{e:gd},
\begin{eqnarray}
  \nsp{2}&& \nind\oneon{K} \log \expect{Z^u(\Y,\S)} \nn \\
  \nsp{2}&=&\nsp{2} \oneon{K} \log \expect{ \expbr{ L \, \left( \Gu(\Q)
      + \mathcal{O}\left(K^{-1}\right) \right) } } \label{e:zgo} \\
\nsp{2}&=&\nsp{2} \oneon{K} \log \int \expbr{ \frac{K}{\beta} \Gu(\Q) }
  \intd \muuK(\Q) + \mathcal{O}\Big(\oneon{K}\Big)
  \label{e:gmu}
\end{eqnarray}
where the expectation over the replicated symbols is rewritten as an
integral over the probability measure of the correlation matrix $\Q$,
which is expressed as
\begin{equation} \label{e:mu}
  \muuK(\Q) = \expect{ \prodab \delta
    \left(\sumK \snr_k X_{ak} X_{bk} - KQ_{ab} \right) }
\end{equation}
where $\delta(\cdot)$ is the Dirac function.  Note that the limit in
$K$ and the expectation can be exchanged from \eref{e:zgo} to
\eref{e:gmu} by Lebesgue's dominated convergence theorem since $\expb{
  \Gu(\Q) }$ is bounded by a function of $u$ independent of $\Q$.

By Cram\'er's theorem \cite[Theorem II.4.1]{Ellis85}, the probability
measure of the empirical means $Q_{ab}$ defined by \eref{e:qab}
satisfies, as $K\rightarrow\infty$, the large deviations property with
some rate function $\Iu(\Q)$.  Let the moment generating function be
defined as
\begin{equation}
  \wu{\tQ} = \expect{ \expbr{ \snr \XT\tQ\X } }
  \label{e:mgf}
\end{equation}
where $\tQ$ is a $(u+1)\times (u+1)$ symmetric matrix, $\X=\tran{ [X_0,
X_1,\dots, X_u]}$, and the expectation in \eref{e:mgf} is taken over
independent random variables $\snr\sim P_\snr$, $X_0\sim p_X$ and
$X_1,\dots, X_u\sim q_X$.  The rate of the measure $\muuK$ is given by
the Legendre-Fenchel transform\index{Legendre-Fenchel transform} of
the cumulant generating function (logarithm of the moment generating
function) \cite{Ellis85}:
\begin{equation}  \label{e:iuq}
  \Iu(\Q) = \sup_{\tQ} \left[ \trace{\tQ\Q} - \log \wu{\tQ} \right]
\end{equation}
where the supremum is taken with respect to the symmetric matrix
$\tQ$.

Note the factor $K$ in the exponent in the integral in \eref{e:gmu}.
As $K\rightarrow\infty$, the integral is dominated by the maximum of
the overall effect of the exponent and the rate of the measure on
which the integral takes place.  Precisely, by Varadhan's theorem
\cite[Theorem II.7.1]{Ellis85},\index{Varadhan's Theorem}
\begin{equation}  \label{e:zgi}
  \liminfty{K} \oneon{K} \log \expect{Z^u(\Y,\S)}
  = \sup_{\Q} \left[ \oneon{\beta} \Gu(\Q)-\Iu(\Q) \right]
\end{equation}
where the supremum is over all (symmetric) valid correlation matrices.

\begin{figure*}[!t]
\normalsize
\setcounter{mytempeqncnt}{\value{equation}}
\setcounter{equation}{116}
\begin{eqnarray}
  \wut{\tQ^*}
  &=& \expect{ \expbr{ \snr \left( 2d \sumu X_0 X_a
      + 2f \sum^u_{0<a<b} X_a X_b + c X_0^2 + g \sumu X_a^2 \right) } }
        \label{e:wui} \\
  &=& \Exp\left\{ \exp \left[
        \snr \left( \frac{d}{\sqrt{f}} X_0 + \sqrt{f}\sumu X_a \right)^2
        + \left(c-\frac{d^2}{f} \right) \snr X_0^2
        + (g-f) \snr \sumu X_a^2 \right] \right\},
    \label{e:wua}
\end{eqnarray}
\setcounter{equation}{\value{mytempeqncnt}}
\hrulefill
\vspace*{4pt}
\end{figure*}
\setcounter{equation}{\value{mytempeqncnt}}

\begin{figure*}[t]
\normalsize
\setcounter{mytempeqncnt}{\value{equation}}
\setcounter{equation}{119}
\begin{equation}
    \wut{\tQ^*}
    =\Exp \left\{ \sqrt{\frac{d^2}{f\pi}} \int \exp \left[
      - \frac{d^2}{f} z^2 + 2 \sqrt{\snr} \bigg( \frac{d^2}{f}
      X_0 + d \sumu X_a \bigg) z 
      + \bigg(c-\frac{d^2}{f} \bigg) \snr X_0^2
      + (g-f) \snr \sumu X_a^2 \right] \intd z \right\}.
  \label{e:wuq}
\end{equation}
\begin{equation}
  \begin{split}
    & \Iu \left( \Q^* \right)
    = rc + upg + 2umd + u(u-1) qf \\
    & - \log \Exp \Biggl\{ \int \sqrt{\frac{d^2}{f\pi}}
    \expcnd{ \expbr{ -\frac{d^2}{f} \left(z - \sqrt{\snr} X_0 \right)^2
        + c \snr X_0^2 } }{\snr}
    \Big[ \expqcnd{ \expbr{ 2d \sqrt{\snr} X z
      + (g-f) \snr X^2 } }{\snr} \Big]^u \intd z
    \vphantom{\frac{d^2}{f}} \Biggr\}.
  \end{split}
\label{e:iu}
\end{equation}
\setcounter{equation}{\value{mytempeqncnt}}
\hrulefill
\vspace*{4pt}
\end{figure*}
\setcounter{equation}{\value{mytempeqncnt}}

By \eref{e:zgi}, \eref{e:iuq} and \eref{e:guq}, one has
\begin{eqnarray}
  \nsp{5}&& \nsp{3} \oneon{K} \log \expect{Z^u(\Y,\S)} \nn \\
  \nsp{5}&=&\nsp{1.5} \sup_{\Q} \left[ \oneon{\beta} \Gu(\Q) -
        \sup_{\tQ} \left[ \trace{\tQ\Q} - \log \wu{\tQ} \right]
         \right] \\
  \nsp{5}&=&\nsp{1.5} \sup_{\Q} \, \inf_{\tQ} \, T^{(u)}(\Q,\tQ)
  \label{e:zqq}
\end{eqnarray}
where
\begin{equation}
  \begin{split}
    T^{(u)}(\Q,\tQ) =& -\oneon{2\beta} \log \det(\I+\mSigma\Q)
    - \trace{\tQ\Q} \\
    & + \log \expect{ \expbr{ \snr \XT\tQ\X } } \\
    & - \oneon{2\beta} \log\left(1+\frac{u}{\sigma^2}\right)
    - \frac{u}{2\beta} \log\left( 2\pi\sigma^2 \right).
  \end{split}
  \label{e:tu}
\end{equation}
For an arbitrary $\Q$, we first seek the point of zero gradient with
respect to $\tQ$ and find that for any given $\Q$, the extremum in
$\tQ$ satisfies
\begin{equation}
    \Q = \frac{
        \expect{ \snr \X\XT \expbr{ \snr\XT\tQ\X } } }
        { \expect{ \expbr{ \snr\XT\tQ\X } } }.
    \label{e:QtQ}
\end{equation}
Let $\tQ^*(\Q)$ denote the solution to \eref{e:QtQ}.  We then seek the
point of zero gradient of $T^{(u)} \left(\Q,\tQ^*(\Q)\right)$ with
respect to $\Q$.\footnote{The following identities are useful:
  \[
  \frac{\partial \log\det \Q}{\partial x} =\trace{ \inv{\Q}
  \frac{\partial \Q}{\partial x} }, \quad
  \frac{\partial \inv{\Q}}{\partial x} = -\inv{\Q}
  \frac{\partial \Q}{\partial x} \inv{\Q}.
  \]}
By virtue of the relationship \eref{e:QtQ}, one finds that the
derivative of $\tQ^*$ with respect to $\Q$ is multiplied by 0 and
hence inconsequential.  Therefore, the extremum in $\Q$ satisfies
\begin{equation}
  \tQ = -\beta^{-1} \inv{ \left( \I + \mSigma\Q \right) } \mSigma.
  \label{e:tQQ}
\end{equation}
It is interesting to note from the resulting joint equations
\eref{e:QtQ}--\eref{e:tQQ} that the order in which the supremum and
infimum are taken in \eref{e:zqq} can be exchanged.  The solution
$\left(\Q^*, \tQ^*\right)$ is in fact a saddle point of $T^{(u)}$.
Notice that \eref{e:QtQ} can also be expressed as
\begin{equation}
  \Q = \expcnd{ \snr \X\XT }{ \tQ }
  \label{e:xxq}
\end{equation}
where the expectation is over an appropriately defined conditional
Gaussian measure $p_{\X,\snr|\tQ}$.

Solving joint equations \eref{e:QtQ} and \eref{e:tQQ} directly is
prohibitive except in the simplest cases such as $q_X$ being Gaussian.
In the general case, because of symmetry in the matrix $\mSigma$
\eref{e:sab}, we postulate that the solution to the joint equations
satisfies {\em replica symmetry},\index{Replica method!replica
symmetry} namely, both $\Q^*$ and $\tQ^*$ are invariant if two
(nonzero) replica indexes are interchanged.  In other words, the
extremum can be written as
\begin{subequations}  \label{e:rs}
\begin{equation}
  \Q^* = \;
  \begin{bmatrix}
  r & m & m & \dots & m \cr \-
  m & p & q & \dots & q \cr
  m & q & p & \ddots & \vdots \cr
  \vdots & \vdots  & \ddots & \ddots & q \cr
  m & q & \dots & q & p \cr
  \end{bmatrix},
\end{equation}
\begin{equation}
  \tQ^* =\!
  \begin{bmatrix}
    \;c\; & d & d & \dots & d\; \\
  \;d\; & g & f & \dots & f\; \\
  \;d\; & f & g & \ddots & \vdots \\
  \;\vdots\; & \vdots & \ddots & \ddots & f\; \\
  \;d\; & f & \dots & f & g\;
  \end{bmatrix}
\end{equation}
\end{subequations}
where $r,m,p,q,c,d,f,g $ are some real numbers.  Under replica
symmetry, \eref{e:guq} is evaluated to obtain
\begin{eqnarray}
  \nsp{3} && G^{(u)} \left( \Q^* \right)
    = -\frac{u}{2} \log\tps
    - \frac{u-1}{2} \log\left[ 1 + \frac{\beta}{\sigma^2}(p-q) \right] \nn \\
    \nsp{3} && \; - \half \log \left[1+\frac{\beta}{\sigma^2}(p-q)
      + \frac{u}{\sigma^2} (1+\beta(r-2m+q)) \right].
  \label{e:gu}
\end{eqnarray}
The moment generating function \eref{e:mgf} is evaluated as
\eref{e:wui}--\eref{e:wua} where $X_0\sim p_X$ while $X_a\sim q_X$ are
all independent.  The expectation \eref{e:wua} with respect to the
symbols can be decoupled using the unit area property of Gaussian
density:\footnote{Equation \eref{e:hs} is a variant of the
  Hubbard-Stratonovich transform \cite{Hubbar59PRL}.}
\addtocounter{equation}{2} %
\begin{equation}
  e^{x^2} = \sqrt{\frac{\eff}{2\pi}} \int \expbr{
    -\frac{\eff}{2} z^2 + \sqrt{2\eff}\,xz } \intd z, \;\;\; \forall x, \eff.
  \label{e:hs}
\end{equation}
Using \eref{e:hs} with $\eff=2d^2/f$, \eref{e:wua} becomes
\eref{e:wuq}.  Since $X_0,\dots,X_u$ and $\snr$ are independent, the
rate of the measure \eref{e:iuq} under replica symmetry is obtained
from \eref{e:wuq} as \eref{e:iu}.  Let $\Q^*$ be the replica-symmetric
solution to \eref{e:QtQ}--\eref{e:tQQ}.  The free energy is then found
by \eref{e:fku} and \eref{e:zgi}:
\addtocounter{equation}{2} %
\begin{equation}
  \fe = - \limzero{u} \frac{\partial}{\partial u}
  \left[ \beta^{-1} G^{(u)}\left( \Q^* \right)
        - \Iu\left(\Q^*\right) \right].
  \label{e:fep}
\end{equation}

The eight parameters $(r,m,p,q,c,d,f,g)$ that define $\Q^*$ and
$\tQ^*$ are the solution to the joint equations
\eref{e:QtQ}--\eref{e:tQQ} under replica symmetry.  It is interesting
to note that as functions of $u$, the derivative of each of the eight
parameters with respect to $u$ vanishes as $u\rightarrow0$.  Thus for
the purpose of the free energy \eref{e:fep}, it suffices to find the
extremum of $\left[ \beta^{-1} G^{(u)} - \Iu \right]$ at $u=0$.  Using
\eref{e:tQQ}, it can be shown that at $u=0$,
\begin{subequations}
  \begin{eqnarray}
    c &=& 0, \label{e:c} \\
    d &=& \oneon{2[ \sigma^2 + \beta(p-q)]}, \label{e:d} \\
    f &=& \frac{ 1+\beta(r-2m+q) }{ 2[\sigma^2 + \beta(p-q)]^2 },\label{e:f} \\
    g &=& f - d. \label{e:g}
  \end{eqnarray}
  \label{e:cdfg}%
\end{subequations}
The parameters $r,m,p,q$ can be determined from \eref{e:xxq} by
studying the measure $p_{\X,\snr|\tQ}$ under replica symmetry and
$u\rightarrow0$.  For that purpose, define two useful parameters:
\begin{equation} \label{e:exdf}%
    \eff = \frac{2d^2}{f} \quad \text{and} \quad \xi = 2d.
\end{equation}
Noticing that $c=0$, $g-f=-d$, \eref{e:wuq} can be written as
\begin{equation} \label{e:wtq}
  \begin{split}
    & \wut{\tQ^*}
    =\Exp \left\{ \sqrt{\frac{\eff}{2\pi}} \int
   \expbr{ -\frac{\eff}{2} \left( z-\sqrt{\snr}X_0 \right)^2 } \right. \\
      & \; \times \left. \left[ \expqcnd{ \expbr{ -\frac{\xi}{2} z^2
    -\frac{\xi}{2} \left(z-\sqrt{\snr} X\right)^2 } }{\snr} \right]^u
   \nsp{1} \intd z \right\}.
  \end{split}
\end{equation}
It is clear that the limit of \eref{e:wtq} as $u\rightarrow0$ is 1.
Hence by \eref{e:QtQ}, as $u\rightarrow0$,
\begin{eqnarray}
  Q^*_{ab} &=& \expcnd{ \snr X_a X_b }{ \tQ^* } \\
  &\rightarrow& \expect{ \snr X_a X_b \expbr{ \XT\tQ^*\X } }.
  \label{e:qs}
\end{eqnarray}
We now give a useful representation for the parameters $r,m,p,q$
defined in \eref{e:rs}.  Consider for instance $a=0$ and $b=1$.  Note
that as $u\rightarrow0$,
\begin{equation}
  \begin{split}
    & \nsp{3}\expect{ \snr X_0 X_1 \expbr{ \XT\tQ^*\X } } \\
    =& \Exp \Biggl\{ \snr X_0 \, \int \sqrt{\frac{\eff}{2\pi}}
    \expbr{ -\frac{\eff}{2} \left( z-\sqrt{\snr}X_0 \right)^2 } \\
    & \times \frac{ X_1 \, \sqrt{\frac{\xi}{2\pi}}
    \expbr{ -\frac{\xi}{2} \left( z-\sqrt{\snr}X_1 \right)^2 } }
    { \expqcnd{ \sqrt{\frac{\xi}{2\pi}}
    \expbr{ -\frac{\xi}{2} \left( z-\sqrt{\snr}X_1 \right)^2 } }{\snr} }
    \,\intd z \Biggr\}.
  \end{split}
  \label{e:exx}
\end{equation}
Let two single-user Gaussian channels be defined as in Section
\ref{s:main}, i.e., $p_{Z|X,\snr;\eff}$ given by \eref{e:pzx} and
$q_{Z|X,\snr;\xi}$ by \eref{e:qzx}.  Assuming that the input
distribution to the channel $q_{Z|X,\snr;\xi}$ is $q_X$, a posterior
probability distribution $q_{X|Z,\snr;\xi}$ is induced, which defines
a retrochannel.\index{Retrochannel} Let $X_0$ be the input to the
channel $p_{Z|X,\snr;\eff}$ and $X=X_1$ be the output of the
retrochannel $q_{X|Z,\snr;\xi}$.  The posterior mean with respect to
the measure $q$, denoted by $\sm{X}_q$, is given by \eref{e:xq}.  The
Gaussian channel $p_{Z|X,\snr;\eff}$, the retrochannel
$q_{X|Z,\snr;\xi}$ and the PME, all in the single-user setting, are
depicted in Figure \ref{f:sur}.  Then, \eref{e:exx} can be understood
as an expectation over $X_0$, $X$ and $Z$ to obtain
\begin{eqnarray}
  Q^*_{01} \nsp{1}&=&\nsp{1} \expect{ \snr X_0 X_1 \expbr{ \XT\tQ^*\X } } \\
  \nsp{1}&=&\nsp{1} \Exp\bigg\{
  \snr X_0 \int \expqcnd{X}{Z=z,\snr;\xi} \nn \\
    && \qquad\quad
    \times p_{Z|X,\snr;\eff}(z|X_0,\snr;\eff) \,\intd z \bigg\} \\
        \nsp{1}&=&\nsp{1} \expect{ \snr \, X_0 \sm{X}_q }.
\end{eqnarray}
Similarly, \eref{e:qs} can be evaluated for all indexes $(a,b)$
yielding together with \eref{e:rs}:
\begin{subequations}  \label{e:rmpq}%
  \begin{eqnarray}
    r \nsp{1}&=&\nsp{1} Q^*_{00} = \expect{ \snr\, X_0^2 } = \expect{\snr}, \label{e:r} \\
    m \nsp{1}&=&\nsp{1} Q^*_{01} = \expect{ \snr\, X_0 \sm{X}_q }, \label{e:m} \\
    p \nsp{1}&=&\nsp{1} Q^*_{11} = \expect{ \snr\, X^2 }, \label{e:p} \\
    q \nsp{1}&=&\nsp{1} Q^*_{12} = \expect{ \snr\, ( \sm{X}_q )^2 }.\label{e:q}
  \end{eqnarray}
\end{subequations}%
In summary, under replica symmetry, the parameters $c,d,f,g$ are given
by \eref{e:cdfg} as functions of $r,m,p,q$, which are in turn
determined by the statistics of the two channels \eref{e:pzx} and
\eref{e:qzx} parameterized by $\eff=2d^2/f$ and $\xi=2d$ respectively.
It is not difficult to see that
\begin{subequations}
  \begin{eqnarray}
    r - 2m + q &=& \expect{ \snr\,\left(X_0-\sm{X}_q\right)^2 },\\
    p - q &=& \expect{ \snr\,\left(X-\sm{X}_q\right)^2 }.
\end{eqnarray}
\end{subequations}
Using \eref{e:cdfg} and \eref{e:exdf}, it can be checked that
\begin{subequations} \label{e:epr}
  \begin{eqnarray}
    r - 2m + q &=& \oneon{\beta} \left( \oneon{\eff} - 1 \right), \\
    p - q &=& \oneon{\beta} \left( \oneon{\xi} - \sigma^2 \right).
\end{eqnarray}
\end{subequations}
Thus $\Gu$ and $\Iu$ given by \eref{e:gu} and \eref{e:iu} can be
expressed in $\eff$ and $\xi$.  Using \eref{e:fep} and \eref{e:epr},
the free energy is found as \eref{e:fel}, where $(\eff,\xi)$ satisfies
fixed-point equations
\begin{subequations}
  \begin{eqnarray}
  \eff^{-1} &=& 1\; + \beta \, \expect{ \snr
    \left( X_0 - \sm{X}_q \right)^2 },
  \label{e:ee} \\
  \xi^{-1} &=& \!\sigma^2 + \beta\,\expect{ \snr
    \left( X - \sm{X}_q \right)^2 }.
  \label{e:xx}
  \end{eqnarray}
  \label{e:eexx}%
\end{subequations}
Because of \eref{e:zgi}, in case of multiple solutions to
\eref{e:eexx}, $(\eff,\xi)$ is chosen as the solution that gives the
minimum free energy $\fe$.  By defining $\mse(\snr;\eff,\xi)$ and
$\vrc(\snr;\eff,\xi)$ as in \eref{e:gmse} and \eref{e:vrc}, the
coupled equations \eref{e:cdfg} and \eref{e:rmpq} can be summarized to
establish the key fixed-point equations \eref{e:ex}.  It will be shown
in Section \ref{s:jm} that, from an individual user's viewpoint, the
multiuser PME and the multiuser retrochannel, parameterized by
arbitrary $(q_X,\sigma)$, have an equivalence as a single-user PME and
a single-user retrochannel.

Finally, for the purpose of the total spectral efficiency, we set the
postulated measure $q$ to be identical to the actual measure $p$
(i.e., $q_X=p_X$ and $\sigma=1$).  The inverse noise variances
$(\eff,\xi)$ satisfy joint equations but we choose the
replica-symmetric solution $\eff=\xi$ as argued in Section
\ref{s:main}.  Using \eref{e:cf}, the total spectral efficiency is
\begin{equation}
  \begin{split}
    \Cjnt =& - \beta\, \expect{\int \pz{z} \, \log \pz{z} \intd\,z } \\
    & \quad -\frac{\beta}{2} \log\frac{2\pi e}{\eff} + \half(\eff-1-\log\eff),
  \label{e:ctj}
  \end{split}
\end{equation}
where $\eff$ satisfies
\begin{equation}
  \eff + \eff\,\beta\, \expect{\snr \, \left[ 1 - \int
      \frac{ \left[ \pzi{1} \right]^2 }{ \pz{z} } \intd\,z \right] } = 1.
  \label{e:fpe}
\end{equation}
The optimal spectral efficiency of the multiuser channel is thus
found.

\subsection{Joint Moments}
\label{s:jm}

Consider again the Gaussian channel, the PME and the retrochannel in
the multiuser setting depicted in Figure \ref{f:rch}.  The joint
moments \eref{e:mmt} are of interest here.  For simplicity, we first
study joint moments of the input symbol and the retrochannel output,
which can be obtained as expectations under the replicated system
\cite[Lemma 3.1]{Guo04PhD}:
\begin{equation} \label{e:xij}
  \expect{ X_{0k}^i X_k^j } = \expect{ X_{0k}^i X_{mk}^j }, \quad
  m = 1, \dots, u.
\end{equation}
It is then straightforward to calculate \eref{e:mmt} by following the
same procedure.

The following lemma allows us to determine the expected value of a
function of the symbols and their replicas by considering a modified
partition function akin to~\eref{e:zu}.
\begin{lemma} \label{lm:ef}
  Given an arbitrary function $f(\X_0,\Xa)$, where $\Xa=[\X_1,\dots,
  \X_u]$, define
\begin{equation}
  \begin{split}
    & Z^{(u)}(\y,\S,\x_0;h) \\
    & = \expqcnd{ 
    \expbr{h\, f(\x_0,\Xa)} \, \produ q_{\Y|\X,\S}(\y|\X_a,\S) }{ \S },
  \label{e:zuh}
  \end{split}
\end{equation}
where $\X_a$ has i.i.d.\ entries with distribution $q_X$.  If\,
$\expcnd{ f(\X_0,\Xa) } {\Y,\S,\X_0}$ is not dependent on $u$, then
\begin{equation}
  \expect{ f(\X_0,\Xa) }
  = \limzero{u} \frac{\partial}{\partial h}
  \log \left. \expect{ Z^{(u)}(\Y,\S,\X_0;h) } \right|_{h=0}.
\label{e:ef}
\end{equation}
\end{lemma}
\begin{proof}
It is easy to see that
\begin{equation}
  \left. Z^{(u)}(\Y,\S,\X_0;h) \right|_{h=0} = Z^u(\Y,\S).
\end{equation}
By taking the derivative and letting $h=0$, the right hand side
of \eref{e:ef} is
\begin{equation}
  \begin{split}
    \oneon{K} \limzero{u} & \Exp\Bigg\{ \Exp_q\bigg\{ f(\X_0,\Xa') \\
    & \quad \times \produ
    q_{\Y|\X,\S}(\Y|\X_a',\S) \bigg| { \Y,\S,\X_0 } \bigg\}\Bigg\},
  \label{e:fda}
  \end{split}
\end{equation}
where $\Xa'$ has the same statistics as $\Xa$ (i.e., contains i.i.d.\ 
entries with distribution $q_X$) but independent of $(\X_0,\Y,\S)$.
Also note that
\begin{equation}
  \begin{split}
    & q_{\Xa|\Y,\S} (\Xa\,|\,\Y,\S) \\
    & \quad = Z^{-u}(\Y,\S) \, q_{\Xa}(\Xa) \, \produ
    q_{\Y|\X,\S}(\Y\,|\,\X_a,\S).
  \end{split}
\end{equation}
One can change the expectation over the replicas $\Xa'$ independent of
$(\Y,\S,\X_0)$ to an expectation over $\Xa$ conditioned on
$(\Y,\S,\X_0)$.  Hence \eref{e:fda} can be further written as
\begin{eqnarray}
  && \nind \oneon{K} \limzero{u} \expect{
    \expect{ f(\X_0,\Xa) | \Y,\S,\X_0 } \, Z^u(\Y,\S) } \nn \\
  &=& \oneon{K} \expect{ \expect{ f(\X_0,\Xa) | \Y,\S,\X_0 } }
  \label{e:eef} \\
  &=& \oneon{K} \expect{ f(\X_0,\Xa) }
\end{eqnarray}
where $Z^u(\Y,\S)$ can be dropped as $u\rightarrow0$ in \eref{e:eef}
since the conditional expectation is not dependent on $u$ by the
assumption in the lemma.
\end{proof}

For the function $f(\X_0,\Xa)$ to have influence on the free energy,
it must grow at least linearly with $K$.  Assume that $f(\X_0,\Xa)$
involves users 1 through $K_1=\alpha_1 K$ where $0<\alpha_1<1$ is
fixed as $K\rightarrow\infty$:
\begin{equation}
  f(\X_0,\Xa) = \sumKo X_{0k}^i X_{mk}^j
\end{equation}
where $m$ is an arbitrary replica number in $\{1,\dots,u\}$.  Without
loss of generality, we calculate \eref{e:xij} for a user $\kappa
\in\{1,\dots, K_1\}$.  It is also assumed that user 1 through $K_1$
take the same signal-to-noise ratio $\snr$.  We will finally take the
limit $\alpha_1\rightarrow0$ so that the equal-power constraint for
the first $K_1$ users becomes superfluous.

Clearly, the moments \eref{e:xij} for user $\kappa$ can be rewritten
as
\begin{eqnarray}
  \expect{ X_{0\kappa}^i X_{m\kappa}^j }
  &=& \oneon{K_1} \, \sumKo \expect{ X_{0k}^i \, X_{mk}^j } \label{e:xmk} \\
  &=& \oneon{K_1} \, \expect{ f(\X_0,\Xa) }.
  \label{e:mf}
\end{eqnarray}
Note that
\begin{equation}
  \expcnd{ f(\X_0,\Xa) }{\Y,\S,\X_0} = 
  \expcnd{ \sumKo X_{0k}^i \, X_k^j }{\Y,\S,\X_0}
\end{equation}
is not dependent on $u$.  By Lemma \ref{lm:ef}, the moments
\eref{e:mf} can be obtained as
\begin{equation}
  \limzero{u} \frac{\partial}{\partial h} \oneon{\alpha_1 K}
  \log \left. \expect{ Z^{(u)}(\Y,\S,\X_0;h) } \right|_{h=0}
  \label{e:zh}
\end{equation}
where
\begin{equation}  \label{e:zxh}
  \begin{split}
    Z^{(u)}(\y,\S,\x_0;h)
    &= \tps^{-\frac{uL}{2}}
    \Exp_q\left\{ \expbr{h \sumKo x_{0k}^j X_{mk}^i} \right. \\
        & \times \left.\left. \produ \expbs{\ySXa} \right| \S \right\}.
  \end{split}
\end{equation}
Regarding \eref{e:zxh} as a partition function for some random system
allows the same techniques in Section \ref{s:fe} to be used to write
\begin{equation}
  \begin{split}
    \liminfty{K} & \oneon{K} \log \expect{ Z^{(u)}(\Y,\S,\X_0;h) } \\
    & = \sup_{\Q} \left[ \beta^{-1} \Gu(\Q) - \Iu(\Q;h) \right]
  \label{e:zhq}
  \end{split}
\end{equation}
where $\Gu(\Q)$ is given by \eref{e:guq} and $\Iu(\Q;h)$ is the rate
of the following measure (cf.\ \eref{e:mu})
\begin{equation}
  \begin{split}
    \nsp{1} \mu_K^{(u)}(\Q;h) = & \Exp\Bigg\{ \prodab \delta \left(\sumK
    \snr_k X_{ak} X_{bk} - KQ_{ab} \right) \\
  & \qquad\quad \times \expbr{h \sumKo X_{0k}^i X_{mk}^j} \Bigg\}.
  \label{e:muh}
  \end{split}
\end{equation}
By the large deviations property, one finds the rate
\begin{equation}
  \begin{split}
    & \Iu(\Q;h)
    = \sup_{\tilde{\Q}} \bigg[ \trace{\tQ\Q} - \log \wut{\tQ} \\
      &\; - \alpha_1 \left( \log \wut{\tQ,\snr;h}
        - \log \wut{\tQ,\snr;0} \right) \bigg]
  \end{split}
\label{e:ih}
\end{equation}
where $\wut{\tQ}$ is defined in \eref{e:mgf}, and
\begin{equation}
  \begin{split}
    & \wut{\tQ,\snr;h} \\
 &\quad = \expcnd{ \expbr{h \, X_0^i X_m^j} \expbr{ \snr \XT\tQ\X } }{\snr}.
  \label{e:xuph}
  \end{split}
\end{equation}
From \eref{e:zhq} and \eref{e:ih}, taking the derivative in
\eref{e:zh} with respect to $h$ at $h=0$ leaves only one term
\begin{equation}
  \begin{split}
    \frac{\partial}{\partial h} \log & \wut{\tQ,\snr;h} \bigg|_{h=0} \\
    & = \frac{ \expect{ X_0^i X_m^j \expbr{ \snr \XT\tQ\X } } }
  { \expect{ \expbr{ \snr \XT\tQ\X } } }.
  \label{e:xd}
  \end{split}
\end{equation}
Since
\begin{equation}
  \left. Z^{(u)}(\Y,\S,\X_0;h) \right|_{h=0} = Z^u(\Y,\S),
\end{equation}
the $\tQ$ in \eref{e:xd} that give the supremum in \eref{e:ih} at
$h\rightarrow0$ is exactly the $\tQ$ that gives the supremum of
\eref{e:iuq}, which is replica-symmetric by assumption.  By
introducing the parameters $(\eff,\xi)$ the same as in Section
\ref{s:fe}, and by definition of $q_i$ and $p_i$ in \eref{e:qzi} and
\eref{e:pzi} respectively, \eref{e:xd} can be further evaluated as
\begin{equation}
    \frac{
      \int \Big( \sqrt{\frac{2\pi}{\xi}} e^\frac{\xi z^2}{2} \Big)^u
      p_i(z,\snr;\eff) \, q_0^{u-1}(z,\snr;\xi)
      \, q_j(z,\snr;\xi) \intd z }
    { \int \Big( \sqrt{\frac{2\pi}{\xi}} e^\frac{\xi z^2}{2} \Big)^u
         p_0(z,\snr;\eff) q_0^u(z,\snr;\xi) \intd z }.
       \label{e:ff}
\end{equation}
Taking the limit $u\rightarrow0$, one has from
\eref{e:mf}--\eref{e:ff} that as $K\rightarrow \infty$,
\begin{equation} \label{e:pqz}
  \oneon{K_1} \, \sumKo \expect{ X_{0k}^i \, X_{mk}^j }
  \rightarrow \int p_i(z,\snr;\eff) \, \frac{ q_j(z,\snr;\xi) }
    { q_0(z,\snr;\xi) } \intd z.
\end{equation}
Let $X_0\sim p_X$ be the input to the single-user Gaussian channel
$p_{Z|X,\snr;\eff}$ and $Z$ be its output (see Figure \ref{f:sur}).
Let $X$ be the corresponding output of the companion retrochannel
with $Z$ as its input.  Then $X_0$--$Z$--$X$ is a Markov chain.  By
definition of $p_i$ and $q_i$, the right hand side of \eref{e:pqz} is
\begin{equation} \label{e:ize}
  \begin{split}
    \int p_0(z,\snr;\eff) & \, \frac{ p_i(z,\snr;\xi) }{ p_0(z,\snr;\xi) }
  \, \frac{ q_j(z,\snr;\xi) }{ q_0(z,\snr;\xi) } \intd z \\
  & = \expect{ \expcnd{X_0^i}{Z} \expcnd{X^j}{Z} }.  
  \end{split}
\end{equation}
Letting $K_1\rightarrow1$ (thus $\alpha_1\rightarrow0$) so that the
requirement that the first $K_1$ users take the same SNR becomes
unnecessary, we have proved by \eref{e:xij}, \eref{e:xmk},
\eref{e:pqz} and \eref{e:ize} that for every SNR distribution and
every user $k\in \{1,\dots,K\}$
\begin{equation}
  \expect{ X_{0k}^i \, X_k^j } \rightarrow \expect{ X_0^i X^j }
         \quad \text{as}\; K\rightarrow \infty.
  \label{e:xk}
\end{equation}

Since the moments \eref{e:xk} are uniformly bounded, the distribution
is thus uniquely determined by the moments due to Carleman's Theorem
\cite[p.~227]{Feller71}.\index{Carleman's Theorem} Therefore, for
every user $k$, the joint distribution of the input $X_{0k}$ to the
multiuser channel and the output $X_k$ of the multiuser retrochannel
converges to the joint distribution of the input $X_0$ to the
single-user Gaussian channel $p_{Z|X,\snr;\eff}$ and the output $X$ of
the single-user retrochannel $q_{X|Z,\snr;\xi}$.

Applying the same methodology as developed thus far in this subsection,
one can also calculate the joint moments \eref{e:mmt} by letting
\begin{equation}
  f(\X_0,\Xa) = \sumKo X_{0k}^i X_{mk}^j \prod^l_{a=1} X_{ak}
\end{equation}
where it is assumed that $m>l$.  The rationale is that
$\X_0$--$(\Y,\S)$--$\X_a$ is a Markov chain and $X_a$'s are i.i.d.\
conditioned on $(\Y,\S)$; hence \eref{e:mmt} can be calculated as
expectations under the replicated system:
\begin{eqnarray}
  && \nind \expect{ X_{0k}^i \, X_k^j \, \sm{X_k}_q^l } \nn \\
  &=& \expect{ X_{0k}^i \, X_{mk}^j \, \prod^l_{a=1}
    \expcnd{X_{ak}}{\Y,\S} } \\
  &=& \expect{ f(\X_0,\Xa) }. \label{e:efa}
\end{eqnarray}
It is straightforward by Lemma \ref{lm:ef} to calculate \eref{e:efa}
and obtain that, as $K\rightarrow \infty$,
\begin{equation} \label{e:efi}
  \begin{split}
  & \expect{ f(\X_0,\Xa) } \rightarrow \\
  & \quad \int p_i(z,\snr;\eff) \, 
  \frac{ q_j(z,\snr;\xi) }{ q_0(z,\snr;\xi) } \,
  \left( \frac{ q_1(z,\snr;\xi) }{ q_0(z,\snr;\xi) } \right)^l
  \intd z.
  \end{split}
\end{equation}
Let $\sm{X}_q$ be the single-user PME output as seen in Figure
\ref{f:sur}, which is a function of the Gaussian channel output $Z$.
Then the right hand side of \eref{e:efi} represents a joint moment
and thus
\begin{equation}
  \expect{ X_{0k}^i \, X_k^j \, \sm{X_k}_q^l }
  \rightarrow \expect{ X_0^i X^j \sm{X}_q^l }.
\end{equation}
Again, by Carleman's Theorem, the joint distributions of $(X_{0k},
X_k, \sm{X_k}_q)$ converge to that of $(X_0,X,\sm{X}_q)$.  Indeed,
from the viewpoint of user $k$, the multiuser setting is equivalent to
the single-user setting in which the SNR suffers a degradation $\eff$
(compare Figures \ref{f:sur} and \ref{f:rch}).  Hence we have proved
the decoupling principle\index{Decoupling principle} and Claim
\ref{th:dp}.

In the large-system limit, the transformation from the input $X_{0k}$
to the multiuser detection output $\sm{X_k}_q$ is nothing but a
single-user Gaussian channel $p_{Z|X,\snr;\eff}$ concatenated with a
decision function \eref{e:qdf}.  The decision function can be ignored
from both detection- and information-theoretic viewpoints due to its
monotonicity:
\begin{proposition}
  The decision function \eref{e:qdf} is strictly monotone increasing
  in $z$ for all $\snr$ and $\xi$.
\end{proposition}
\begin{proof}
  Let $(\cdot)'$ denote derivative with respect to $z$.  One can show
  that for $i=0,1,\dots$,
  \begin{equation}
    q_i'(z,\snr;\xi) = 
    \xi \sqrt{\snr}\, \qzi{i+1} - \xi z\, \qzi{i}.
  \end{equation}
  Clearly,
  \begin{equation}
    \begin{split}
    & \left[ \frac{\qzi{1}}{\qzi{0}} \right]' \\
    & \quad = \xi \sqrt{\snr} \, \frac{ \qzi{2} \qzi{0}
      - q_1^2(z,\snr;\xi) }{ q_0^2(z,\snr;\xi) }.
    \label{e:fr}
    \end{split}
  \end{equation}
  The numerator in \eref{e:fr} is nonnegative by the Cauchy-Schwartz
  inequality.  For the numerator in \eref{e:fr} to be 0, $X$ must be a
  constant, which contradicts the assumption that $X$ has zero mean
  and unit variance.  Therefore, \eref{e:qdf} is strictly increasing.
\end{proof}

We may now conclude that the equivalent single-user channel is an
additive Gaussian noise channel with input signal-to-noise ratio
$\snr$ and noise variance $\eff^{-1}$ as depicted in Figure
\ref{f:sur}.  Corollaries \ref{cr:dp} and \ref{cr:i} are thus proved.
In the special case that the postulated measure $q$ is identical to
the actual measure $p$, Claim \ref{th:dp} reduces to Claim \ref{th:c}.

The single-user mutual information is now simply that of a Gaussian
channel with input distribution $p_X$,
\begin{equation}  \label{e:80}
  \begin{split}
  I(\eff\,\snr) = & - \int \pz{z} \log \pz{z} \intd z \\
  & \quad -\half \log\frac{2\pi e}{\eff},
  \end{split}
\end{equation}
which is as defined in \eref{e:id}.  The overall spectral efficiency
under separate decoding is
\begin{equation}
  \Csep = \beta\, \expect{ I(\eff\,\snr) }.  
  \label{e:cs}
\end{equation}
Hence the proof of \eref{e:csep}.  Claim \ref{th:x} is proved by
comparing \eref{e:cs} to \eref{e:ctj}.

\section{Complex-valued Channels}
\label{s:cc}

Until now the discussion is based on a real-valued setting of the
multiuser system, namely, both the inputs $X_k$ and the spreading
chips $S_{nk}$ take real values.  In practice, particularly in
carrier-modulated communications where spectral
efficiency\index{Spectral efficiency} is a major concern, transmission
in the complex domain must be addressed.  Either the input symbols or
the spreading chips or both can take values in the complex number set.
In the complex-valued setting, the channel model \eref{e:sch} is
equivalent to the following real-valued one:
\begin{equation}
  \begin{bmatrix}
    \Y^\sr \\
    \Y^\si
  \end{bmatrix}
  =
  \begin{bmatrix}
    \S^\sr & -\S^\si \\
    \S^\si & \S^\sr
  \end{bmatrix}
  \begin{bmatrix}
    \X^\sr \\
    \X^\si
  \end{bmatrix}
  +
  \begin{bmatrix}
    \N^\sr \\
    \N^\si
  \end{bmatrix},
  \label{e:ch2}
\end{equation}
where the superscripts $\sr$ and $\si$ denote real and imaginary
components respectively.  Note that the previous analysis does not
apply to \eref{e:ch2} since the entries of the channel state matrix
are not i.i.d.\ in this case.

If the inputs take complex values but the spreading is real-valued
($\S^\si=0$), the channel can be regarded as two uses of the
real-valued channel $\S=\S^\sr$, where the inputs $\X^\sr$ and
$\X^\si$ to the two channels may be dependent.  Since independent
inputs maximize the channel capacity, there is little reason to
transmit dependent signals in the two subchannels.  Thus the analysis
of the real-valued channel in previous sections also applies to the
case of independent in-phase and quadrature components, while the only
change is that the spectral efficiency is the sum of that of the two
subchannels.

We can also compare the real-valued and the complex-valued channels
assuming the same real-valued input distribution.  Under the
complex-valued channel,
\begin{equation}
  \begin{bmatrix}
    \Y^\sr \\
    \Y^\si
  \end{bmatrix}
  =
  \begin{bmatrix}
    \S^\sr \\
    \S^\si
  \end{bmatrix}
  \X
  +
  \begin{bmatrix}
    \N^\sr \\
    \N^\si
  \end{bmatrix},
  \label{e:cri}
\end{equation}
which is equivalent to transmitting the same real-valued $\X$ twice
over the two component real-valued channels.  This is equivalent to
having a real-valued channel with the load $\beta$ halved.

If both the symbols and the spreading chips are complex-valued, the
analysis in the previous sections can be modified to take this into
account.  For convenience it is assumed that the real and imaginary
components of spreading chips, $S_{nk}^\sr$, $S_{nk}^\si$ are i.i.d.\
with zero mean and unit variance.  The noise vector has i.i.d.\
circularly symmetric Gaussian entries, i.e., $\expect {\N\herm{\N}}
=2\I$.  Thus the conditional probability density function of the
actual multiuser channel is
\begin{equation}
  p_{\Y|\X,\S}(\y|\x,\S) = (2\pi)^{-L} \expb{ -\frac{\ySx}{2} },
  \label{e:pyc}
\end{equation}
whereas that of the postulated channel is
\begin{equation}
  q_{\Y|\X,\S}(\y|\x,\S) = \tps^{-L} \expb{ -\frac{ \ySx }{2\sigma^2} }.
  \label{e:qyc}
\end{equation}
Also, the actual and the postulated input distributions $p_X$ and
$q_X$ have both zero-mean and unit variance, $\expect{|X|^2}
=\Exp_q\left\{ |X|^2\right\}=1$.  Note that the in-phase and the
quadrature components are intertwined due to complex spreading.

The replica analysis can be carried out in parallel to that in Section
\ref{s:prf}.  In the following we highlight the major differences.
Given $(\Snr,\XX)$, the variables $V_a$ defined in \eref{e:va} have
asymptotically independent real and imaginary components.  Thus,
$\Gu_K$ can be evaluated to be twice that under real-valued channels
with
\begin{equation}
  Q_{ab} = \oneon{K}\sumK \snr_k \Re\big\{ X_{ak} X_{bk}^* \big\},
  \quad a,b = 0,\dots,u.
\end{equation}
The rate $\Iu$ of the measure $\muuK$ of $\Q$ is obtained as
\begin{equation}
  \Iu(\Q) = \sup_{\tQ} \left[ \trace{\tQ\Q} -
    \log_e \expect{ \expbr{ \snr \XH\tQ\X } } \right].
\end{equation}
As a result, the fixed-point joint equations for $\Q$ and $\tQ$ are
\begin{subequations}
  \begin{eqnarray}
    \tQ &=& -\frac{2}{\beta} \inv{ \left( \inv{\mSigma}+\Q \right) },
    \label{e:tQQc} \\
    \Q &=& \frac{
        \expect{ \snr \X\XH \expbr{ \snr\XH\tQ\X } } }
        { \expect{ \expbr{ \snr\XH\tQ\X } } }.
    \label{e:QtQc}
  \end{eqnarray}
  \label{e:qtqc}%
\end{subequations}
Under replica symmetry \eref{e:rs}, the parameters $(c,d,f,g)$ are
found to be 2 times the corresponding values given in \eref{e:cdfg},
and $(r,m,p,q)$ are found the same as in \eref{e:rmpq} except that all
squares are replaced by squared norms.  By defining two parameters
(which differ from \eref{e:exdf} by a factor of 2):
\begin{equation}
  \eff = \frac{d^2}{f} \quad \text{and} \quad  \xi = d,
  \label{e:exdfc}%
\end{equation}
we have the following result.

\begin{claim}  \label{th:cc}
  Let the  multiuser posterior mean estimate of the
  complex-valued multiple-access channel \eref{e:pyc} with
  complex-valued spreading be $\sm{\X}_q$ parameterized by a
  postulated input distribution $q_X$ and noise level $\sigma$.  Then,
  in the large-system limit, the distribution of the multiuser
  detection output $\sm{X_k}_q$ conditioned on $X_k=x$ being
  transmitted with signal-to-noise ratio $\snr_k$ is identical to the
  distribution of the  estimate $\sm{X}_q$ of a
  single-user complex Gaussian channel
\begin{equation}
  Z = \sqrt{\snr} \, X + \oneon{\sqrt{\eff}} \, N
  \label{e:ccc}
\end{equation}
conditioned on $X=x$ being transmitted with $\snr=\snr_k$, where $N$
is circularly symmetric Gaussian with unit variance, $\expect{
  |N|^2}=1$.  The multiuser efficiency\index{Multiuser efficiency}
$\eff$ and the inverse noise variance\index{Inverse noise variance}
$\xi$ of the postulated single-user channel \eref{e:qyc} satisfy the
coupled equations \eref{e:ex}, where the mean-square error
$\mse(\snr;\eff,\xi)$ of the posterior mean estimate and the variance
$\vrc(\snr;\eff,\xi)$ of the retrochannel are defined similarly as
that of the real-valued channel, with the squares in \eref{e:gmse} and
\eref{e:vrc} replaced by squared norms.  In case of multiple solutions
to \eref{e:ex}, $(\eff,\xi)$ are chosen to minimize the free energy:
\begin{equation}
  \begin{split}
    \nsp{1}\fe =& - \expect{ \int\nsp{1} \pz{z} \log \qz{z} \intd z } \\
    & \quad + \oneon{\beta} [(\xi-1)\loge-\log\xi]
    + \log\frac{\xi}{\pi} - \frac{\xi}{\eff}\loge \\
    & \quad + \frac{\sigma^2\xi(\eff-\xi)}{\beta\eff}\loge
    + \oneon{\beta}\log(2\pi) + \frac{\xi}{\beta\eff}\loge.
  \end{split}
\end{equation}
\end{claim}

\begin{corollary}
  For the complex-valued channel \eref{e:pyc}, the mutual information
  of the single-user channel seen at the multiuser posterior mean
  estimator output for a user with signal-to-noise ratio $\snr$ takes
  the same formula as \eref{e:id}:
\begin{equation}
  I(\eff\,\snr) = \Dzx.
  \label{e:cec}
\end{equation}
where $\eff$ is the multiuser efficiency given by Claim \ref{th:cc}
and $p_{Z|\snr;\eff}$ is the marginal probability distribution of the
output of channel \eref{e:ccc}.  The overall spectral efficiency under
suboptimal separate decoding is $\Csep(\beta) = \beta\, \expect{
  I(\eff\,\snr) }.$
\label{cor:cc}
\end{corollary}

\begin{claim}
The optimal spectral efficiency under joint decoding is
\begin{equation}
  \Cjnt(\beta) = \beta\, \expect{ I(\eff\,\snr) } + (\eff-1)\loge -\log\eff,
  \label{e:cjc}
\end{equation}
where $\eff$ is the optimal multiuser efficiency determined by Claim
\ref{th:cc} by postulating a measure $q$ that is identical to $p$.
\end{claim}

It is interesting to compare the performance of the real-valued
channel and that of the complex-valued channel.  We assume the
in-phase and quadrature components of the input symbols are
independent with identical distribution $p_X'$ which has a variance of
$\half$.  By Claim \ref{th:cc}, the equivalent single-user channel
\eref{e:ccc} can also be regarded as two independent subchannels.  The
mean-square error and the variance in \eref{e:ex} are the sum of those
of the subchannels.  It can be checked that the performance of each
subchannel is identical to that of the real-valued channel with input
distribution $p_X'$ normalized to unit variance.  Note, however, that
the total transmit energy in case of complex spreading take twice the
energy of their real counterparts.  In all, the error performance
under complex-valued spreading is exactly the same as those under
real-valued spreading.  This result simplifies the analysis of
complex-valued channels such as those arise in multiantenna systems.
If we have control over the channel state matrix, as in CDMA systems,
complex-valued spreading should be avoided due to higher complexity
with no direct performance gain.

\section{Numerical Results}
\label{s:nr}

\begin{figure}
  \begin{center} \includegraphics[width=\myfigwidth]{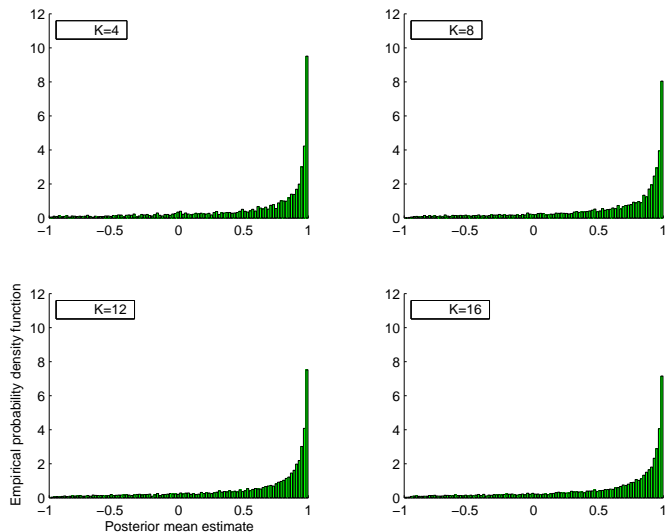}
  \caption{The empirical probability density functions of the posterior
    mean estimates with binary input conditioned on ``+1'' being
    transmitted.}
  \label{f:xp}
  \end{center}
\end{figure}

\begin{figure}
  \begin{center} \includegraphics[width=\myfigwidth]{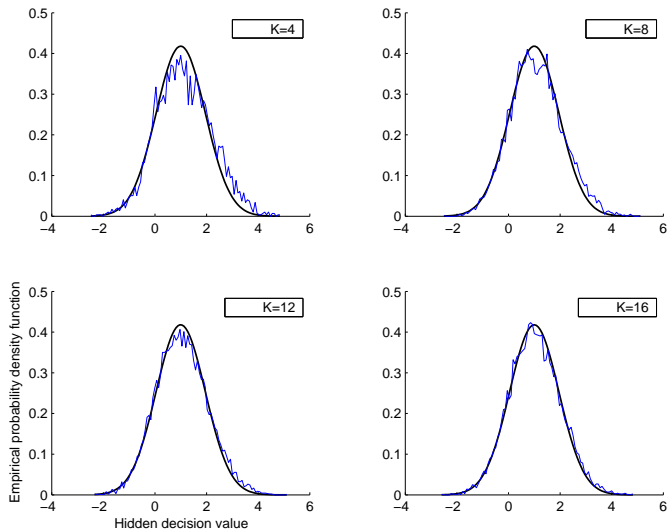}
  \caption{The empirical probability density functions of the
    ``hidden'' Gaussian statistic with binary input conditioned on
    ``+1'' being transmitted.  The asymptotic Gaussian distribution
    predicted by the decoupling principle is also plotted for
    comparison.}
  \label{f:zp}
  \end{center}
\end{figure}

Figures \ref{f:xp}--\ref{f:zp} plot the simulated distribution of the
posterior mean estimate and its corresponding ``hidden'' Gaussian
statistic.  Equal-power users with binary input are considered.  We
simulate CDMA systems of 4, 8, 12 and 16 users respectively.  The load
is fixed to $\beta=2/3$ and the SNR is 2 dB.  Let $X_k=1$ be
transmitted by all users.  We collect the output decision statistics
of the posterior mean estimator (i.e., the soft output of the
individually optimal detector, $\sm{X_k}$) out of 1000 trials.  A
histogram of the statistic is obtained and then scaled to plot an
estimate of the probability density function in Figure \ref{f:xp}.  We
also apply the inverse nonlinear decision function to recover the
``hidden'' Gaussian decision statistic (normalized so that its
conditional mean is equal to $X_k=1$), which in this case is
\begin{equation}
        \tilde{Z}_k = \frac{ \tanh^{-1}(\sm{X_k}) }{ \eff\,\snr_k }.
\end{equation}
The probability density function of $\tilde{Z}_k$ estimated from its
histogram is then compared to the theoretically predicted Gaussian
density function in Figure \ref{f:zp}.  It is clear that even though
the PME output $\sm{X_k}$ takes a non-Gaussian distribution, the
equivalent statistic $\tilde{Z}_k$ converges to a Gaussian
distribution centered at $X_k$ as $K$ becomes large.  This result is
particularly desirable considering that the ``fit'' to the Gaussian
distribution is quite good even for a system with merely 8 users.


\begin{figure*}
  \begin{center}
    \subfigure[]{
    \includegraphics[width=\myhalfwidth]{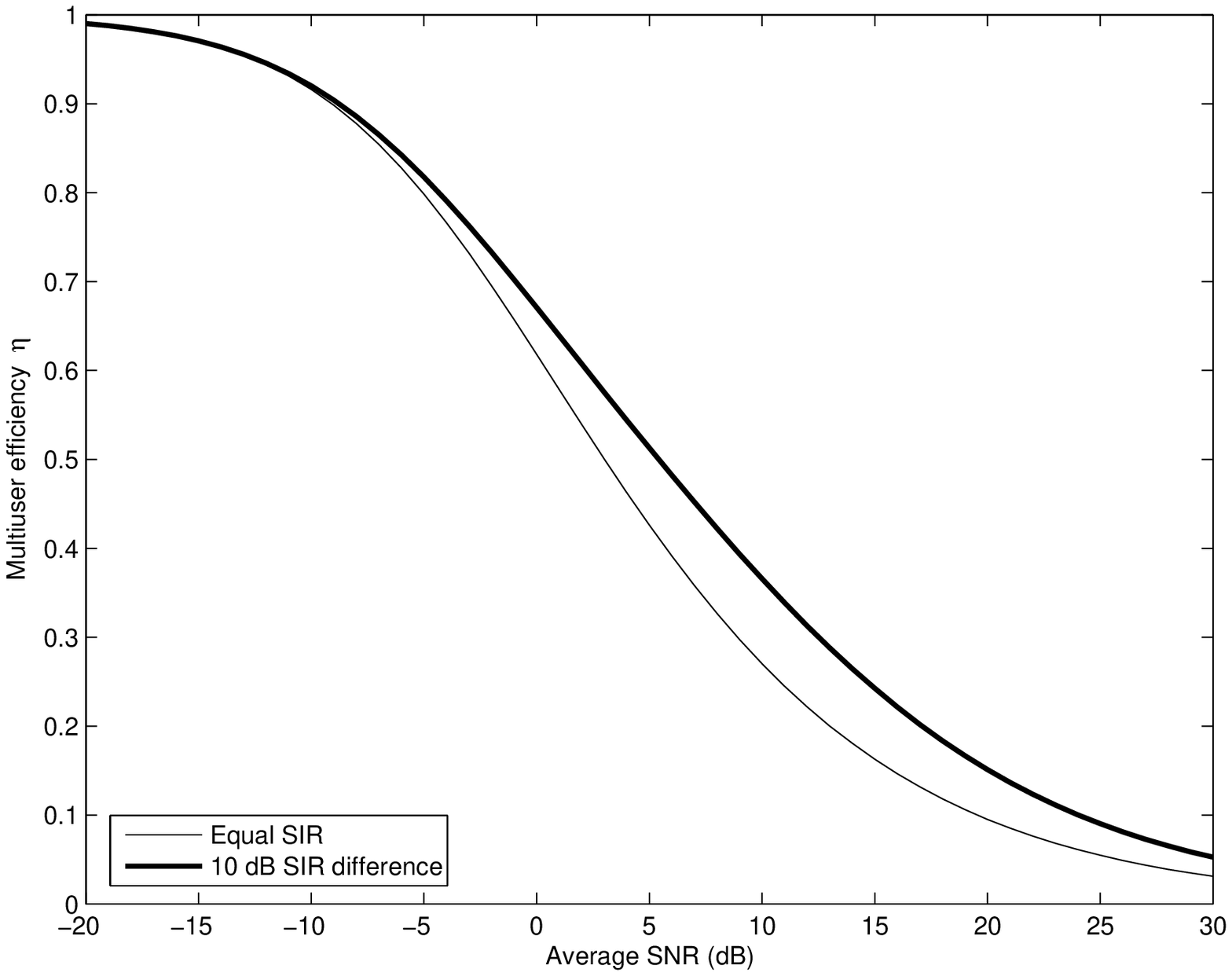}
    \label{f:men1}
    }
    \subfigure[]{
    \includegraphics[width=\myhalfwidth]{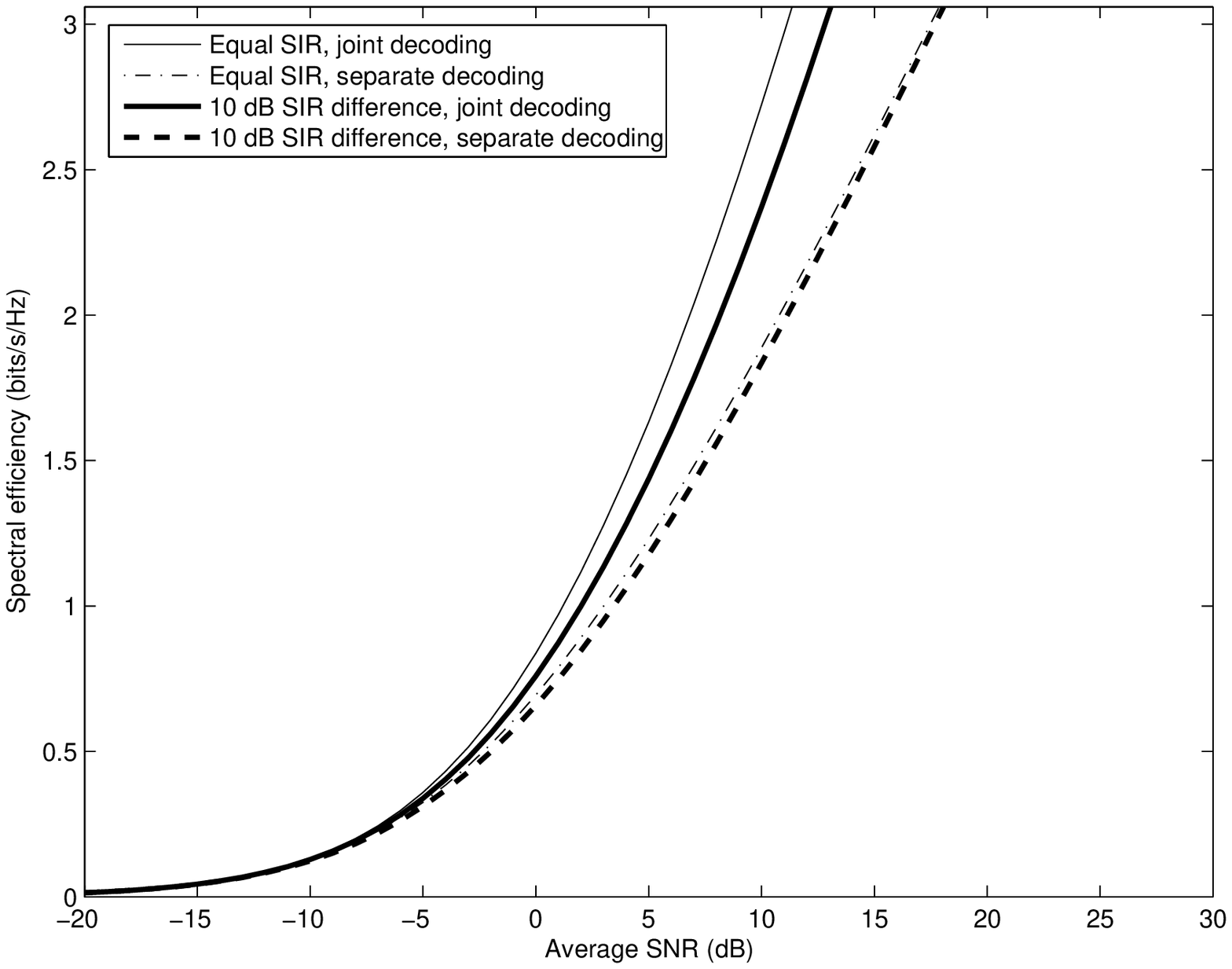}
    \label{f:sen1}
    }
    \subfigure[]{
    \includegraphics[width=\myhalfwidth]{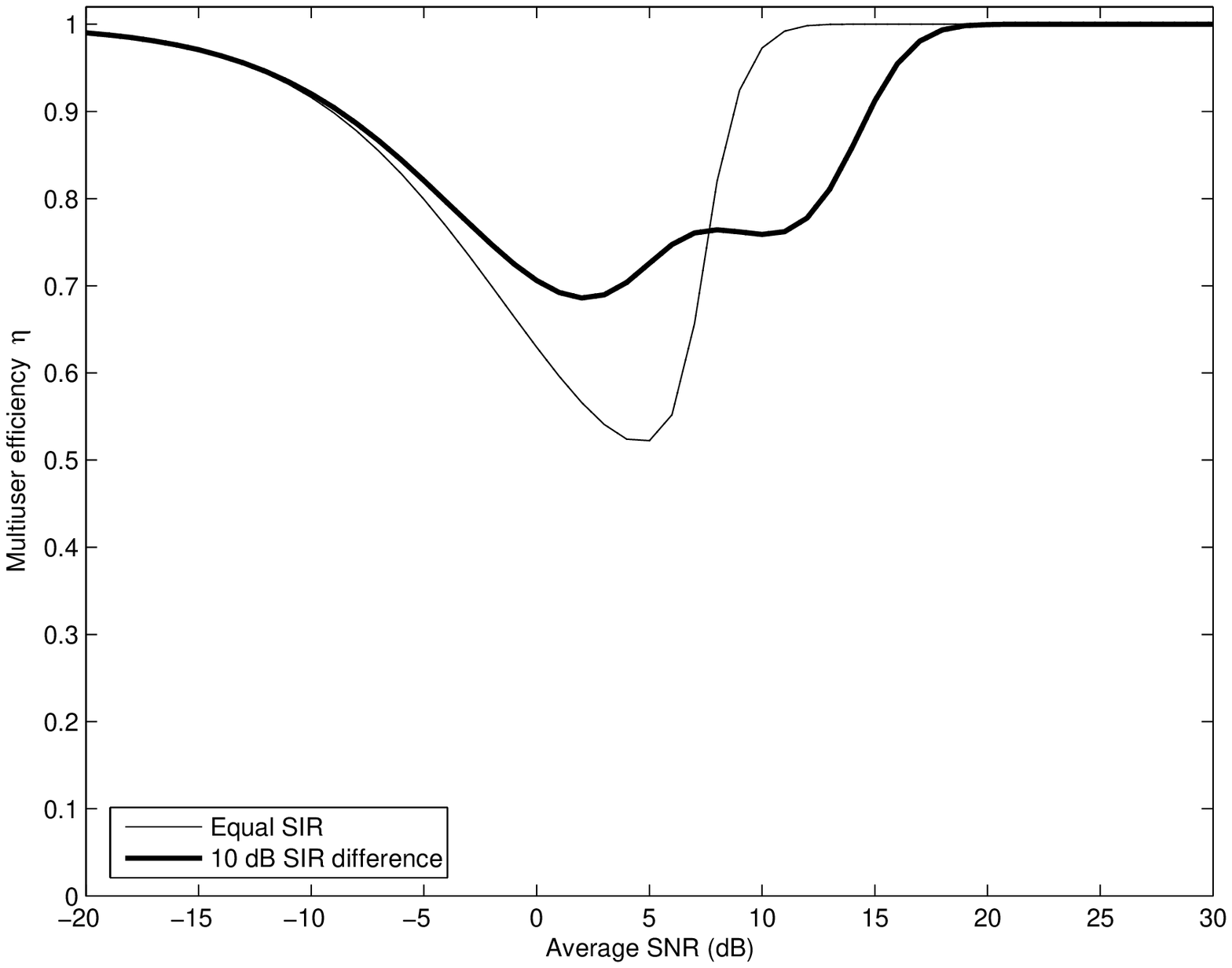}
    \label{f:meb1}
    }
    \subfigure[]{
    \includegraphics[width=\myhalfwidth]{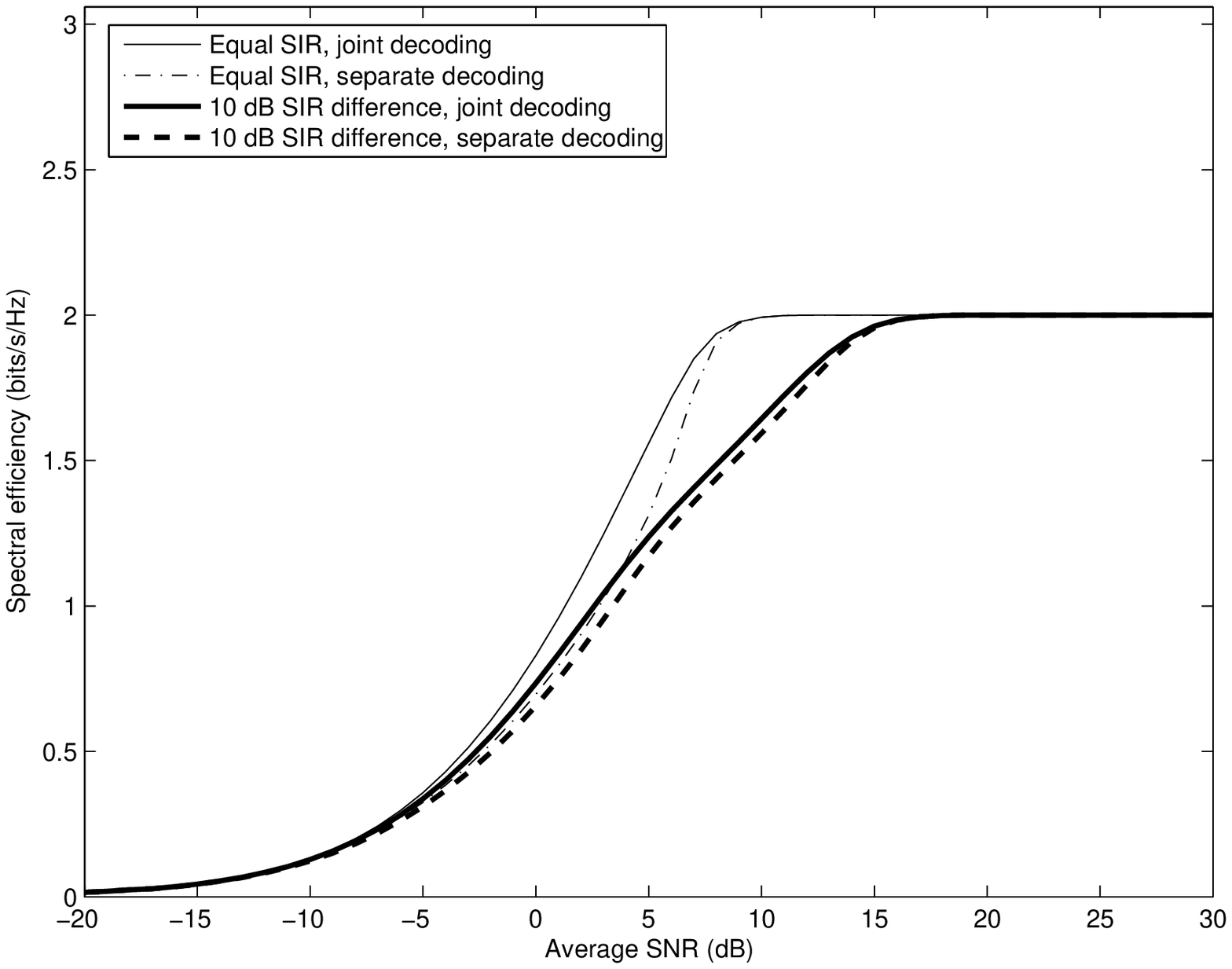}
    \label{f:seb1}
    }
    \subfigure[]{
    \includegraphics[width=\myhalfwidth]{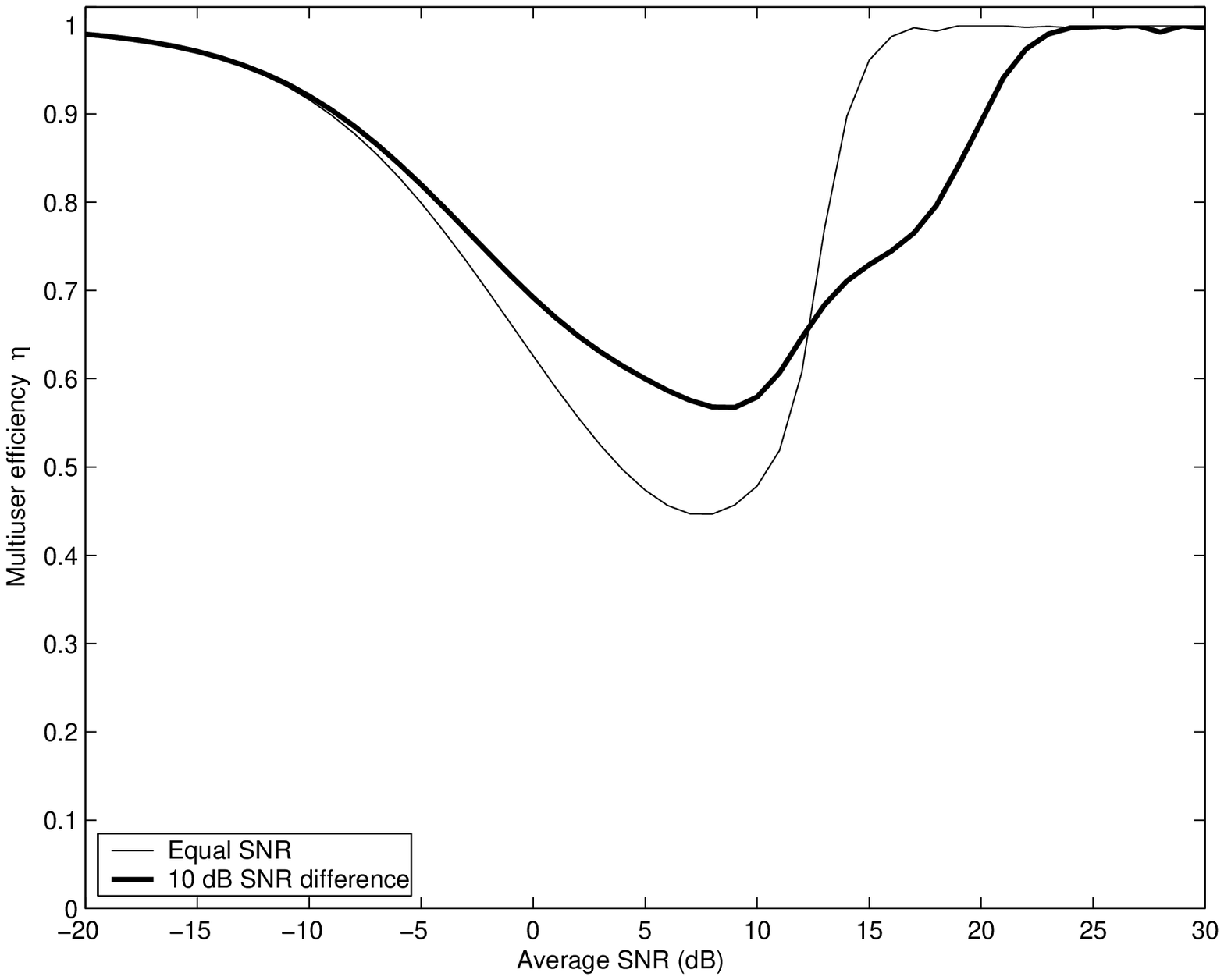}
    \label{f:me81}
    }
    \subfigure[]{
    \includegraphics[width=\myhalfwidth]{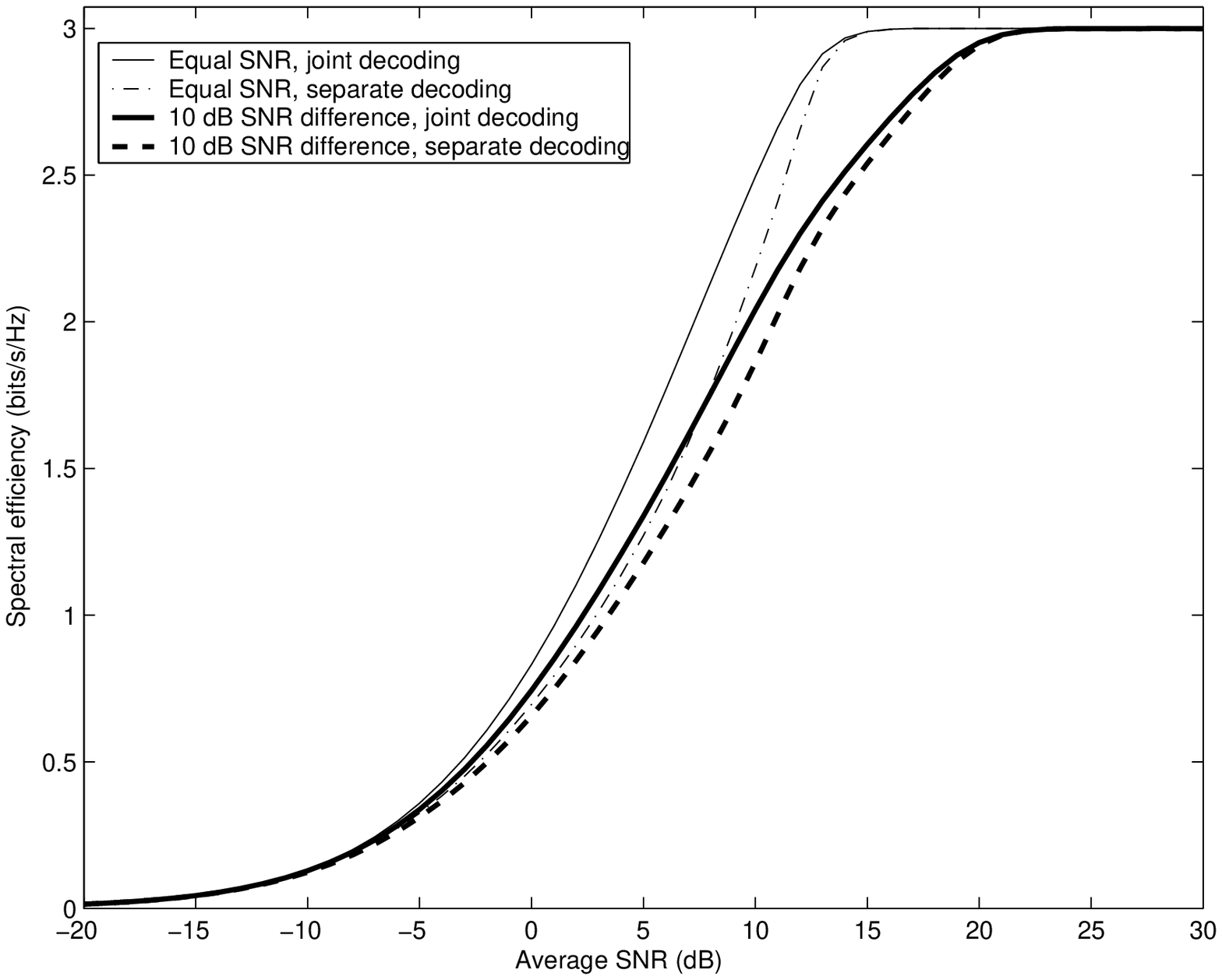}
    \label{f:se81}
    }
    \caption{Multiuser efficiency and spectral efficiency as
      functions of SNR.  The load is $\beta=1$.  (a) Multiuser
      efficiency, complex Gaussian inputs.  (b) Spectral efficiency,
      complex Gaussian inputs.  (c) Multiuser efficiency, QPSK inputs.
      (d) Spectral efficiency, QPSK inputs.  (e) Multiuser efficiency,
      8PSK inputs.  (f) Spectral efficiency, 8PSK inputs.}
    \label{f:1}
  \end{center}
\end{figure*}

\begin{figure*}
  \begin{center}
    \subfigure[]{
    \includegraphics[width=\myhalfwidth]{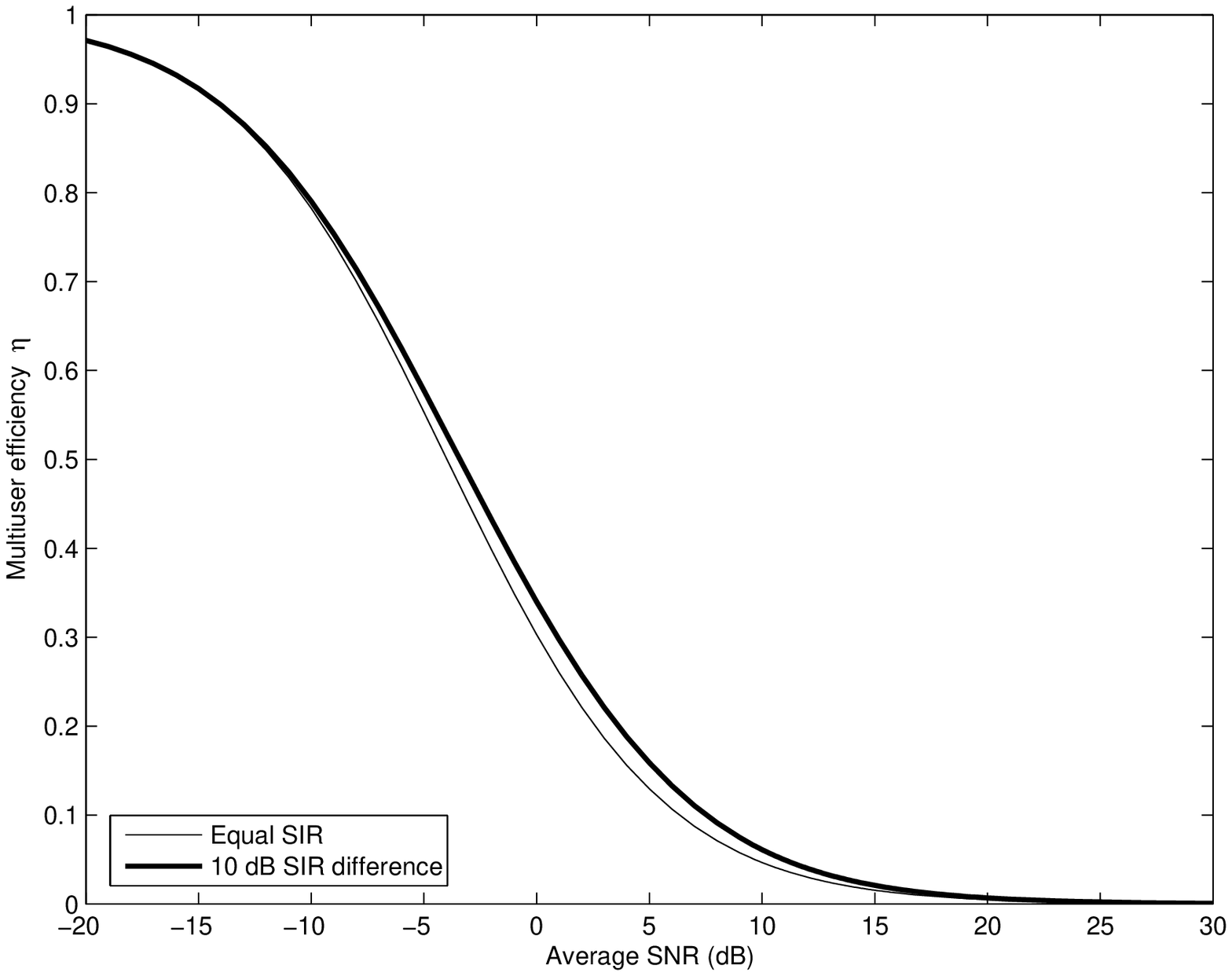}
  \label{f:men3}
    }
    \subfigure[]{
    \includegraphics[width=\myhalfwidth]{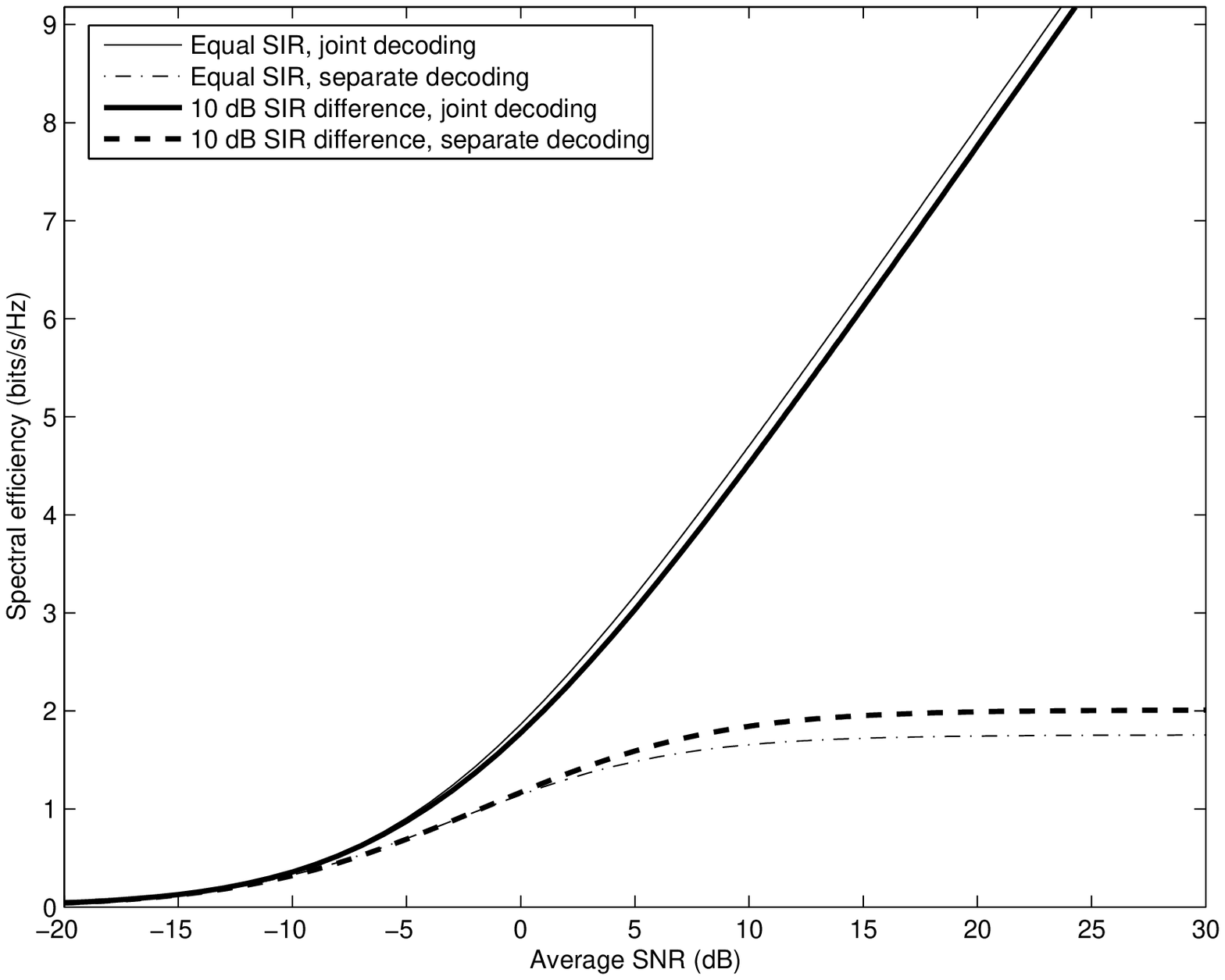}
  \label{f:sen3}
    }
    \subfigure[]{
    \includegraphics[width=\myhalfwidth]{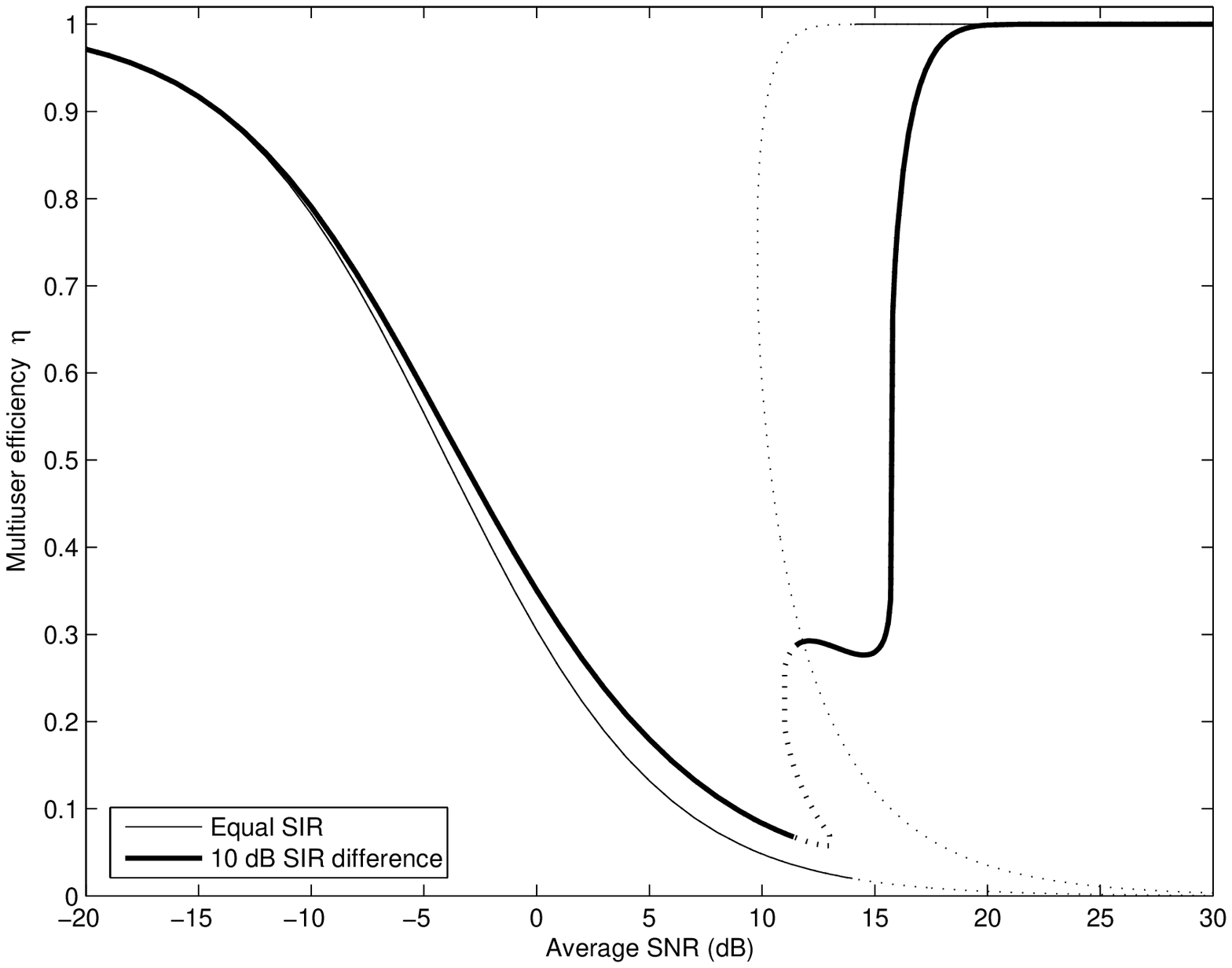}
  \label{f:meb3}
    }
    \subfigure[]{
    \includegraphics[width=\myhalfwidth]{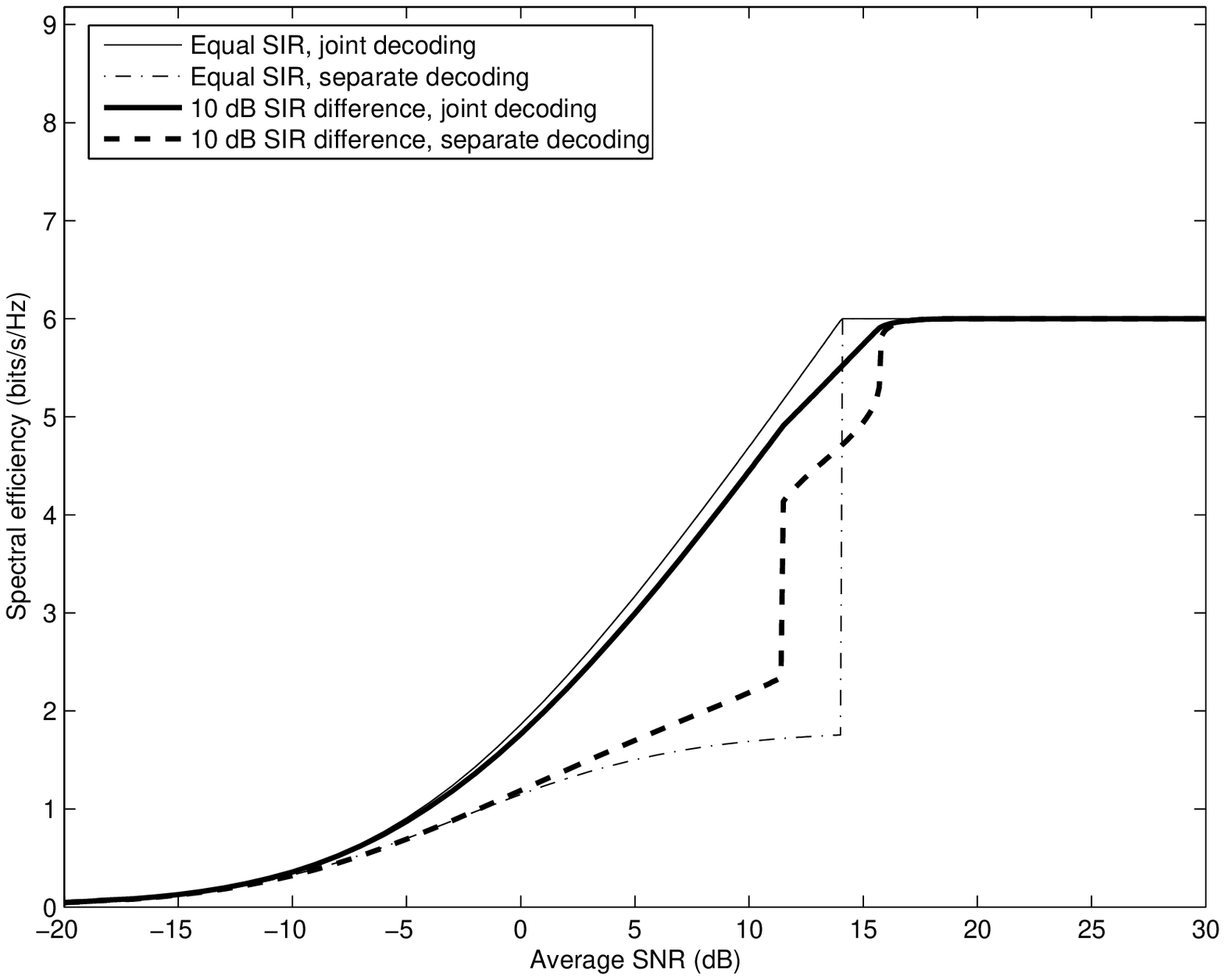}
  \label{f:seb3}
    }
    \subfigure[]{
    \includegraphics[width=\myhalfwidth]{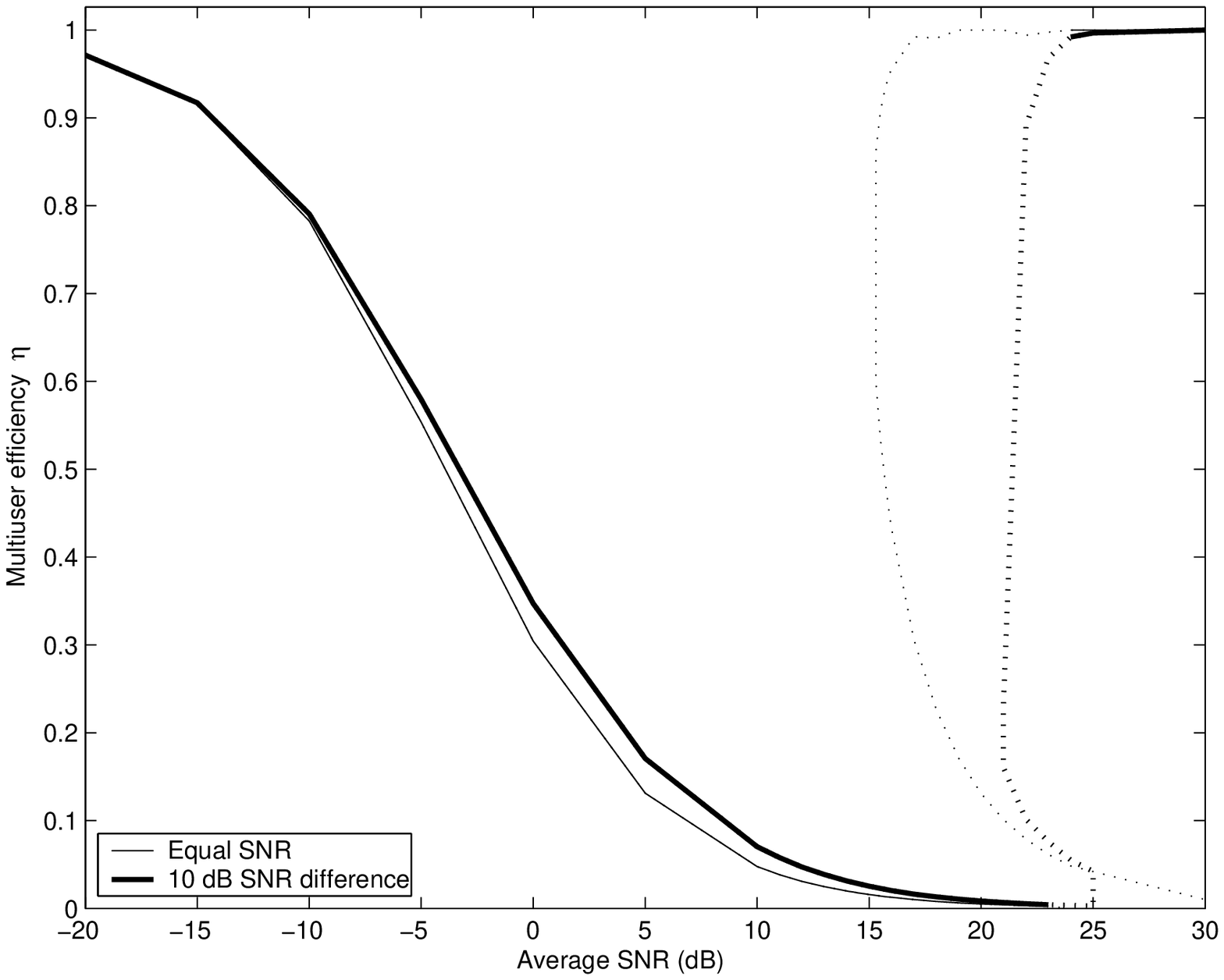}
    \label{f:me83}
    }
    \subfigure[]{
    \includegraphics[width=\myhalfwidth]{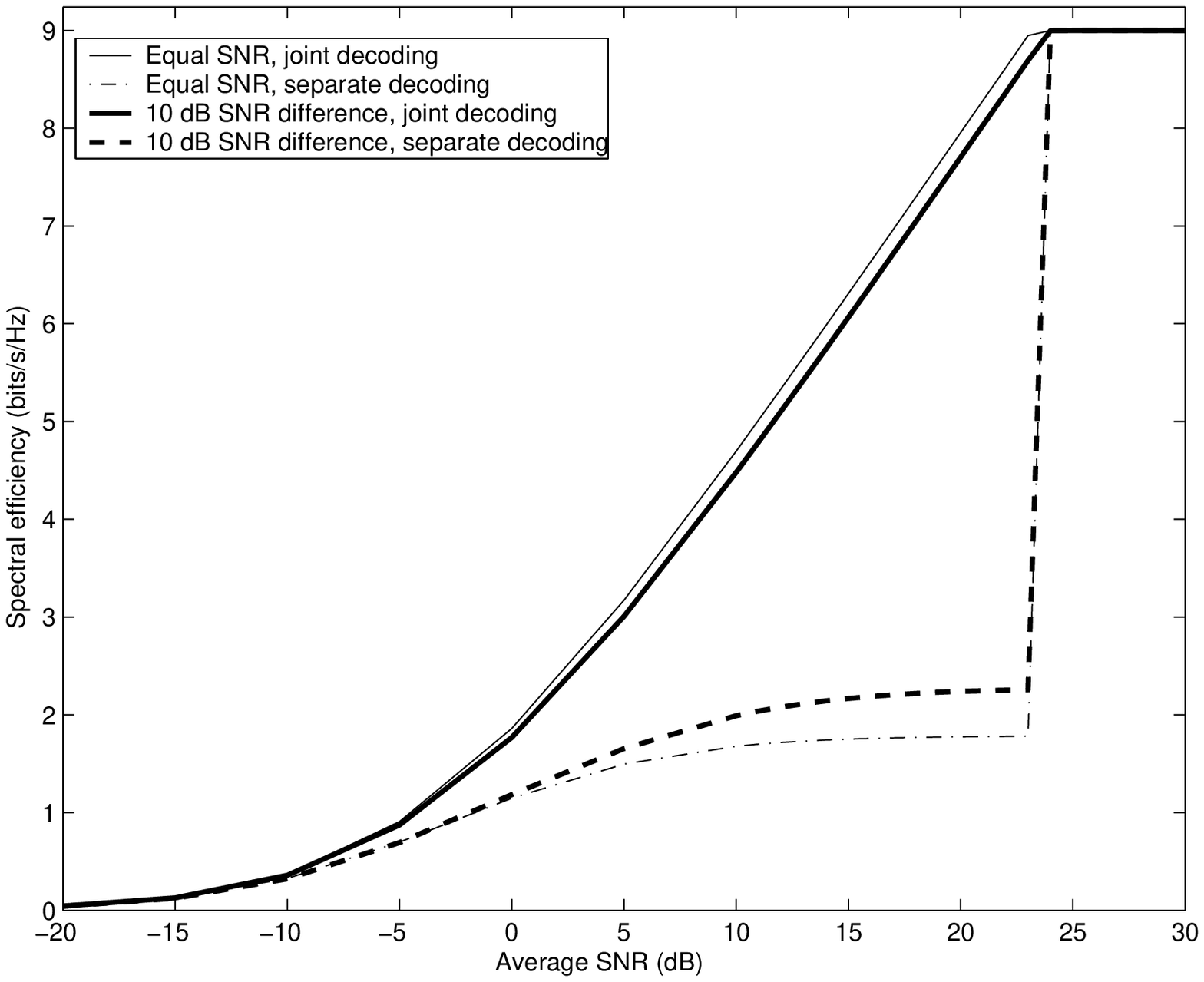}
    \label{f:se83}
    }
    \caption{Multiuser efficiency and spectral efficiency as
      functions of SNR.  The load is $\beta=3$.  (a) Multiuser
      efficiency, complex Gaussian inputs.  (b) Spectral efficiency,
      complex Gaussian inputs.  (c) Multiuser efficiency, QPSK inputs.
      (d) Spectral efficiency, QPSK inputs.  (e) Multiuser efficiency,
      8PSK inputs.  (f) Spectral efficiency, 8PSK inputs.}
    \label{f:3}
  \end{center}
\end{figure*}

In Figures \ref{f:1}--\ref{f:3}, multiuser efficiency\index{Multiuser
  efficiency} and spectral efficiency\index{Spectral efficiency} are
plotted as functions of the average SNR.  We consider three input
distributions, namely, QPSK, 8PSK, and complex Gaussian inputs.
Complex-valued spreading is assumed, where the multiuser efficiency
and the spectral efficiency are given by Claim \ref{th:cc} and
Corollary \ref{cor:cc} respectively.  We also consider two SNR
distributions:\index{SNR distribution} 1) identical SNRs for all users
(perfect power control), and 2) two groups of users of equal
population with a power difference of 10 dB.  We first assume a system
load of $\beta=1$ and then redo the experiments with $\beta=3$.

In Figure \ref{f:men1}, multiuser efficiency under complex Gaussian
inputs and linear MMSE detection is plotted as a function of the
average SNR.  The load is $\beta=1$.  We find the multiuser
efficiencies decrease from 1 to 0 as the SNR increases.  The
monotonicity can be easily verified by inspecting the Tse-Hanly
equation \eref{e:th}.\index{Tse-Hanly equation} Transmission with
unbalanced power improves the multiuser efficiency.  The corresponding
spectral efficiencies of the system are plotted in Figure
\ref{f:sen1}.  Both joint decoding and separate decoding are
considered.  The gain in the spectral efficiency due to joint decoding
is small for low SNR but significant for high SNR.  Unbalanced SNR
reduces the spectral efficiency, where under separate decoding the
loss is almost negligible.

Multiuser efficiency with QPSK inputs and nonlinear MMSE (individually
optimal) detection is plotted in Figure \ref{f:meb1}.  Note that this
function is not monotonic: it converges to 1 for both vanishing SNR
and infinite SNR.  While for vanishing SNR this follows directly from
the definition of multiuser efficiency, the convergence to unity as
the SNR goes to infinity was shown in \cite{TseVer00IT} for the case
of binary inputs.  A single dip is observed for the case of identical
SNRs while two dips are observed in the case of two SNRs of equal
population with 10 dB difference in SNR (the gap is about 10 dB).  The
corresponding spectral efficiencies are plotted in Figure
\ref{f:seb1}.  The spectral efficiencies saturate to 1 bit/s/dimension
at high SNR.  The difference between joint decoding and separate
decoding is quite small for both very low and very high SNRs while it
can be 30\% at around 6 dB.

Multiuser efficiency under 8PSK inputs and nonlinear MMSE
detection is plotted in Figure \ref{f:me81}.  The multiuser efficiency
curve is slightly better than that for QPSK inputs.  The corresponding
spectral efficiencies are plotted in Figure \ref{f:se81}.  The spectral
efficiencies saturate to 3 bit/s/dimension at hight SNR.

In Figure \ref{f:3}, we redo the previous experiments only with a
different system load $\beta=3$.  The results are to be compared with
those in Figure \ref{f:1}.

For Gaussian inputs, the multiuser efficiency curves in Figure
\ref{f:men3} take a similar shape as in Figure \ref{f:men1}, but are
significantly lower due to higher load.  The corresponding spectral
efficiencies are shown in Figure \ref{f:sen3}.  It is clear that
higher load results in higher spectrum usage under joint decoding.
Separate decoding, however, is interference limited and the spectral
efficiency saturates under high SNR (cf.\ \cite[Figure
1]{VerSha99IT}).

Figure \ref{f:meb3} plots multiuser efficiency under QPSK inputs.
All solutions to the fixed-point equation \eref{e:ecme} of the
multiuser efficiency are shown.  Under equal SNR, multiple solutions
coexist for an average SNR of 10 dB or higher.  If two groups of users
with 10 dB difference in SNR, multiple solutions are seen between 11
to 13 dB.  The solution that minimizes the free energy is valid and is
shown in solid lines, while invalid solutions are plotted using dotted
lines.  An almost 0 to 1 jump is observed under equal SNR and a much
smaller jump is seen under unbalanced SNRs.  This is known as phase
transition.\index{Phase transition} The asymptotics under equal SNR
can be shown by taking the limit $\snr\rightarrow\infty$ in
\eref{e:meb}.  Essentially, if $\eff\snr\rightarrow\infty$, then
$\eff\rightarrow1$; while if $\eff\snr\rightarrow\tau$ where $\tau$ is
the solution to
\begin{equation}
  \tau \int \oostp e^{-\frac{z^2}{2}} [ 1 - \tanh(\tau-z\sqrt{\tau}) ]
  \intd z = \oneon{\beta},
  \label{e:tau}
\end{equation}
then $\eff\rightarrow0$.  If $\beta>2.085$, there exists a solution to
\eref{e:tau} so that two solutions coexist for large SNR.

The spectral efficiency under QPSK inputs and $\beta=3$ is shown in
Figure \ref{f:seb3}.  As a result of phase transition, one observes a
jump to saturation in the spectral efficiency under equal-power
inputs.  The gain due to joint decoding can be significant in moderate
SNRs.  In case of two groups of users with 10 dB difference in SNR,
the spectral efficiency curve also shows one jump and the loss due to
separate decoding is reduced significantly for a small window of SNRs
around the areas of phase transition (11--13 dB).  Therefore, perfect
power control may not be the best strategy in such cases.

Under 8PSK inputs, the multiuser efficiency and spectral efficiency
curves in Figure \ref{f:me83} and \ref{f:se83} take similar shape as
the curves under QPSK inputs.  Phase transition causes jumps in both
the multiuser efficiency and the spectral efficiency.

In Figures \ref{f:seb3} and \ref{f:se83} a sharp bend upward is
observed at the point of phase transition.  This is known as
``spinodal'' in statistical physics.

A comparison of Figures \ref{f:sen3}, \ref{f:seb3} and \ref{f:se83}
shows that under separate decoding, the spectral efficiency under
Gaussian inputs saturates well below that of QPSK and 8PSK inputs.

\section{Conclusion}
\label{s:con}

The main contribution of this paper is a simple characterization of
the performance of CDMA multiuser detection under arbitrary input
distribution and SNR (and/or flat fading) in the large-system limit.
A broad family of multiuser detectors is studied under the name of
posterior mean estimators, which includes well-known detectors such as
the matched filter, decorrelator, linear MMSE detector, maximum
likelihood (jointly optimal) detector, and the individually optimal
detector.

A key conclusion is the decoupling of a Gaussian multiuser channel
concatenated with a generic multiuser detector front end.  It is found
that the multiuser detection output for each user is a deterministic
function of a hidden Gaussian statistic centered at the transmitted
symbol.  Hence the single-user channel seen at the multiuser detection
output is equivalent to a Gaussian channel in which the overall effect
of multiple-access interference is a degradation in the effective
signal-to-interference ratio.  The degradation factor, known as the
multiuser efficiency, is the solution to a pair of coupled fixed-point
equations, and can be easily computed numerically if not analytically.

Another set of results, tightly related to the decoupling principle,
lead to general formulas for the large-system spectral efficiency of
multiuser channels expressed in terms of the multiuser efficiency,
both under joint and separate decoding.  It is found that the
decomposition of optimum spectral efficiency as a sum of single-user
efficiencies and a joint decoding gain applies under more general
conditions than shown in \cite{ShaVer01IT}, thereby validating
M\"uller's conjecture \cite{Muller02WCIT}.  A relationship between the
spectral efficiencies under joint and separate decoding is one of the
applications of a recent basic formula that links the mutual
information and the MMSE \cite{GuoSha05IT}.

From a practical viewpoint, this paper presents new results on the
efficiency of CDMA communication under arbitrary input signaling such
as $m$-PSK and $m$-QAM with an arbitrary power profile.  More
importantly, the results in this paper allow the performance of
multiuser detection to be characterized by a single parameter, the
multiuser efficiency.  The efficiency of spectrum usage is also easily
quantified by means of this parameter.  Thus, the results offer
convenient performance measures and valuable insights in the design
and analysis of CDMA systems, e.g., in power control
\cite{MesGuo04SPWC}.

The linear system in our study also models multiple-input
multiple-output channels under various circumstances.  The results can
thus be used to evaluate the output SNR or spectral efficiency of
high-dimensional MIMO channels (such as multiple-antenna systems) with
arbitrary signaling and various detection techniques.  Some of the
results in this paper have been generalized to MIMO channels with
spatial correlation at both transmitter and receiver sides
\cite{WenWon04ISIT}.

\section{Acknowledgements}

The authors are grateful to the anonymous referees for their helpful
reviews.  Thanks also to Ralf M\"uller, Toshiyuki Tanaka and Chih-Chun
Wang for interesting discussions.


\end{document}